\documentclass[longauth]{aa} 

\usepackage{graphicx}
\usepackage{txfonts}
\usepackage{natbib}
\usepackage{longtable}
\usepackage{lscape}
\usepackage{color}

\providecommand{\sorthelp}[1]{}

\begin{document}
 
\title{
{\it Herschel} and SCUBA-2 observations of dust emission in a sample of {\it Planck} cold clumps
\thanks{
{\it Planck} \emph{(http://www.esa.int/Planck)} is a project of the European Space
Agency -- ESA -- with instruments provided by two scientific consortia funded by ESA
member states (in particular the lead countries: France and Italy) with contributions
from NASA (USA), and telescope reflectors provided in a collaboration between ESA and a
scientific consortium led and funded by Denmark.
}
\thanks{{\it Herschel} is an ESA space observatory with science instruments provided
by European-led Principal Investigator consortia and with important
participation from NASA.}
}

\author{Mika     Juvela\inst{1},
Jinhua      He\inst{2,3,4,5},
Katherine   Pattle\inst{6},
Tie         Liu\inst{7,8},
George      Bendo\inst{9},
David J.    Eden\inst{10},
Orsolya     Feh\'er\inst{11,12},
Michel      Fich\inst{13},
Gary        Fuller\inst{9},
Naomi       Hirano\inst{14},
Kee-Tae     Kim\inst{7},
Di          Li\inst{15,16},
Sheng-Yuan  Liu\inst{14},
Johanna     Malinen\inst{17},
Douglas J.  Marshall\inst{18},        
Deborah     Paradis\inst{19,20}
Harriet     Parsons\inst{8},
Veli-Matti  Pelkonen\inst{1},
Mark G.     Rawlings\inst{8},
Isabelle    Ristorcelli\inst{19,20},
Manash R.   Samal\inst{21},
Ken'ichi    Tatematsu\inst{22},
Mark        Thompson\inst{23},
Alessio     Traficante\inst{24},
Ke          Wang\inst{25},
Derek       Ward-Thompson\inst{6},
Yuefang     Wu\inst{26},
Hee-Weon    Yi\inst{27},
Hyunju      Yoo\inst{28}
}

\institute{
Department of Physics, P.O.Box 64, FI-00014, University of Helsinki,
Finland, {\em mika.juvela@helsinki.fi}
\and 
Yunnan Observatories, Chinese Academy of Sciences, 396
Yangfangwang, Guandu District, Kunming, 650216, China
\and 
Chinese Academy of Sciences South America Center for Astronomy,
China-Chile Joint Center for Astronomy, Camino El Observatorio \#1515,
Las Condes, Santiago, Chile
\and 
Key Laboratory for the Structure and Evolution of Celestial
Objects, Chinese Academy of Sciences, 396 Yangfangwang, Guandu
District, Kunming, 650216, China
\and 
Center for Astronomical Mega-Science, Chinese Academy of Sciences,
20A Datun Road, Chaoyang District, Beijing, 100012, China
\and 
Jeremiah Horrocks Institute, University of Central Lancashire,
Preston, PR1 2HE, United Kingdom
\and 
Korea Astronomy and Space Science Institute, 776
Daedeokdaero, Yuseong-gu, Daejeon 34055, Republic of Korea
\and 
East Asian Observatory, 660 N. A'oh\={o}k\={u} Place, 
Hilo, Hawaii 96720-2700, USA
\and 
UK ALMA Regional Centre Node, Jodrell Bank Centre for Astrophysics,
The University of Manchester, Oxford Road, Manchester M13 9PL, UK 
\and 
Astrophysics Research Institute, Liverpool John Moores University,
Ic2, Liverpool Science Park, 146 Brownlow Hill, Liverpool, L3 5RF, UK
\and 
Konkoly Observatory, Research Centre for Astronomy and Earth Sciences,
Hungarian Academy of Sciences, H-1121 Budapest, Konkoly Thege Mikl\'os
\'ut 15-17, Hungary
\and 
E\"ov\"os Lor\'and University, Department of Astronomy, P\'azm\'any P\'eter
s\'et\'any 1/A, 1117 Budapest, Hungary
\and 
Department of Physics and Astronomy, University of Waterloo, Waterloo,
Ontario, Canada N2L 3G1
\and 
Institute of Astronomy and Astrophysics, Academia Sinica.
11F of Astronomy-Mathematics Building, AS/NTU No.1, Sec. 4, Roosevelt
Rd, Taipei 10617, Taiwan, R.O.C.
\and 
National Astronomical Observatories, Chinese Academy of
Sciences, Beijing, 100012, China
\and 
Key Laboratory of Radio Astronomy, Chinese Academy of
Science, Nanjing 210008, China
\and 
Institute of Physics I, University of Cologne, Germany
\and 
Laboratoire AIM, IRFU/Service d’Astrophysique - CEA/DSM - CNRS -
Université Paris Diderot, Bât. 709, CEA-Saclay, F-91191,
Gif-sur-Yvette Cedex, France
\and 
Universit\'e de Toulouse, UPS-OMP, IRAP, F-31028 Toulouse cedex 4, France   
\and 
CNRS, IRAP, 9 Av. colonel Roche, BP 44346, F-31028 Toulouse cedex 4, France  
\and 
Graduate Institute of Astronomy, National Central University 300,
Jhongli City, Taoyuan County 32001, Taiwan
\and 
Nobeyama Radio Observatory, National Astronomical Observatory of
Japan, National Institutes of Natural Sciences, 462-2 Nobeyama,
Minamimaki, Minamisaku, Nagano 384-1305, Japan
\and 
Centre for Astrophysics Research, School of Physics
Astronomy \& Mathematics, University of Hertfordshire, College Lane,
Hatfield, AL10 9AB, UK
\and 
IAPS - INAF, via Fosso del Cavaliere, 100, I-00133 Roma, Italy
\and 
European Southern Observatory, Karl-Schwarzschild-Str.2,
D-85748 Garching bei M\"{u}nchen, Germany
\and 
Department of Astronomy, Peking University, 100871,
Beijing China
\and 
School of Space Research, Kyung Hee University, 1732, Deogyeong-daero,
Giheung-gu, Yongin-si, Gyeonggi-do, Republic of Korea
\and 
Department of Astronomy and Space Science, Chungnam National
University, 99 Daehak-ro,
Yuseong-gu, Daejeon 34134, Korea
}

\authorrunning{M. Juvela et al.}

\date{Received September 15, 1996; accepted March 16, 1997}

\abstract { 
Analysis of all-sky $Planck$ submillimetre observations and the IRAS
100\,$\mu$m data has led to the detection of a population of Galactic
cold clumps. The clumps can be used to study star formation and dust
properties in a wide range of Galactic environments.
} 
{
Our aim is to measure dust spectral energy distribution (SED)
variations as a function of the spatial scale and the wavelength.
}
{
We examine the SEDs at large scales using IRAS, $Planck$, and
$Herschel$ data. At smaller scales, we compare with JCMT/SCUBA-2
850\,$\mu$m maps with $Herschel$ data that are filtered using the
SCUBA-2 pipeline. Clumps are extracted using the Fellwalker method and
their spectra are modelled as modified blackbody functions.
}
{
According to IRAS and $Planck$ data, most fields have dust colour
temperatures $T_{\rm C}\sim 14-18$\,K and opacity spectral index
values of $\beta=1.5-1.9$. The clumps/cores identified in SCUBA-2 maps
have $T\sim 13$\,K and similar $\beta$ values. There are some
indications of the dust emission spectrum becoming flatter at
wavelengths longer than 500\,$\mu$m. In fits involving $Planck$ data,
the significance is limited by the uncertainty of the corrections for
CO line contamination. The fits to the SPIRE data give a median
$\beta$ value slightly above 1.8. In the joint SPIRE and SCUBA-2
850\,$\mu$m fits the value decreases to $\beta \sim$1.6. Most of the
observed $T$-$\beta$ anticorrelation can be explained by noise.
}
{ 
The typical submillimetre opacity spectral index $\beta$ of cold
clumps is found to be $\sim$1.7. This is above the values of diffuse
clouds but lower than in some previous studies of dense clumps. There
is only tentative evidence of $T$-$\beta$ anticorrelation and $\beta$
decreasing at millimetre wavelengths.
}

\keywords{
ISM: clouds -- Infrared: ISM -- Submillimetre: ISM -- dust, extinction -- Stars:
formation -- Stars: protostars
}

\maketitle

\section{Introduction} \label{sect:intro}

The all-sky survey of the {\it Planck} satellite \citep{Tauber2010},
assisted with the IRAS 100\,$\mu$m data, made it possible to identify
and to catalogue cold interstellar clouds on a Galactic scale. The
angular resolution of $\sim5\arcmin$ is sufficient to identify compact
sources that can be directly associated with the earliest stages of
star formation. Analysis of the {\it Planck} and IRAS data led to the
creation of the Planck Catalogue of Cold Clumps \citep[PGCC,][]{PGCC},
which lists the properties of over 13000 sources. At distances up to
$\sim$500\,pc, the {\it Planck} observations are able to resolve
sub-parsec structures. For the more distant sources, up to 7\,kpc, the
observed signal may correspond to the detection of entire clouds.
Because the PGCC detection method is based on the relative cold
temperature of the sources compared to their environment
\citep{Montier2010}, the low temperatures (typically $T\la14$\,K) are
a proxy for high column densities and also exclude the possibility of
very strong internal heating. Even when spatially unresolved, the {\it
Planck} clumps are thus likely to contain higher density substructure,
including gravitationally bound pre-stellar and protostellar cores.

Some PGCC objects have already been studied in detail using the {\it
Herschel} follow-up observations made in the {\it Herschel} key
programme {\em Galactic Cold Cores} (GCC). In GCC, 116 fields were
selected from an early version of the {\it Planck} clump catalogue
\citep{planck2011-7.7b} and the fields were mapped with {\it Herschel}
PACS and SPIRE instruments \citep{Pilbratt2010, Poglitsch2010,
Griffin2010} between 100 and 500\,$\mu$m \citep{Juvela2010, PlanckII,
GCC-III}. The size of each {\it Herschel} map is $\sim 40 \arcmin$.
This means that the fields typically contain a few $Planck$ clumps but
also a significant area of the surrounding cloud and even diffuse
medium. This has enabled, for example, the study of the filamentary
structure of the clouds around the clumps \citep{GCC-VII, GCC-VIII,
Malinen2016}. 

To study the variations in the submillimetre dust opacity,
\citet{GCC-V} compared the 250\,$\mu$m optical depths to the
near-infrared (NIR) extinctions. The typical value of $\tau(250\mu{\rm
m})/\tau({\rm J})$ in the fields was found to be $1.6\times 10^{-3}$,
which is more than twice the values in diffuse medium
\citep{Planck2014A-XI-allsky, planck2013-XVII} but consistent with
other $Planck$ studies of molecular clouds \citep{planck2013-XIV}
\citep[see also, e.g.][]{Martin2011, Roy2013, Lombardi2014}. The
submillimetre opacity was also found to be correlated with column
density (and anticorrelated with temperature), with individual cores
reaching values still higher by a factor of two. 

\citet{GCC-VI} studied the value and variations of the opacity
spectral index $\beta$ in GCC fields. Based on the IRAS and {\it
Planck} data, the average value was found to be $\beta=1.84$, somewhat
higher than the typical {\it Planck} values for molecular clouds at
larger scales \citep[e.g.][]{Planck2011b, planck2013-XIV,
Planck2014A-XI-allsky}.
The analysis of {\it Planck} data down to 217\,GHz suggested a
flattening of the dust emission spectrum \citep[cf][]{planck2013-XIV}.
There is strong evidence of spectral index variations within the
individual fields and the cold cores were associated with
submillimetre spectral index values of $\beta\sim 2.0$ or even higher.
Thus, the results are qualitatively consistent with many earlier
observations of the $T-\beta$ -anticorrelation \citep{Dupac2003,
Desert2008, Rodon2010, Veneziani2010, Paradis2010, Etxaluze2011,
PlanckII, planck2011-6.4b, planck2011-7.0, planck2013-XVII,
planck2013-XIV}.
In such studies one must remember that the colour temperature and the
apparent $\beta$ values are affected by both noise and temperature
variations \citep{Shetty2009a, Shetty2009b, Juvela2012_bananasplit,
Malinen2011, Juvela2012_Tmix, Ysard2012, Juvela2013_TBmethods,
Pagani2015}. In particular, to estimate the effect of noise on the
apparent $T-\beta$ -anticorrelation one needs to have very precise
knowledge of all sources of error.

The above studies were limited to wavelengths below 500\,$\mu$m or did
include longer wavelengths but at a much lower resolution. The dust
spectrum of cores can be further constrained by ground-based
millimetre wavelength observations. However, concerning the spectral
index values, the results have shown a very wide range of $\beta$
estimates, from $\sim 1$ to values far above 2.0 \citep{Shirley2005,
Friesen2005, Schnee2010, Sadavoy2013, Ward-Thompson2016, Chen2016,
Sadavoy2016}. The large scatter is partly explained by the different
wavelengths used, the spatial filtering resulting from sky noise
reduction (when ground-based and space-borne observations are
combined), and the varying nature of the sources (e.g. low values in
regions with large temperature variations). Because of the additional
problems in correlating the changes of the apparent dust spectra with
the intrinsic dust properties, it is safe to say that our
understanding of dust evolution across the star formation process is
still very incomplete. This has implications for the estimates of the
mass and the detailed structure of star-forming clouds. Accurate
estimates of dust properties could also be related to laboratory
measurements, to determine how dust properties vary according to the 
environment \citep{Boudet2005, Coupeaud2011, Demyk2017}.

To this end, we started in late 2015 a legacy survey SCOPE (SCUBA-2
Continuum Observations of Pre-protostellar Evolution) to map about
1000 PGCCs in 850\,$\mu$m continuum emission with the SCUBA-2
instrument at the James Clerk Maxwell Telescope (JCMT)
\citep{Liu_2017_SCOPE, Eden_2017_SCOPE}. Thousands
of dense cores have been identified by the SCOPE survey, and most of
them are either starless cores or protostellar cores with very young
(Class 0/I) objects \citep[see][]{Liu2016ApJS222_7, Tatematsu2017,
Kim2017}
Those SCOPE sources are prime candidates for probing how
prestellar/protostellar cores form and evolve, and for studying the very
early stages of star formation across a wide variety of galactic
environments. In addition, the 850\,$\mu$m data are important in
providing high resolution information of the dust emission at the scales
of individual cores and at a wavelength that, according to previous
studies, lies in the regime between the steeper far-infrared part and
the flatter millimetre part of the dust emission spectrum
\citep{Planck2013_XVII, Paradis2012_500um}.

In this paper we examine dust properties in a set of some 90 fields
associated to PGCC sources. The fields were mapped at 850\,$\mu$m with
the SCUBA-2 instrument, as part of the SCOPE pilot programme. The
present sample contains the targets of the pilot programme for which
{\it Herschel} photometric observations were available. The goal of the
paper is to examine dust SED variations as a function of the spatial
scale and as a function of the wavelength.

The structure of the paper is the following. The observations and data
reduction are described in Sect.~\ref{sect:obs} and the analysis
methods are explained in Sect.~\ref{sect:methods}. The main results
are presented in Sect.~\ref{sect:results}. We discuss the results in
Sect.~\ref{sect:discussion} before listing the main conclusions in
Sect.~\ref{sect:conclusions}.

\section{Observational data}  \label{sect:obs}

\subsection{SCUBA-2 observations} \label{sect:obs_SCUBA}

The target selection for SCUBA-2 observations was based on PGCC
catalogue but also employed {\it Herschel} follow-up observations to
pinpoint the exact locations of the column density maxima. Thus, all
SCUBA-2 maps are not centred on PGCC coordinates and one of the
fields, G150.4+3.9A2, is not intersected by any PGCC clumps. The
fields are listed in Table~\ref{table:fields}.

The SCUBA-2 \citep{Holland2013} maps at 850\,$\mu$m were observed at
JCMT between April and October 2015 (projects M15AI05 and M15BI061,
PI: Tie Liu). The spatial resolution of the data is 14$\arcsec$.
All observations employed the CV Daisy mode \citep{Bintley2014},
resulting in maps with an approximate diameter of 12$\arcmin$ and a
roughly constant integration time within the innermost 3$\arcmin$. The
observations of all fields were of equal length with a total observing
time of some twenty minutes per field and a median exposure time of
some 80 seconds per map position. The observing conditions typically
corresponded to $\tau(225\,{\rm GHz})=0.10-0.15$. Maps with a
4.0$\arcsec$ pixel size were created using the {\tt makemap} task of
the Starlink SMURF package \citep{Jenness2011, Chapin2013}. The final
noise values are very similar in all fields. The approximate rms noise
values are 5.6\,mJy/beam for the entire maps and 3.4\,mJy/beam for the
central area within a 5$\arcmin$ radius.


In the map making, we used filtering scales of $\theta_{\rm
F}$=200$\arcsec$ and $\theta_{\rm F}$=500$\arcsec$.
Classically SCUBA-2 data are reduced using external masking (or auto
masking method) where source masks are derived from the SCUBA-2 data
\citep{Sadavoy2013, Moore2015, Pattle2015}. However, we rely mainly on
masks derived from {\it Herschel} maps, which provide a more complete
picture of the cloud structure at lower column densities.
The {\it Herschel} 500\,$\mu$m maps were high-pass filtered with a
Gaussian filter with FWHM equal to 300$\arcsec$ and the remaining
emission was initially thresholded at 50\,MJy\,sr$^{-1}$,
which corresponds approximately to the 1-$\sigma$ noise of the SCUBA-2
maps (for typical values of dust temperature and spectral index). The
resulting masks are thus significantly more extended than in auto
masking where the threshold would be several times higher.
If necessary, the mask was further adjusted so that it covers 10--30\%
of the map area and each distinct masked area is at least 40$\arcsec$
in size (larger than the {\it Herschel} beam). Alternative masks were
created the traditional way, by thresholding preliminary 850\,$\mu$m
signal-to-noise maps at a level of five. The masks were expanded by
15$\arcsec$ because they are also used when {\it Herschel} data (with
larger beam sizes) are processed through the SCUBA-2 pipeline, in
order to match the spatial filtering that affects SCUBA-2 maps
\citep{Pattle2015}.
To check how the results depend on the mask size, we use a second
set of {\it Herschel}-based source masks that are 50\% smaller. In the
following, these large and small {\it Herschel} masks are called LM
and SM, respectively. The masks created using SCUBA-2 data are referred
to as M850. The masks are shown in the figures of
Appendix~\ref{sect:allmaps}.

In the following, we assume for SCUBA-2 a 10\% relative uncertainty.
This covers the uncertainty of the calibration and the uncertainty of
the contamination by CO(3-2) line emission. Although the CO
contribution can in the 850\,$\mu$m measurements sometimes reach tens
of percent \citep{Drabek2012}, it is usually below 10\%
\citep[e.g.][]{Moore2015, Mairs2016}
and can be expected to be lower for cold clumps. The uncertainties 
caused by CO emission are discussed in Appendix~\ref{sect:CO}.


\longtab{
\begin{longtable}{lrrcccccc}
\caption[]{List of SCUBA-2 fields. The columns are: 
(1) name of the field, 
(2)-(5) centre coordinates, 
(6) {\it Herschel} observation ID, 
(7) fraction of SCUBA-2 map covered by {\it Herschel} observations,
(8) median 250\,$\mu$m surface brightness, 
(9) reference to CO observations.} \\
\hline\hline
Field     &   $l$             &  $b$                &    RA      & DEC        &  Obs.Id.  &
Cover     &  S(250\,$\mu$m)   & CO$^a$  \\
          &                   &                     &  (J2000.0) & (J2000.0)  &           &
  (\%)    &  (MJy\,sr$^{-1}$) &                      \\
\hline
\endfirsthead
\caption{continued.}\\
\hline\hline
Field  &   $l$  &  $b$  &    RA      & DEC        &  Obs.Id. &
Cover   &  S(250\,$\mu$m)  &  CO$^a$  \\
       &        &       &  (J2000.0) & (J2000.0)  &
       &  (\%)       &  (MJy\, sr$^{-1}$)  & \\
\hline
\endhead
\hline
\endfoot
G003.7+18.3A1 &    3.74 &  18.34  & 16:48:48.2 &  -15:34:40.1  & 1342267754, 1342267755  & 100 &  86.6 &   \\
G004.4+15.9A1 &    4.41 &  15.91  & 16:58:34.7 &  -16:28:49.3  & 1342267756, 1342267757  &  46 &  27.9 &   \\
G007.5+21.1A1 &    7.54 &  21.08  & 16:48:13.6 &  -11:04:32.1  & 1342267724, 1342267725  & 100 &  67.3 &   \\
G007.8+21.1A1 &    7.82 &  21.10  & 16:48:44.7 &  -10:51:01.3  & 1342267724, 1342267725  & 100 &  61.9 &   \\
G038.3-00.9A1 &   38.35 &  -0.95  & 19:04:44.7 &  +04:23:03.4  & 1342207026, 1342207027  & 100 &  314.7 &   \\
G070.4-01.5A1 &   70.55 &  -1.58  & 20:15:08.3 &  +32:02:07.5  & 1342219987  & 100 &  267.5 &   \\
G070.7-00.6A1 &   70.75 &  -0.66  & 20:12:02.8 &  +32:42:38.8  & 1342219987  &  99 &  241.3 &   \\
G074.1+00.1A1 &   74.13 &   0.11  & 20:18:01.0 &  +35:57:18.8  & 1342244190, 1342244191  & 100 &  286.8 &   \\
G111.6+20.2A1 &  111.68 &  20.20  & 20:57:37.8 &  +77:35:36.9  & 1342198861, 1342198862  & 100 &  64.1 &   \\
G113.4+16.9A1 &  113.43 &  17.01  & 21:59:43.3 &  +76:35:35.4  & 1342188679, 1342188680  & 100 &  58.8 &   \\
G115.9+09.4A1 &  115.94 &   9.47  & 23:24:09.3 &  +71:08:30.0  & 1342222598  & 100&   47.1 & Z \\
G127.67+2.65 &  127.69 &   2.65  & 01:36:47.8 &  +65:06:12.1  & 1342203610  & 100 &  68.6 &   \\
G127.88+2.68 &  127.89 &   2.68  & 01:38:47.2 &  +65:05:40.0  & 1342203610  & 100 &  64.4 & Z \\
G130.1+11.0A1 &  130.11 &  11.12  & 02:28:27.1 &  +72:37:29.9  & 1342216918  &  78 &  56.7 &   \\
G130.1+11.0A2 &  130.22 &  11.05  & 02:29:21.2 &  +72:31:26.1  & 1342216918  &  93 &  40.2 &   \\
G130.3+11.2A1 &  130.40 &  11.27  & 02:32:39.3 &  +72:39:26.4  & 1342216918  & 100 &  68.0 &   \\
G131.7+09.7A1 &  131.75 &   9.68  & 02:40:04.2 &  +70:40:26.8  & 1342203609  & 100 &  43.5 & Z  \\
G132.0+08.8A1 &  132.10 &   8.77  & 02:39:25.3 &  +69:41:36.2  & 1342216919  & 100 &  50.6 & Z  \\
G132.0+08.9A1 &  132.05 &   8.97  & 02:39:49.3 &  +69:53:40.0  & 1342216919  & 100 &  65.3 & Z  \\
G144.6+00.1A1 &  144.71 &   0.22  & 03:37:14.4 &  +55:54:57.1  & 1342226995, 1342226996  & 100 &  152.0 & Z  \\
G144.8+00.7A1 &  144.88 &   0.78  & 03:40:34.9 &  +56:15:53.3  & 1342226995, 1342226996  & 100 &  103.7 & Z  \\
G148.2+00.4A1 &  148.19 &   0.39  & 03:57:01.7 &  +53:54:06.9  & 1342239047, 1342239048  & 100 &  119.9 & Z  \\
G149.2+03.0A2 &  149.28 &   3.05  & 04:14:51.2 &  +55:09:11.6  & 1342216920  &  39 &  114.9 &  Z \\
G149.41+3.38 &  149.42 &   3.39  & 04:17:12.6 &  +55:17:47.9  & 1342216920  & 100 &  94.0 &  Z \\
G149.60+3.45 &  149.60 &   3.45  & 04:18:28.4 &  +55:12:57.0  & 1342216920  & 100 &  103.0 & Z  \\
G149.68+3.56 &  149.69 &   3.57  & 04:19:27.0 &  +55:14:29.1  & 1342216920  & 100 &  99.2 &  Z \\
G150.2+03.9B1 &  150.29 &   3.92  & 04:24:11.1 &  +55:03:34.2  & 1342214702  & 100 &  105.3 & Z  \\
G150.2+03.9B3 &  150.21 &   3.96  & 04:23:56.4 &  +55:08:45.8  & 1342214702  & 100 &  88.2 & Z  \\
G150.4+03.9A1 &  150.44 &   4.04  & 04:25:29.1 &  +55:02:11.9  & 1342214702  & 100 &  91.7 & Z  \\
G150.4+03.9A2 &  150.36 &   3.97  & 04:24:45.1 &  +55:03:06.4  & 1342214702  & 100 &  100.7 & Z  \\
G151.4+03.9A1 &  151.45 &   3.96  & 04:29:56.1 &  +54:15:17.9  & 1342203607  & 100 &  112.6 & Z  \\
G154.0+05.0A1 &  154.07 &   5.09  & 04:47:21.8 &  +53:03:08.3  & 1342216921  & 100 &  59.0 & Z  \\
G155.45-14.60 &  155.46 & -14.60  & 03:35:51.2 &  +37:40:49.4  & 1342226626  & 100 &  74.6 &   \\
G156.9-08.4A1 &  156.94 &  -8.50  & 04:01:46.7 &  +41:28:24.5  & 1342203615  & 100 &  64.0 & M  \\
G157.12-8.72 &  157.13 &  -8.71  & 04:01:46.9 &  +41:11:26.9  & 1342203615  & 100 &  85.2 & M  \\
G157.93-2.51 &  157.93 &  -2.50  & 04:28:35.8 &  +45:06:46.8  & 1342216923  & 100 &  58.8 &   \\
G158.8-21.6A1 &  158.85 & -21.67  & 03:27:40.2 &  +30:06:00.4  & 1342190326, 1342190327  & 100 &  81.3 &   \\
G158.8-34.1A1 &  158.95 & -34.19  & 02:55:58.1 &  +19:49:49.5  & 1342239263, 1342239264  & 100 &  62.9 &   \\
G159.0-08.4A1 &  159.03 &  -8.46  & 04:09:59.0 &  +40:06:15.4  & 1342239276, 1342239277  & 100 &  80.9 &  M \\
G159.0-08.4A3 &  159.02 &  -8.33  & 04:10:25.0 &  +40:12:37.2  & 1342239276, 1342239277  & 100 &  69.1 &  M \\
G159.1-08.7A1 &  159.16 &  -8.76  & 04:09:24.3 &  +39:47:52.0  & 1342239276, 1342239277  & 100 &  55.3 &   \\
G159.21-34.28 &  159.22 & -34.27  & 02:56:33.9 &  +19:38:18.7  & 1342239263, 1342239264  & 100 &  91.4 &   \\
G159.23-34.51 &  159.23 & -34.50  & 02:56:03.6 &  +19:26:14.7  & 1342239263, 1342239264  & 100 &  98.9 &   \\
G159.4-34.3A1 &  159.39 & -34.39  & 02:56:47.0 &  +19:27:44.7  & 1342239263, 1342239264  & 100 &  68.4 &   \\
G159.6-19.6A1 &  159.67 & -19.59  & 03:36:26.0 &  +31:15:50.3  & 1342214504, 1342214505  & 100 &  117.5 & M  \\
G159.6-19.6A2 &  159.60 & -19.67  & 03:35:57.2 &  +31:14:56.7  & 1342214504, 1342214505  & 100 &  93.9 &  M \\
G159.6-19.6A3 &  159.72 & -19.71  & 03:36:14.9 &  +31:08:42.9  & 1342214504, 1342214505  & 100 &  104.8 & M  \\
G159.7-19.6A1 &  159.77 & -19.63  & 03:36:40.6 &  +31:10:47.0  & 1342214504, 1342214505  & 100 &  118.9 &   \\
G160.6-16.7A1 &  160.62 & -16.72  & 03:48:19.7 &  +32:54:57.5  & 1342214504, 1342214505  & 100 &  74.0 &  M \\
G160.8-09.4A1 &  160.83 &  -9.42  & 04:13:12.6 &  +38:10:54.4  & 1342203616  & 100 &  76.5 &   \\
G160.8-09.4A2 &  160.81 &  -9.49  & 04:12:52.9 &  +38:09:04.1  & 1342203616  & 100 &  74.3 &   \\
G161.3-09.3A1 &  161.34 &  -9.32  & 04:15:22.3 &  +37:54:36.1  & 1342203616  & 100 &  77.0 &   \\
G161.56-9.29 &  161.57 &  -9.28  & 04:16:19.1 &  +37:46:23.8  & 1342203616  & 100 &  71.1 &   \\
G162.4-08.7A1 &  162.45 &  -8.70  & 04:21:30.3 &  +37:34:12.0  & 1342239278, 1342239279  & 100&   84.5 &  M \\
G163.32-8.41 &  163.32 &  -8.40  & 04:25:37.4 &  +37:09:28.0  & 1342205047, 1342205048  & 100 &  73.8 & M  \\
G163.68-8.33 &  163.68 &  -8.31  & 04:27:10.3 &  +36:57:29.4  & 1342205047, 1342205048  & 100 &  79.3 &   \\
G163.82-8.33 &  163.82 &  -8.32  & 04:27:36.5 &  +36:50:59.8  & 1342205047, 1342205048  & 100 &  79.7 &   \\
G164.1-08.8A1 &  164.15 &  -8.88  & 04:26:41.8 &  +36:13:41.1  & 1342205047, 1342205048  & 100&   63.6 &  M \\
G164.11-8.17 &  164.12 &  -8.15  & 04:29:12.2 &  +36:45:06.7  & 1342205047, 1342205048  & 100 &  81.1 &   \\
G164.26-8.39 &  164.26 &  -8.37  & 04:28:54.5 &  +36:29:47.3  & 1342205047, 1342205048  & 100 &  87.7 &   \\
G165.1-07.5A1 &  165.18 &  -7.53  & 04:35:02.3 &  +36:23:41.8  & 1342240279, 1342240314  & 100&   60.1 &   \\
G165.3-07.5A1 &  165.36 &  -7.50  & 04:35:44.8 &  +36:17:24.3  & 1342240279, 1342240314  & 100&   56.1 &   \\
G165.6-09.1A1 &  165.71 &  -9.12  & 04:30:59.7 &  +34:56:20.8  & 1342240279, 1342240314  & 100&   68.9 &   \\
G167.2-15.3A1 &  167.24 & -15.33  & 04:14:30.7 &  +29:35:06.5  & 1342190616, 1342202090  & 100&   33.9 &   \\
G168.1-16.3A1 &  168.15 & -16.40  & 04:13:50.0 &  +28:12:25.8  & 1342190616, 1342202090  & 100&   111.6 & M  \\
G169.1-01.1A1 &  169.16 &  -1.15  & 05:12:20.1 &  +37:19:43.6  & 1342250233, 1342266671  &  81&   70.8 & Z  \\
G170.0-16.1A1 &  169.96 & -16.17  & 04:19:58.2 &  +27:06:38.3  & 1342190616, 1342202090  & 100 &  66.6 & M  \\
G170.1-16.0A1 &  170.14 & -16.06  & 04:20:51.7 &  +27:03:12.9  & 1342190616, 1342202090  & 100 &  61.9 &   \\
G170.8-18.3A1 &  170.86 & -18.36  & 04:15:27.9 &  +24:59:13.5  & 1342190617, 1342190618  & 100 &  51.6 &   \\
G170.9-15.8A1 &  171.03 & -15.79  & 04:24:22.2 &  +26:36:40.1  & 1342190616, 1342202090  & 100 &  58.6 &   \\
G171.14-17.57 &  171.15 & -17.57  & 04:18:51.3 &  +25:19:33.0  & 1342204860, 1342204861  & 100 &  60.7 &   \\
G171.8-15.3A1 &  171.80 & -15.38  & 04:27:56.8 &  +26:19:33.8  & 1342239280, 1342239281  & 100 &  53.2 &   \\
G172.8-14.7A1 &  172.91 & -14.74  & 04:33:12.6 &  +25:56:30.2  & 1342239280, 1342239281  & 100 &  72.9 &  M \\
G173.1-13.3A1 &  173.14 & -13.32  & 04:38:41.7 &  +26:41:30.4  & 1342202252, 1342202253  & 100 &  66.8 &   \\
G173.3-16.2A1 &  173.42 & -16.29  & 04:29:27.8 &  +24:33:29.3  & 1342190654, 1342190655  & 100 &  119.6 &   \\
G173.9-13.7A1 &  173.96 & -13.75  & 04:39:29.4 &  +25:48:17.9  & 1342202252, 1342202253  & 100 &  159.0 &   \\
G174.0-15.8A1 &  174.06 & -15.83  & 04:32:44.8 &  +24:23:28.1  & 1342190654, 1342190655  & 100 &  94.8 &  M \\
G174.4-15.7A1 &  174.41 & -15.76  & 04:33:57.7 &  +24:10:46.4  & 1342190654, 1342190655  & 100 &  63.9 &   \\
G174.7-15.4A1 &  174.73 & -15.50  & 04:35:42.2 &  +24:06:39.4  & 1342190652, 1342190653  & 100 &  105.2 &   \\
G174.8-17.1A1 &  174.88 & -17.08  & 04:30:49.5 &  +22:58:36.2  & 1342190652, 1342190653  & 100 &  46.2 &   \\
G175.2+01.2A2 &  175.17 &   1.29  & 05:38:48.5 &  +33:43:18.2  & 1342250332, 1342250333  & 100 &  38.8 & Z  \\
G175.4-16.8A2 &  175.51 & -16.78  & 04:33:30.1 &  +22:43:05.0  & 1342190652, 1342190653  & 100 &  85.2 &   \\
G177.6-20.3A1 &  177.62 & -20.35  & 04:27:16.9 &  +18:53:32.9  & 1342202250, 1342202251  & 100 &  79.3 &   \\
G181.8+00.3A1 &  181.92 &   0.35  & 05:51:28.1 &  +27:28:02.5  & 1342250798, 1342250799  & 100 &  82.7 &   \\
G201.2+00.4A1 &  201.27 &   0.44  & 06:31:32.0 &  +10:33:54.8  & 1342204419, 1342204420  & 100 &  123.7 &   \\
G201.83+2.83 &  201.83 &   2.83  & 06:41:13.5 &  +11:10:09.6  & 1342228342  & 100 &  48.6 &   \\
G202.00+2.65 &  202.00 &   2.65  & 06:40:52.1 &  +10:56:00.0  & 1342228342  & 100 &  69.2 &   \\
G202.54+2.46 &  202.53 &   2.46  & 06:41:10.9 &  +10:22:34.6  & 1342228342  & 100 &  131.3 &   \\
G204.4-11.3A2 &  204.50 & -11.36  & 05:55:37.5 &  +02:11:15.3  & 1342205074, 1342205075  & 100 &  100.5 &   \\
G204.8-13.8A1 &  204.85 & -13.89  & 05:47:22.9 &  +00:41:23.2  & 1342215982, 1342215983  & 100 &  148.5 &   \\
G210.90-36.55 &  210.89 & -36.55  & 04:35:09.3 &  -14:14:41.4  & 1342216940  & 100 &  57.2 &   \\
G215.44-16.38 &  215.43 & -16.38  & 05:57:02.5 &  -09:32:28.3  & 1342203631  & 100 &  43.0 &   \\
G216.76-2.58 &  216.75 &  -2.57  & 06:49:13.7 &  -04:34:27.1  & 1342219956  & 100  & 71.4 &   \\
G219.13-9.72 &  219.12 &  -9.72  & 06:27:43.0 &  -09:52:51.8  & 1342227708  & 100  & 58.4 &   \\
G219.22-10.02 &  219.22 & -10.02  & 06:26:47.7 &  -10:05:36.5  & 1342227708  & 100 &  59.8 &   \\
G219.36-9.71 &  219.36 &  -9.71  & 06:28:10.3 &  -10:04:58.6  & 1342227708  & 100  & 84.3 &   \\
\label{table:fields}  
\end{longtable}
\tablefoot{
\tablefoottext{a}{M=\citet{Meng2013}, Z=\citet{Zhang2016}}
}
} 

\subsection{{\it Herschel} observations} \label{sect:obs_Herschel}

The {\it Herschel} SPIRE data at 250\,$\mu$m, 350\,$\mu$m, and
500\,$\mu$m were taken from the {\it Herschel} Science Archive
(HSA)\footnote{http://archives.esac.esa.int/hsa/whsa/}. We use the
level 2.5 maps produced by the standard data reduction pipelines and
calibrated for extended emission (the so-called Photometer Extended
Map Product). The quality of the SPIRE maps in HSA is good and we find
no need for manual data reduction. Out of the 96 fields, 43 were
observed as part of the GCC project \citep{Juvela2010}. For
consistency, we use HSA pipeline data also for these fields.

The resolutions of the SPIRE observations are 18.4$\arcsec$,
25.2$\arcsec$, and 36.7$\arcsec$ for the 250\,$\mu$m, 350\,$\mu$m, and
500\,$\mu$m bands,
respectively\footnote{The Spectral and Photometric Imaging
Receiver (SPIRE) Handbook,
http://herschel.esac.esa.int/Docs/SPIRE/spire\_handbook.pdf}.
The maps were convolved to 40$\arcsec$ and fitted with modified
blackbody (MBB) curves with $\beta=1.8$. These initial SED estimates
were used to colour correct the SPIRE maps to obtain monochromatic
values at the nominal wavelengths. The
SCUBA-2 data were later colour corrected using the same SEDs. In the
temperature range of $T=10-20$\,K, the corrections are 0-2\% for both
SPIRE and SCUBA-2.  The effect on the SED shapes and especially on the
$\beta$ estimates are even smaller because the corrections are
correlated between the bands (see Appendix~\ref{sect:CC}). We adopt
for the SPIRE bands a relative uncertainty of 4\% with $\rho=0.5$
correlation between the bands \citep{Bendo2013}.
The assumed correlation between the photometric values decreases the
formal $T$ and $\beta$ uncertainty in the fits to SPIRE data
\citep{Galametz2012}. However, when SPIRE is combined with SCUBA-2,
the situation may no longer be true. If the 250-500\,$\mu$m data have
errors preferentially in the same direction, the effect on the SED
shape is large when SPIRE is combined with the (presumably)
uncorrelated 850\,$\mu$m data point.

Main analysis is done without PACS data, which are only compared to
the SED fits of longer wavelength measurements. This is because PACS
observations cover only a small fraction of the SCUBA-2 fields and
because the inclusion of shorter wavelengths can bias estimates of the
spectral index, especially in the presence of embedded radiation
sources or emission from very small grains \citep[e.g.][]{Shetty2009a,
Malinen2011, Juvela2012_Tmix}. Out of the 96 fields, 37 are fully or
partially covered by PACS observations (for observation numbers, see
Table~\ref{table:PACS}). 
We use the UNIMAP versions of the 160\,$\mu$m maps provided in the
{\it Herschel} science archive \citep{Piazzo2015}. We assume for PACS
a relative uncertainty of 10\% to cover the 5\% calibration
uncertainty (estimated for point sources) \citep{Balog2014} and the
map-making uncertainty.

\begin{table}
\centering 
\caption[]{List of PACS observations.}
\label{table:PACS}
\begin{tabular}{ll}
\hline\hline
Field & PACS observation numbers \\
\hline
G070.4-01.5A1 & 1342220091, 1342220092 \\
G111.6+20.2A1 & 1342198861, 1342198862 \\
G113.4+16.9A1 & 1342197673, 1342197674 \\
G115.9+09.4A1 & 1342220665, 1342220666 \\
G127.67+2.65  & 1342216511 \\
G127.67+2.65  & 1342216512 \\
G130.3+11.2A1 & 1342218718, 1342218719 \\
G131.7+09.7A1 & 1342218640 \\
G131.7+09.7A1 & 1342223882 \\
G132.0+08.9A1 & 1342223880, 1342223881 \\
G144.8+00.7A1 & 1342226995, 1342226996 \\
G149.60+3.45  & 1342217532, 1342217533 \\
G150.4+03.9A2 & 1342217530, 1342217531 \\
G151.4+03.9A1 & 1342205042 \\
G151.4+03.9A1 & 1342205043 \\
G154.0+05.0A1 & 1342217528, 1342217529 \\
G157.12-8.72  & 1342216415 \\
G157.12-8.72  & 1342216416 \\
G157.93-2.51  & 1342217526, 1342217527 \\
G159.0-08.4A1 & 1342239276, 1342239277 \\
G159.21-34.28 & 1342239263, 1342239264 \\
G160.6-16.7A1 & 1342216420, 1342216421 \\
G162.4-08.7A1 & 1342239278, 1342239279 \\
G163.82-8.33  & 1342205047, 1342205048 \\
G171.14-17.57 & 1342204860, 1342204861 \\
G172.8-14.7A1 & 1342228001, 1342228002, \\
              & 1342228003, 1342228004 \\
G173.9-13.7A1 & 1342228174, 1342228175 \\
G174.0-15.8A1 & 1342228005, 1342228006 \\
G177.6-20.3A1 & 1342202250, 1342202251 \\
G201.2+00.4A1 & 1342269260, 1342269261,\\
              & 1342269262, 1342269263 \\
G202.00+2.65  & 1342228371, 1342228372 \\
G204.8-13.8A1 & 1342216450, 1342216451 \\
G210.90-36.55 & 1342225212, 1342225213 \\
G215.44-16.38 & 1342204305 \\
G215.44-16.38 & 1342204306 \\
G216.76-2.58  & 1342219406, 1342219407 \\
G219.13-9.72  & 1342219402, 1342219403 \\
\hline
\end{tabular}
\end{table}

\subsection{{\it Planck} and IRAS data} \label{sect:obs_Planck}

We use IRAS and {\it Planck} observations to examine the dust emission
at large $\sim 10\arcmin$ scales. We use the IRIS version of the
100\,$\mu$m IRAS data \citep{Miville2005} and the 857\,GHz, 545\,GHz,
353\,GHz, and 217\,GHz {\it Planck} maps taken from the Planck Legacy
Archive\footnote{https://www.cosmos.esa.int/web/planck/pla} and
correspond to the 2015 data release \citep{Planck2015} where the CMB
emission has been subtracted.
We subtract from {\it Planck} maps the estimated levels of the cosmic
infrared background (CIB). Because the targets have high column
densities, the CIB correction ($\sim$1\,MJy\,sr$^{-1}$ or less in all
bands) is relatively unimportant \citep{Planck2016_HFI}. The
amount of emission from stochastically heated very small grains (VSG)
at 100\,$\mu$m is unknown. However, it could amount to some tens of
percent in cold clouds \citep{Li2001, Compiegne2011}.
We make no correction for the VSG emission. Thus, the combined SED of
IRAS and {\it Planck} data is expected to correspond to a higher
colour temperature (and slightly altered and probably lower spectral
index values) than the SED of the large grains alone \citep{GCC-VI}.

{\it Planck} 353\,GHz and 217\,GHz bands are contaminated by CO
$J=2-1$ and $J=3-2$ line emission, which is a significant source of
uncertainty in the estimates of the long wavelength dust emission
spectrum.  The effect is more significant at 217\,GHz, because 
$J=2-1$ is usually the stronger of the two CO lines while dust
emission at 217\,GHz is only about one fifth of the emission at
353\,GHz (assuming $\beta=1.7$). We correct these two bands the same
way as in \citet{GCC-VI}. The correction makes use of the Type 3 CO
maps provided by the {\it Planck} Consortium and is based on the
assumption of line ratios $T_{\rm A}$(2--1)/$T_{\rm A}$(1--0)=0.5 and
$T_{\rm A}$(3--2)/$T_{\rm A}$(1--0)=0.3. The justification and
uncertainty of these values is discussed in \citet{GCC-VI}.

The 217\,GHz band could have additional contribution from free-free
emission, in the case of actively star forming clumps. Therefore,
similar to \citet{GCC-VI}, we subtract a model for the low-frequency
foregrounds\footnote{see
https://wiki.cosmos.esa.int/planckpla/index.php/ \\
CMB\_and\_astrophysical\_component\_maps}, although this is in
practice much smaller than the CO correction.

We assume absolute calibration uncertainties of 10\% for IRAS
100\,$\mu$m, 5\% for the 857\,GHz and 545\,GHz {\it Planck} channels,
and 2\% for the 353\,GHz and 217\,GHz channels. Because of the common
calibration, we furthermore assume a correlation of $\rho=0.5$ between
the first two $Planck$ bands and again between the last two bands. In
the 353\,GHz and 217\,GHz bands we include an additional uncertainty
(added in squares) that corresponds to 30\% of the applied CO
correction.

\subsection{Cloud distances} \label{sect:distances}

Out of the 96 sources, a distance estimate was already given for 86
sources in the PGCC catalogue \citep{PGCC}. These are mostly based on
3D extinction mapping, the comparison between the observed stellar
reddening and the predictions derived from a model of the Galactic
stellar distribution \citep{Marshall2006, Marshall2009}. There is also
considerable overlap between the current source sample and the fields
observed as part of the {\it Herschel} key programme Galactic Cold
Cores (GCC). The distances of the GCC fields were discussed in detail
by \citet{GCC-IV}. With only one exception, those estimates agreed
with the original PGCC values. The distance estimates are listed in
Table~\ref{table:distances}, which also notes some nearby molecular
clouds and their distance estimates, if available in the SIMBAD
database\footnote{http://simbad.u-strasbg.fr/simbad/} or in the cloud
compilation of \citet{Dutra2002}.
The distance estimates typically have uncertainties of several tens of
percent and therefore are a major source of uncertainty when the
masses of the sources are estimated.

We revisited the extinction-based distance estimates using the method
MACHETE \citep{Marshall2017}. 
The method is based on the observed reddening of background stars that
is compared to predictions based on the Besancon model
\citep{Robin2003} of stellar distributions and the optimised model of
the dust distribution along the line of sight.
Apart from the improvements in the method itself, we also make use of
{\it Herschel} data to better separate the high-column-density regions
associated to the SCUBA-2 targets. In practice, we use masks that
cover an area where $\tau(250\mu{\rm m})$ is above its 40\% percentile
value (within each map separately). The masks are typically larger
than the size of the individual PGCC clumps and thus should provide
better statistics for the background stars.

\longtab{
\begin{longtable}{lcccc}
\caption{Distance estimates.}
\label{table:distances} \\
\hline\hline
SCUBA-2 field & $d$(PGCC)$^a$ &   GCC field$^b$ &  $d$(GCC)$^c$ &  Associated clouds$^d$ \\
              & (pc)          &                 &  (pc)         &  (pc) \\
\hline
\endfirsthead
\caption{continued.}\\
\hline\hline
SCUBA-2 field & $d$(PGCC)$^a$ &  GCC field$^b$ & $d$(GCC)$^c$ & Associated clouds$^d$ \\
              & (pc)          &                &  (pc)        & (pc) \\
\hline
\endhead
\hline
\endfoot
  G003.7+18.3A1 &     120 &                - &      -  &  \\
  G004.4+15.9A1 &       - &                - &      -  &  \\
  G007.5+21.1A1 &     120 &                - &      -  &  LDN234: 160, MBM 148 \\
  G007.8+21.1A1 &     120 &                - &      -  &  LDN234: 160 \\
  G038.3-00.9A1 &    1200 &                - &      -  &  \\
  G070.4-01.5A1 &    2090 &      G70.10-1.69 &   2090  &  \\
  G070.7-00.6A1 &    2260 &                - &      -  &  LDN841 \\
  G074.1+00.1A1 &    3450 &                - &      -  &  \\
  G111.6+20.2A1 &       - &                - &      -  &  LDN1228: 170, MBM 162 \\
  G113.4+16.9A1 &       - &                - &      -  &  LDN1241 \\
  G115.9+09.4A1 &    1000 &     G115.93+9.47 &   1000  &  \\
   G127.67+2.65 &     800 &     G127.79+2.66 &    800  &  CB11 \\
   G127.88+2.68 &     800 &     G127.79+2.66 &    800  &  CB12 \\
  G130.1+11.0A1 &     600 &    G130.37+11.26 &    600  &  LDN1340A \\
  G130.1+11.0A2 &     600 &    G130.37+11.26 &    600  &  LDN1340A \\
  G130.3+11.2A1 &     600 &    G130.37+11.26 &    600  &  LDN1340C \\
  G131.7+09.7A1 &    1070 &     G131.65+9.75 &   1070  &  \\
  G132.0+08.8A1 &     850 &     G132.12+8.95 &    850  &  \\
  G132.0+08.9A1 &     850 &     G132.12+8.95 &    850  &  \\
  G144.6+00.1A1 &       - &                - &      -  &  \\
  G144.8+00.7A1 &       - &                - &      -  &  \\
  G148.2+00.4A1 &       - &                - &      -  &  \\
  G149.2+03.0A2 &       - &     G149.67+3.56 &    170  &  \\
   G149.41+3.38 &     170 &     G149.67+3.56 &    170  &  LDN1394, B8\\
   G149.60+3.45 &     170 &     G149.67+3.56 &    170  &  LDN1400H, MLB71, B9\\
   G149.68+3.56 &     170 &     G149.67+3.56 &    170  &  \\
  G150.2+03.9B1 &     170 &     G150.47+3.93 &    170  &  LDN1399: 170, LDN1400 \\
  G150.2+03.9B3 &     170 &     G150.47+3.93 &    170  &  LDN1399: 170 \\
  G150.4+03.9A1 &     170 &     G150.47+3.93 &    170  &  LDN1400  \\
  G150.4+03.9A2 &     170 &     G150.47+3.93 &    170  &  LDN1399: 170, \\
                &         &                  &         &  LDN1400, MLB72, MLB73, MLB74 \\
  G151.4+03.9A1 &     170 &     G151.45+3.95 &    170  &  B12, LDN1407, MLB77: 170, LDN1400 \\
  G154.0+05.0A1 &     170 &     G154.08+5.23 &    170  &  LDN1426: 170, LDN 1426 \\
  G155.45-14.60 &     350 &    G155.80-14.24 &    350  &  LDN1434: 350 \\
  G156.9-08.4A1 &     150 &     G157.08-8.68 &    150  &  \\
   G157.12-8.72 &     150 &     G157.08-8.68 &    150  &  LDN1443 \\
   G157.93-2.51 &    2500 &     G157.92-2.28 &   2500  &  \\
  G158.8-21.6A1 &     140 &                - &      -  &  LDN1452: 280, LDN1455\\
  G158.8-34.1A1 &     325 &    G159.23-34.51 &    325  &  LDN1454: 200, MBM12: 70 \\
  G159.0-08.4A1 &     140 &                - &      -  &  LDN1459: 350 \\
  G159.0-08.4A3 &     140 &                - &      -  &  LDN1459: 350 \\
  G159.1-08.7A1 &     140 &                - &      -  &  LDN1459: 350 \\
  G159.21-34.28 &     325 &    G159.23-34.51 &    325  &  LDN1457: 90, MBM12: 70\\
  G159.23-34.51 &     325 &    G159.23-34.51 &    325  &  LDN1457: 90, MBM12: 70  \\
  G159.4-34.3A1 &     325 &    G159.23-34.51 &    325  &  LDN1457: 90, MBM12: 70 \\
  G159.6-19.6A1 &     140 &                - &      -  &  \\
  G159.6-19.6A2 &     140 &                - &      -  &  \\
  G159.6-19.6A3 &     140 &                - &      -  &  \\
  G159.7-19.6A1 &     140 &                - &      -  &  \\
  G160.6-16.7A1 &     140 &                - &      -  &  LDN1471: 350, B5 \\
  G160.8-09.4A1 &     140 &                - &      -  &  \\
  G160.8-09.4A2 &     140 &                - &      -  &  \\
  G161.3-09.3A1 &     250 &     G161.55-9.30 &    250  &  \\
   G161.56-9.29 &     250 &     G161.55-9.30 &    250  &  \\
  G162.4-08.7A1 &     140 &                - &      -  &  LDN1478: 350 \\
   G163.32-8.41 &     420 &                - &      -  &  LDN1478: 350 \\
   G163.68-8.33 &     420 &     G163.82-8.44 &    420  &  LDN1478: 350 \\
   G163.82-8.33 &     420 &     G163.82-8.44 &    420  &  \\
  G164.1-08.8A1 &     420 &                - &      -  &  \\
   G164.11-8.17 &     420 &     G163.82-8.44 &    420  &  \\
   G164.26-8.39 &     420 &     G163.82-8.44 &    420  &  \\
  G165.1-07.5A1 &     140 &                - &      -  &  LDN1483 \\
  G165.3-07.5A1 &     140 &                - &      -  &  LDN1483 \\
  G165.6-09.1A1 &     140 &                - &      -  &  \\
  G167.2-15.3A1 &     140 &                - &      -  &  \\
  G168.1-16.3A1 &     140 &                - &      -  &  LDN1495: 140 \\
  G169.1-01.1A1 &       - &                - &      -  &  \\
  G170.0-16.1A1 &     140 &                - &      -  &  \\
  G170.1-16.0A1 &     230 &                - &      -  &  B213, LDN1521 \\
  G170.8-18.3A1 &     230 &                - &      -  &  B210, LDN1501: 140 \\
  G170.9-15.8A1 &     230 &                - &      -  &  B216, MLB14: 140 \\
  G171.14-17.57 &     230 &                - &      -  &  LDN1506 \\
  G171.8-15.3A1 &     230 &                - &      -  &  B217, LDN1521 \\
  G172.8-14.7A1 &     230 &                - &      -  &  B19, LDN1521: 140 \\
  G173.1-13.3A1 &     230 &                - &      -  &  \\
  G173.3-16.2A1 &     230 &                - &      -  &  LDN1524: 140 \\
  G173.9-13.7A1 &     230 &                - &      -  &  B22, LDN1528:140, LDN1532, MLB26, \\
                &         &                  &         &  TMC1-A: 140, LDN1534: 140  \\
  G174.0-15.8A1 &     230 &                - &      -  &  B18, LDN1529: 140, TMC-2, MLB23: 140 \\
  G174.4-15.7A1 &     230 &                - &      -  &  B18-3: 140 \\
  G174.7-15.4A1 &     230 &                - &      -  &  LDN1535: 140\\
  G174.8-17.1A1 &     230 &                - &      -  &  \\
  G175.2+01.2A2 &    4590 &                - &      -  &  \\
  G175.4-16.8A2 &     230 &                - &      -  &  LDN1536, MLB24: 140 \\
  G177.6-20.3A1 &     230 &                - &      -  &  LDN1543, MBM107: 140 \\
  G181.8+00.3A1 &       - &                - &      -  &  \\
  G201.2+00.4A1 &       - &                - &      -  &  B37 \\
   G201.83+2.83 &     760 &     G202.02+2.85 &    760  &  \\
   G202.00+2.65 &     760 &     G202.02+2.85 &    760  &  \\
   G202.54+2.46 &     760 &     G202.02+2.85 &    760  &  \\
  G204.4-11.3A2 &     415 &                - &      -  &  \\
  G204.8-13.8A1 &     415 &                - &      -  &  NGC2071-North \\
  G210.90-36.55 &     140 &    G210.90-36.55 &    140  &  MBM20: 110, LDN1642: 100, 140 \\
  G215.44-16.38 &     415 &    G215.44-16.38 &   1450  &  \\
   G216.76-2.58 &    2400 &     G216.76-2.58 &   2400  &  \\
   G219.13-9.72 &     905 &     G219.36-9.71 &    905  &  LDN1652: 830 \\
  G219.22-10.02 &     905 &     G219.36-9.71 &    905  &  LDN1652: 830 \\
   G219.36-9.71 &     905 &     G219.36-9.71 &    905  &  LDN1652: 830 \\
\end{longtable}
\tablefoot{
$^a$Distance estimate in PGCC catalogue.
$^b$Overlapping GCC {\it Herschel} field.
$^c$Distance estimate in \citet{GCC-IV}
$^d$Known clouds within half a degree distance of the
field centre. The references are: \citet{Barnard1927} (B),
\citet{Clemens1988} (CB), \citet{Lynds1962} (LDN), \citet{MBM} (MBM),
and \citet{MLB1983} (MLB). Possible literature distance estimates 
given after a colon.
}
}  

\section{Methods} \label{sect:methods}

\subsection{Clump extraction} \label{sect:clumps}

Clumps were extracted from SCUBA-2 850\,$\mu$m signal-to-noise maps 
with the Fellwalker method \citep{Berry2015}, which is also used by
the SCOPE project \citep{Liu_2017_SCOPE}. 
Clumps were required to contain more than 10 pixels and to rise above
three times the rms value $\sigma$. The other relevant parameters of
the algorithm were {\tt MINDIP}=2$\sigma$ and {\tt MAXJUMP}=5
pixels\footnote{
http://starlink.eao.hawaii.edu/docs/sun255.htx/sun255se2.html}. The
clump extraction was repeated for each version of the reduced data
(e.g., the three different source masks, LM, SM, and M850). The
extraction can result in spurious detections near the noisy map
boundaries but these are rejected later when the analysis is limited
to structures within 5$\arcmin$ of the centre of each field.

\subsection{Clump photometry} \label{sect:photometry}

Clump fluxes are measured from the original {\it Herschel} maps and
from SCUBA-2 850\,$\mu$m and {\it Herschel} maps processed through the
SCUBA-2 pipeline. To make the procedure consistent across maps of
originally different spatial resolution, the photometric measurements
were made on maps that were first all convolved to a common resolution
of 40$\arcsec$. 

The measurement apertures are based on the footprints of the
Fellwalker clumps but are extended by 20$\arcsec$ to reduce the flux
losses resulting from the above mentioned map convolution. The
apertures are shown in the figures of Appendix~\ref{sect:allmaps}.

All flux measurements employ a local background subtraction. The
reference annulus extends from 40$\arcsec$ to 60$\arcsec$ outside the
original Fellwalker clump footprint. The annulus is thus non-circular
and follows the clump shape. All pixels that are closer than
20$\arcsec$ to the footprint of another clump are excluded.
We use the median of the remaining annulus pixels as the background
estimate and do not explicitly interpolate the background over the
aperture.

For comparison and to create a subsample of more compact cores,
alternative photometry is carried out using the same reference annuli
but 50\% smaller apertures. These cover the pixels with 
$\tau(250\mu{\rm m})$ values (estimated using {\it Herschel} data, see
Sect.~\ref{sect:obs_Herschel} and Appendix~\ref{sect:allmaps}) above
the median value of the original aperture. Apertures smaller than
256\,arcsec$^2$ (16 pixels with a size 4$\arcsec \times 4\arcsec$
each) are rejected. In the following, the original and the
high-column-density apertures are called $\tau_{\rm all}$ and
$\tau_{\rm high}$ apertures, respectively.

The statistical errors of the flux values are estimated based on the
standard deviation of the pixel values in the annulus and the sizes of
the annulus and the aperture. We take into account the fact that
the maps have been convolved to a resolution of $40\arcsec$ and scale
the standard deviation of the pixel values with the square root of the
ratio between the area of the measuring aperture (or annulus) and the
effective area of the $40\arcsec$ beam. Because of real surface
brightness fluctuations within the reference annuli, the photometric
errors tend to be overestimated. To reduce this effect, we remove the
signal that is linearly correlated between the bands before estimating
the noise. Each band is in turn fitted against a reference, which
is the average of all the other bands. The correlated signal predicted
by the least square fit and the reference data is then subtracted.
The emission is not perfectly linearly correlated, mainly because of
temperature variations. Therefore, the estimates remain an upper limit
of the actual noise.
In SED fits, the relative weighting of the bands also takes into
account calibration uncertainties (4\% and 10\% for SPIRE and SCUBA-2,
respectively).

\subsection{SED fits} \label{sect:SEDfit}

We fit SEDs using three or four bands, combining SCUBA-2 measurements
and fluxes extracted from SPIRE maps processed through the SCUBA-2
pipeline. We also fit the fluxes from the original, unfiltered SPIRE
maps without SCUBA-2 data, to see how the SPIRE SEDs are affected by
the filtering process itself.

The fitted model of flux density is the modified blackbody (MBB)
function
\begin{equation}
F(\nu) =  F(\nu_0) \frac{B_{\nu}(\nu, T)}{B_{\nu}(\nu_0, T)} 
      \left( \frac{\nu}{\nu_0} \right)^\beta
      =  \kappa(\nu) B_{\nu}(\nu, T) M / d^2,
\label{eq:sed}      
\end{equation}
where $\nu$ is the frequency, $\nu_0$ the selected reference
frequency, $B_{\nu}$ the Planck law, $\beta$ the opacity spectral
index, and $d$ the cloud distance \citep{Hildebrand1983}. The three
fitted parameters are the flux density at the reference frequency
$F(\nu_0)$, the dust emission colour temperature $T$, and the spectra
index $\beta$. The assumption of a value for the dust opacity per unit
mass $\kappa(\nu)$ enables estimation of clump masses $M$. We assume a
value $\kappa=0.1\,(\nu/1\,{\rm THz})^\beta$\,cm$^2$\,g$^{-1}$
\citep{Beckwith1990}, as used, for example, in papers of the Galactic
Cold Cores project \citep{GCC-III, GCC-IV} and {\it Herschel} Gould
Belt Surveys \citep{Andre2010}. 

Equation (~\ref{eq:sed}) is describing the apparent shape of an SED in
terms of its colour temperature and apparent spectral index. If the
SED is assumed to be connected to real dust properties (such as the
intrinsic opacity spectral index) and clump properties (such as the
mass), one implicitly assumes that the emission is optically thin and
dust is characterised by a single temperature and a single $\beta$
value. The impact of these assumption is discussed further in
Sect.~\ref{sect:discussion}.

We use only a subset of all the extracted clumps for the SED analysis
(see Sect.~\ref{sect:clumps} and Sect.~\ref{sect:photometry}. To avoid
clumps near high-noise map boundaries, the centre of the included
clumps is required to reside within 5$\arcmin$ of the field centre. We
also reject clumps whose 850\,$\mu$m flux is below 0.1\,Jy or where
the MBB fit to four bands (850\,$\mu$m and the filtered SPIRE data)
results in poor fits (see below).

\section{Results} \label{sect:results}

We investigate differences of dust emission properties in dense,
potentially prestellar clumps and the surrounding more diffuse clouds.
We examine the values of the opacity spectral index $\beta$ of the
clumps. We seek for evidence of correlated changes between the dust
temperature and $\beta$ and of a potential wavelength dependence of
$\beta$. These could be indications of dust processing during the star
formation process or directly related to the local physical conditions
such as the radiation field and the temperature. We start by examining
large-scale dust emission of the selected fields
(Sect.~\ref{results:large}) before concentrating on the compact
SCUBA-2 clumps and cores (Sect.~\ref{result:small}). We compare the
fits of different wavelength ranges up to the millimetre regime,
examine the effect of the extension to the shorter 160\,$\mu$m
wavelength (Sect.~\ref{sect:PACS}), and check the correlations between
young stellar objects and the clump and dust emission properties
(Sect.~\ref{sect:YSO}).

\subsection{Dust spectrum at large scales}  \label{results:large}

To characterise the dust emission at large scales, over the entire
fields where the PGCC clumps reside in, we analyse IRAS, {\it
Herschel}, and {\it Planck} observations in the wavelength range
$\lambda=100-1380$\,$\mu$m (from 3000\,GHz to 217\,GHz).
By using the IRIS data and the zero-point corrected {\it Herschel}
maps, we can use directly surface brightness values averaged within a
radius of $r=8\arcmin$ of the centre of the SCUBA-2 fields.
Alternatively, we may subtract the background that is estimated as the
mean surface brightness in a selected reference area. In that case, we
add the uncertainty of the background subtraction. This is described
by the covariance matrices that are estimated from the values in the
reference region, from the part of the signal that is not linearly
correlated between the bands. The procedure to exclude the correlated
signal is the same as in Sect.~\ref{sect:photometry}.

To ensure consistent handling of all bands, the maps are convolved to
a common resolution of 6$\arcmin$. We carry out MBB fits using
different band combinations, to check the consistency of the data and
to examine possible wavelength dependencies.

\subsubsection{SEDs of IRAS and {\it Planck} data} \label{sect:SED_IP}

\begin{figure*}
\includegraphics[width=18cm]{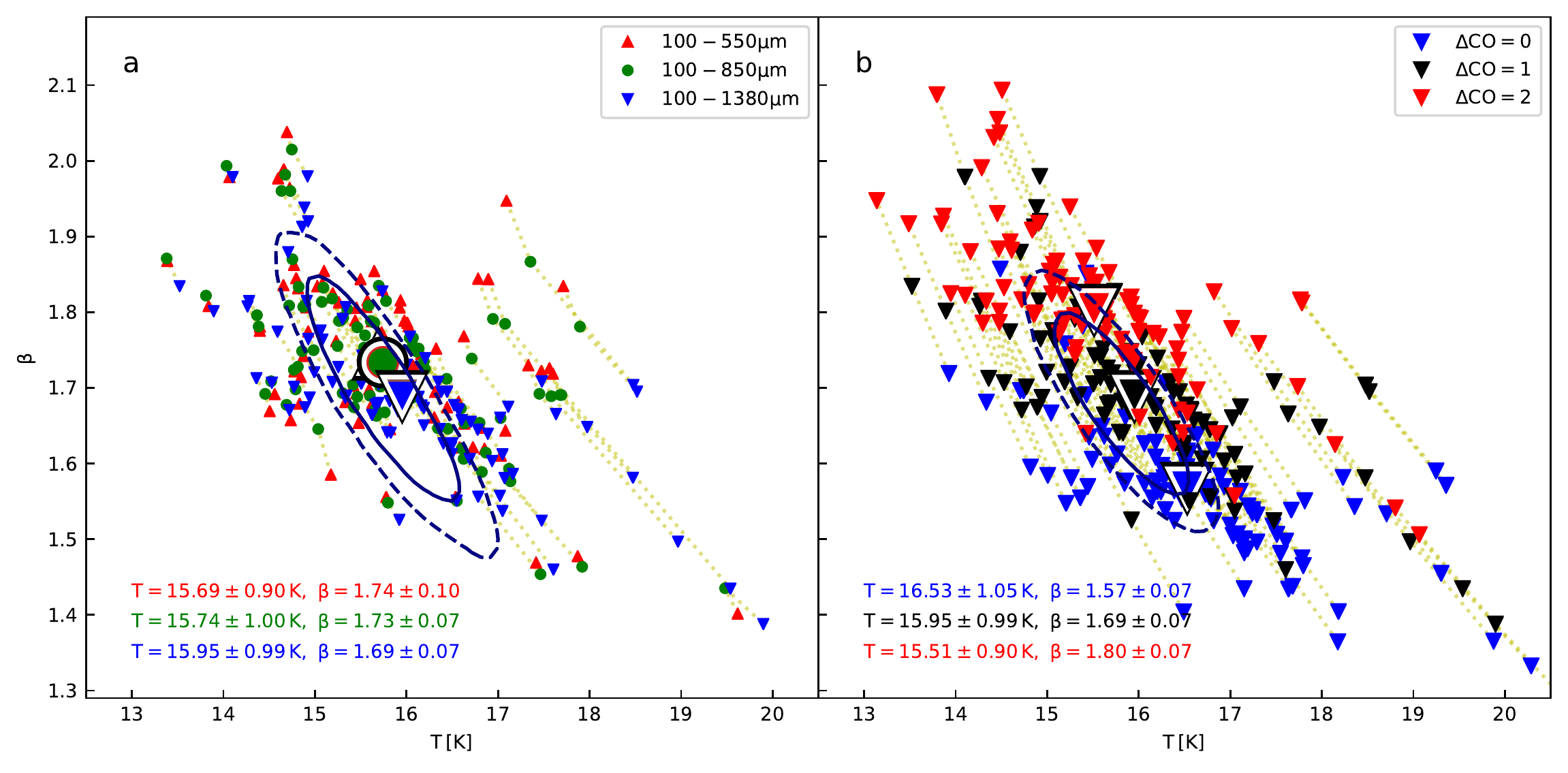}
\caption{
Results of MBB fits to $IRAS$ and $Planck$ data averaged within
8$\arcmin$ of the centre of the SCUBA-2 fields. Frame a shows the fits
completed with 100-550\,$\mu$m, 100-850\,$\mu$m, and 100-1380\,$\mu$m
bands. Each marker corresponds to one field and the dotted lines
connect estimates of the same field. The median values calculated over
all the fields are plotted with large symbols. The median values and
the 1-$\sigma$ dispersion estimated from the interquartile ranges are
quoted in the frame.
Frame b shows the fits to the 100-1380\,$\mu$m data without CO
corrections ($\Delta$CO=0), with the default CO correction
($\Delta$CO=1), and with twice the default correction ($\Delta$CO=2).
The median values are indicated by the three large triangles.
The contours show the 67\% (solid line) and 95\% (dashed line)
confidence regions for a 100-850\,$\mu$m (frame a) and
100-1380\,$\mu$m fit with $\Delta$CO=1 (frame b) calculated using the
median error covariance matrix of the fields.
}
\label{fig:IP_8am}
\end{figure*}

Figure~\ref{fig:IP_8am}a shows the colour temperature $T$ and the
spectral index $\beta$ calculated from the combination of the
100\,$\mu$m IRAS data and the $Planck$ bands at 857\,GHz, 545\,GHz,
353\,GHz, or 217\,GHz (wavelengths 350\,$\mu$m, 550\,$\mu$m,
850\,$\mu$m, and 1380\,$\mu$m, respectively). There is no subtraction
of the local background.
The confidence regions plotted in Fig.~\ref{fig:IP_8am} are calculated
with the Markov chain Monte Carlo (MCMC) method, using flat priors
with $8\,{\rm K} < T < 25\,{\rm K}$ and $0.5 < \beta < 3$. 

The median values of the 100-850\,$\mu$m fit are 
$T=15.74$\,K and $\beta=1.73$. 
The addition of longer wavelengths up to 1380\,$\mu$m increases the
temperature to $T=15.95$\,K and lowers the spectral index to
$\beta=1.69$. 
Because long wavelengths are less sensitive to line-of-sight
temperature variations, one could have expected the colour temperature
to decrease rather than increase (assuming that the opacity spectral
index does not change with wavelength).
However, differences are not significant compared to the statistical
uncertainties ($\sim$1\,K in temperature and $\sim 0.1$ in the
spectral index, see Fig.~\ref{fig:IP_8am}) and could easily be caused
by small systematic errors. 
Figure~\ref{fig:IP_8am}b illustrates the effect of
CO corrections on the 100-1380\,$\mu$m fits. The difference in
Fig.~\ref{fig:IP_8am}a could be explained by the CO corrections being
underestimated by some 10\%. Although there is no particular reason to
suspect that the correction would be underestimated (see below), the
evidence for a wavelength dependence of $\beta$ is in
Fig.~\ref{fig:IP_8am}a at most tentative.

\begin{figure}
\includegraphics[width=8.8cm]{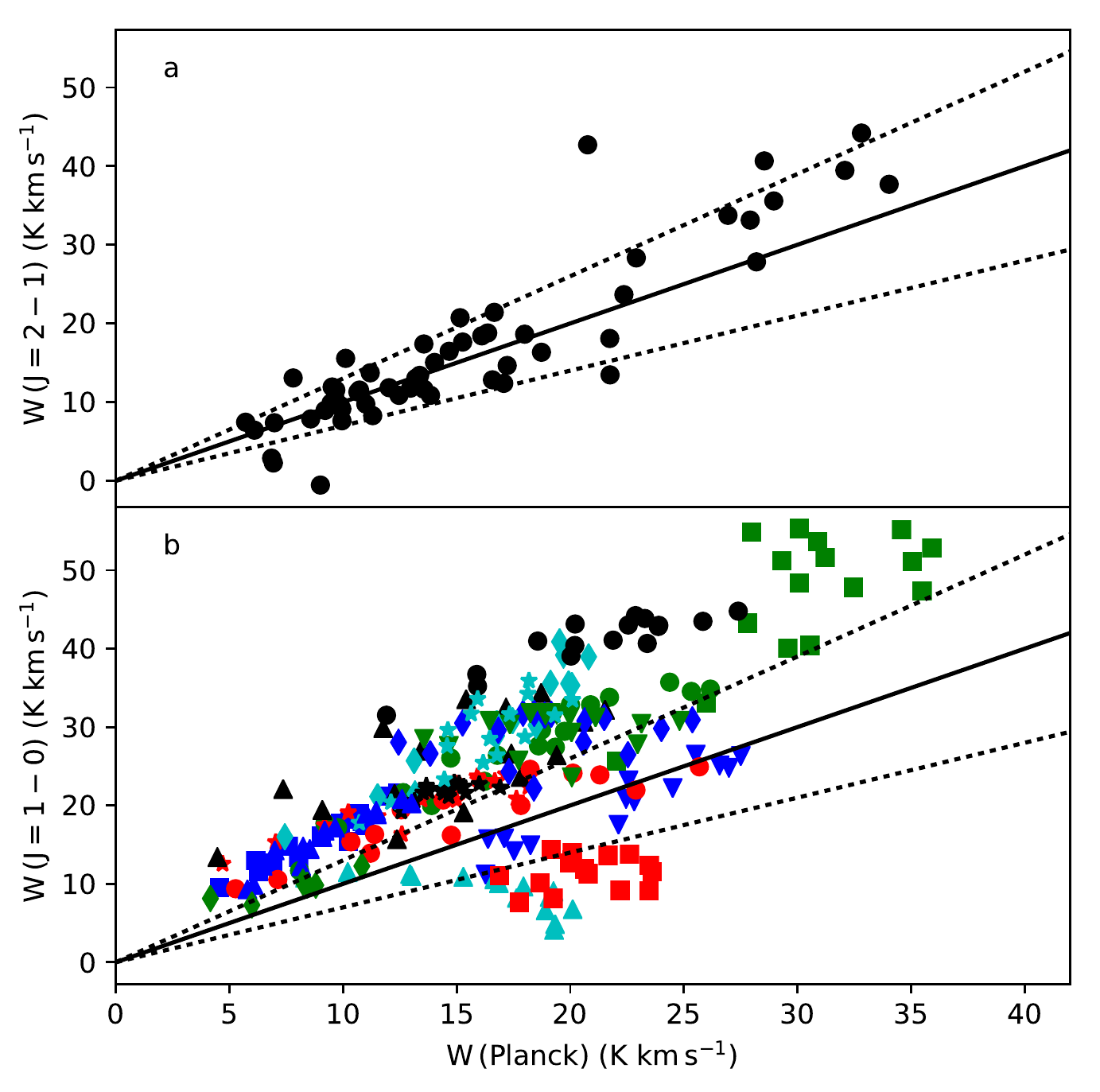}
\caption{
Comparison of CO estimates from {\it Planck} Type 3 CO maps and
ground-based observations. Upper frame compares the sum of
$^{12}$CO(2--1) and $^{13}$CO(2--1) values of 56 SMT maps to {\it
Planck} CO estimates for the assumed line ratio of (2-1)/(1-0)=0.5.
Lower frame shows comparison with CO(1--0) observations from the PMO
telescope, with data at 7$\arcmin$ resolution sampled at 3$\arcmin$
steps. Each colour and symbol combination corresponds to one of 17
fields.
Solid lines show the one-to-one relations and dotted lines correspond
to slopes different by $\pm$30\%.
}
\label{fig:compare_CO}
\end{figure}

To further quantify the reliability of the CO corrections, we compare
{\it Planck} CO estimates to ground-based data that provide a partial
coverage of about half of the fields. 
\citet{Meng2013} and \citet{Zhang2016} made CO(1--0) observations at
the Purple Mountain Observatory (PMO) radio telescope, fully or
partially covering 17 of our SCUBA-2 fields. In
Fig.~\ref{fig:compare_CO}b we show the correlation with the {\it
Planck} CO estimates once the PMO observations have been convolved to
the same 7$\arcmin$ resolution and scaled with the assumed main beam
efficiency of $\eta_{\rm MB}$=0.5. The correlation is generally good
but the ground-based values are on average higher by $\sim$30\%. 

The ongoing SAMPLING\footnote{http://sky-sampling.github.io} survey (PI. Ke Wang) at the Arizona Radio
Observatory 10\,m SMT telescope is mapping a large number of PGCC
targets in the $J=2-1$ transitions of CO isotopomers and thus provide
more direct estimates of the CO contamination of the 217\,GHz band.
There are 56 SMT maps near our targets. Because the maps are only
$\sim 6 \arcmin$ in size, we use in the comparison the {\it Planck} CO
maps at their original resolution of 5.5$\arcmin$. We convolve the SMT
data to this resolution and compare values at the centre positions of
the SMT maps (Fig.~\ref{fig:compare_CO}a). The SMT values are for the
sum of $^{12}$CO(2--1) and $^{13}$CO(2--1) lines and are plotted
against 0.5 times the {\it Planck} CO(1--0) estimates. This
corresponds to the assumption of the (2--1)/(1--0)=0.5 line ratio that
was used in the CO correction of the {\it Planck} 217\,GHz data. The
agreement is very good and shows that the performed CO corrections
were at the appropriate level. The convolution of the SMT data could
be calculated only using the extent of the observed map, which may
cause some overestimation of the SMT values (assuming that the CO
emission decreases with distance from the SCUBA-2 clump). 
If we assume consistent calibration of the SMT and PMO observations,
the comparison of the two frames in Fig.~\ref{fig:compare_CO} shows
that the CO line ratio (2-1)/(1-0) tends to be below 0.5. This would
agree with the interpretation that most of the objects are very cold
and embedded in more extended and CO-bright envelopes. This also
suggests low CO contamination for the SCUBA-2 850\,$\mu$m data.

\subsubsection{SEDs of {\it Herschel} and {\it Planck} data}
\label{result:large}

\begin{figure*}
\sidecaption
\includegraphics[width=18cm]{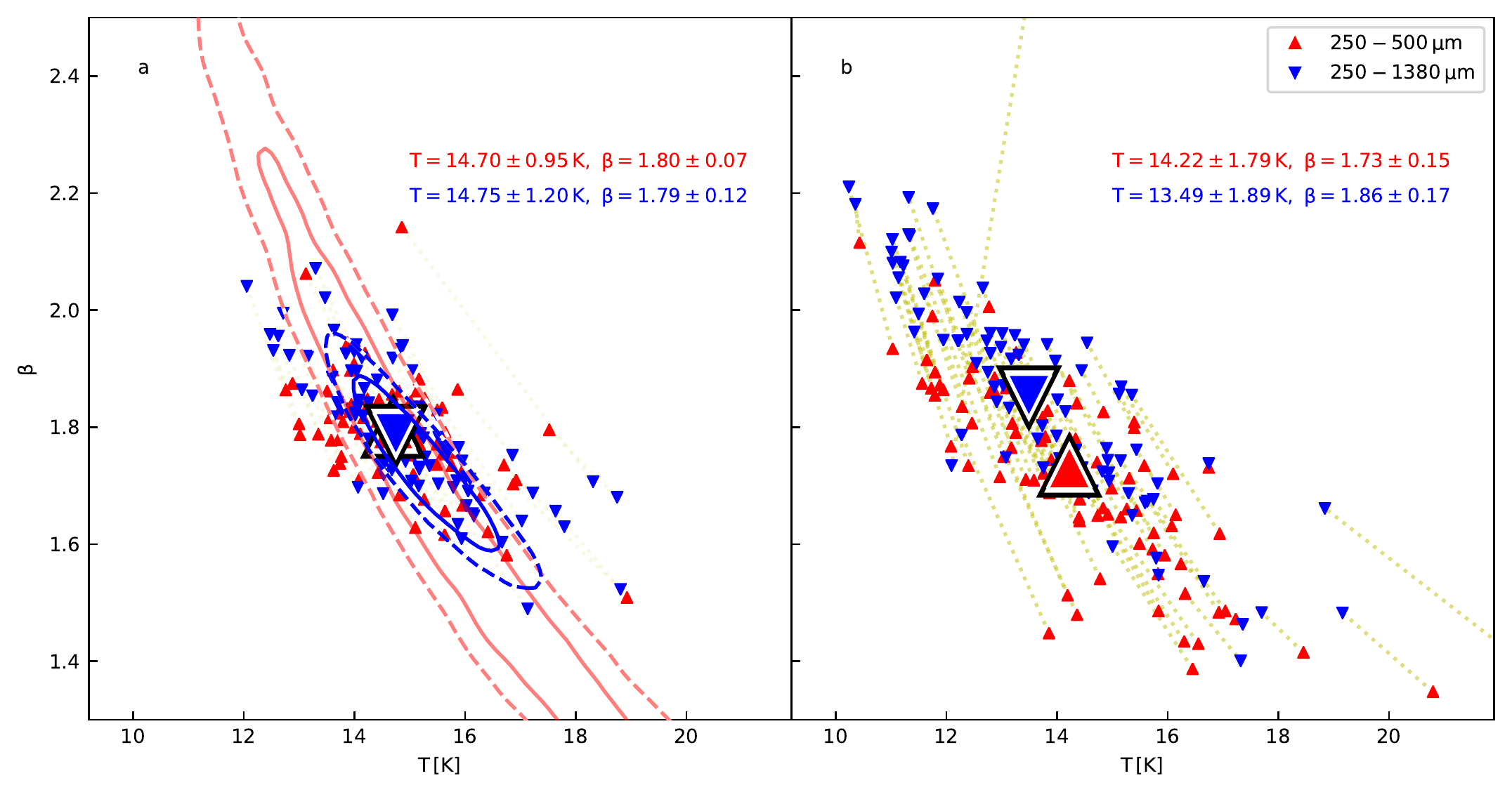}
\caption{
Results of MBB fits to $Herschel$ and $Planck$ data averaged within
8$\arcmin$ of the centre of the SCUBA-2 fields. The results are shown
for absolute surface brightness measurements (frame a) and after
subtraction of the local background emission (frame b). The symbols
correspond to the band combinations listed in the legend in frame b.
The large symbols denote the median values over the fields. In frame
a, the contours show the 67\% (solid lines) and 95\% (dashed lines)
confidence regions for 250-500\,$\mu$m (red) and 250-1380\,$\mu$m
(blue) fits, using error covariance matrices that are the median over
all fields. For the SPIRE-only fits, the adopted error estimates
clearly overestimate the true uncertainty.
}
\label{fig:HP_8}
\end{figure*}

Because the fields were selected based on the availability of {\it
Herschel} data, we can repeat the analysis using only wavelengths
$\lambda \ge$250\,$\mu$m. 
Figure~\ref{fig:HP_8} shows the results from the fits to {\it Herschel}
(250\,$\mu$m, 350\,$\mu$m, and 500\,$\mu$m) and {\it Planck} data. The
maps are convolved to the resolution of 6$\arcmin$ and the surface
brightness is averaged within a radius of $r=8\arcmin$. We have excluded
three fields with only a partial {\it Herschel} coverage. The relative
weighting of the bands follows the uncertainties discussed in
Sect.~\ref{sect:obs_Herschel} and \ref{sect:obs_Planck}, including a
30\% uncertainty for the CO correction of the two lowest frequency {\it
Planck} channels.

Results of Fig.~\ref{fig:HP_8}a are calculated without background
subtraction. Because the intensity zero points of the {\it Herschel}
maps are based on {\it Planck} data, the two data sets are not
completely independent. The median values of the SPIRE fits are 
$T=14.70$\,K and $\beta=1.80$. 
The results are practically identical when the {\it Planck} data are
included,
$T=14.75$\,K and $\beta=1.79$. 
Compared to Fig.~\ref{fig:IP_8am}, the temperatures are lower by 1\,K
and the spectral index values are higher by more than 0.05 units.
These changes are not larger than the estimated uncertainties
given in Figs.~\ref{fig:IP_8am} and \ref{fig:HP_8}.

Figure~\ref{fig:HP_8}a includes examples of confidence regions.
Irrespective of the correctness of the magnitude of the error
estimates, the plot shows the direction along which the $T$ and
$\beta$ values are partly degenerate. 
Figure~\ref{fig:HP_8}b shows the results when estimates of the local
background have been subtracted from the flux values. The background
region is defined by taking a $24 \arcmin \times 24\arcmin$ area
centred on the SCUBA-2 map and rejecting pixels outside or closer than
3$\arcmin$ to {\it Herschel} map borders. The reference area is
defined by the remaining pixels where the 857\,GHz surface brightness
is below its 10\% percentile. We subtract at each wavelength the mean
surface brightness value of the reference area. The scatter of the
($T$, $\beta$) estimates has increased, as expected because of the
smaller residual emission above the background. For the median $\beta$
value is now lower for the 250-500\,$\mu$m fits and higher for the
250-1380\,$\mu$m fits. This would be contrary to the expectation of
$\beta$ decreasing at millimetre wavelengths.

All PGCC clumps with centre coordinates within a radius of 5$\arcmin$
of the SCUBA-2 map centres are marked in the maps of
Appendix~\ref{sect:allmaps}. They are also listed in
Table~\ref{table:PGCC}, including the PGCC values of the spectral
index and the colour temperature (with $\beta=2$ and with free
$\beta$). The PGCC values result from fits to the cold emission
component, after the subtraction of the warm component that is defined
by the IRAS 100\,$\mu$m values and the average SED of the region.
Thus, the PGCC temperatures are significantly lower than in the
analysis above and the spectral index values are correspondingly
higher. For the clumps of Table~\ref{table:PGCC} for which $T$ and
$\beta$ estimates are available, the quartile points (25\%, 50\%, and
75\% percentiles) are 10.7\,K, 11.6\,K, and 12.3\,K for the colour
temperature and 2.01, 2.18, and 2.37 for the spectral index. The
$\beta$ values for the PGCC clumps are thus higher than the values of
the entire fields (Fig.~\ref{fig:IP_8am}). This is not yet a direct
evidence of a change in the intrinsic dust properties. By averaging
emission over large areas, we have also average SEDs with different
temperatures and this can lead to lower apparent $\beta$ values.
Therefore, it is necessary to investigate the spectra also at higher
spatial resolution.

\longtab{
\begin{longtable}{lcccccccccc}
\caption[]{Properties of PGCC clumps in the SCUBA-2 fields according to the PGCC catalogue.} \\
\hline\hline
Field  &   No$^a$  &   RA      & DEC        & $l$   &  $b$  &  $T(\beta=2)$ & $T$  & $\beta$ & $N({\rm H}_2)$  \\
       &       & (J2000.0) & (J2000.0)  & (deg) & (deg) &   (K)         & (K)  &         & ($10^{21}\,{\rm cm}^{-2})$ \\
\hline
\endfirsthead

\caption{continued.}\\
\hline\hline
Field  &   No$^a$  &   RA   & DEC           & $l$   &  $b$  &  $T(\beta=2)$ & $T$  & $\beta$ & $N({\rm H}_2)$  \\
       &       & (J2000.0) & (J2000.0)  & (deg) & (deg) &   (K)         & (K)  &         & ($10^{21}\,{\rm cm}^{-2})$ \\
\hline
\endhead
\hline
\endfoot
     G003.7+18.3A1 & 1 &  16:48:53.8 & -15:35:34 &  3.74 & 18.32 &  12.0 (0.2) &  10.6 (0.7) &  2.4 (0.2) &  10.0 (2.5) \\
     G004.4+15.9A1 & 1 &  16:58:36.1 & -16:28:56 &  4.41 & 15.90 &  14.1 (0.3) &  14.1 (0.7) &  2.0 (0.1) &   - \\
     G007.5+21.1A1 & 1 &  16:48:08.2 & -11:01:20 &  7.57 & 21.13 &  12.9 (0.3) &  12.2 (0.8) &  2.2 (0.2) &  4.2 (1.0) \\
     G007.8+21.1A1 & 1 &  16:48:46.6 & -10:50:58 &  7.82 & 21.10 &  12.8 (0.6) &  11.9 (1.8) &  2.2 (0.4) &  1.9 (1.0) \\
     G038.3-00.9A1 & 1 &  19:04:40.8 & +04:23:45 & 38.35 & -0.93 &  12.2 (1.0) &  - &  - &  7.9 (2.3) \\
     G070.4-01.5A1 & 1 &  20:14:47.2 & +31:56:56 & 70.43 & -1.57 &  13.1 (1.0) &  - &  - &  7.4 (2.0) \\
     G070.7-00.6A1 & 1 &  20:11:52.6 & +32:41:56 & 70.72 & -0.64 &  12.1 (0.6) &  10.7 (1.5) &  2.5 (0.6) &  13.7 (7.1) \\
     G074.1+00.1A1 & 1 &  20:17:51.1 & +35:57:37 & 74.12 &  0.15 &  12.4 (1.8) &  - &  - &  6.4 (3.2) \\
     G111.6+20.2A1 & 1 &  20:57:18.5 & +77:34:05 & 111.65 & 20.20 &  11.6 (0.3) &  11.7 (0.9) &  2.0 (0.2) &  5.1 (1.5) \\
     G113.4+16.9A1 & 1 &  21:59:47.2 & +76:34:00 & 113.41 & 16.98 &  10.7 (0.3) &  9.2 (1.0) &  2.8 (0.4) &  7.7 (3.6) \\
     G115.9+09.4A1 & 1 &  23:24:04.0 & +71:08:41 & 115.93 &  9.47 &  13.4 (0.3) &  12.7 (0.8) &  2.2 (0.2) &  1.4 (0.3) \\
      G127.67+2.65 & 1 &  01:36:40.5 & +65:06:04 & 127.67 &  2.65 &  12.5 (1.3) &  13.7 (3.2) &  1.6 (0.9) &   - \\
      G127.88+2.68 & 1 &  01:38:40.6 & +65:05:32 & 127.88 &  2.68 &  11.5 (0.4) &  11.5 (0.8) &  2.0 (0.2) &  3.2 (0.8) \\
     G130.1+11.0A1 & 1 &  02:28:53.8 & +72:33:05 & 130.17 & 11.06 &  12.7 (0.4) &  13.3 (0.8) &  1.9 (0.2) &  1.8 (0.4) \\
                   & 2 &  02:29:16.1 & +72:43:09 & 130.13 & 11.23 &  12.4 (3.9) &  - &  - &  1.1 (0.9) \\
     G130.1+11.0A2 & 1 &  02:28:53.8 & +72:33:05 & 130.17 & 11.06 &  12.7 (0.4) &  13.3 (0.8) &  1.9 (0.2) &  1.8 (0.4) \\
     G130.3+11.2A1 & 1 &  02:32:22.6 & +72:39:27 & 130.38 & 11.26 &  11.1 (0.4) &  10.2 (1.9) &  2.4 (0.6) &  6.3 (4.4) \\
     G131.7+09.7A1 & 1 &  02:40:04.8 & +70:41:05 & 131.74 &  9.69 &  12.1 (0.3) &  11.7 (0.7) &  2.1 (0.2) &  3.1 (0.7) \\
     G132.0+08.8A1 & 1 &  02:39:13.4 & +69:42:01 & 132.09 &  8.77 &  12.1 (0.5) &  11.8 (1.3) &  2.1 (0.4) &  2.0 (0.8) \\
     G132.0+08.9A1 & 1 &  02:39:43.5 & +69:54:11 & 132.04 &  8.97 &  12.5 (1.3) &  - &  - &  1.6 (0.6) \\
                   & 2 &  02:40:29.1 & +69:51:12 & 132.12 &  8.95 &  -  &  - &  - &   - \\
     G144.6+00.1A1 & 1 &  03:37:00.4 & +55:54:30 & 144.69 &  0.19 &  11.7 (1.3) &  - &  - &  2.2 (0.9) \\
     G144.8+00.7A1 & 1 &  03:40:36.0 & +56:16:09 & 144.88 &  0.78 &  11.8 (1.7) &  - &  - &  2.1 (1.1) \\
     G148.2+00.4A1 & 1 &  03:57:21.5 & +53:52:40 & 148.24 &  0.40 &  11.6 (1.9) &  - &  - &  2.3 (1.3) \\
     G149.2+03.0A2 & 1 &  04:14:52.3 & +55:11:46 & 149.25 &  3.08 &  13.4 (2.4) &  - &  - &  1.9 (1.1) \\
      G149.41+3.38 & 1 &  04:17:08.6 & +55:17:46 & 149.41 &  3.38 &  13.1 (1.3) &  - &  - &  1.5 (0.5) \\
      G149.60+3.45 & 1 &  04:18:24.7 & +55:12:55 & 149.60 &  3.45 &  12.2 (0.5) &  11.1 (1.1) &  2.3 (0.3) &  3.3 (1.3) \\
      G149.68+3.56 & 1 &  04:19:23.0 & +55:14:27 & 149.68 &  3.56 &  13.0 (1.1) &  - &  - &  1.8 (0.6) \\
     G150.2+03.9B1 & 1 &  04:24:00.6 & +55:07:22 & 150.23 &  3.95 &  11.0 (0.5) &  10.1 (2.5) &  2.4 (0.8) &  6.8 (5.8) \\
                   & 2 &  04:23:56.0 & +54:57:45 & 150.34 &  3.83 &  -  &  - &  - &   - \\
     G150.2+03.9B3 & 1 &  04:24:00.6 & +55:07:22 & 150.23 &  3.95 &  11.0 (0.5) &  10.1 (2.5) &  2.4 (0.8) &  6.8 (5.8) \\
     G150.4+03.9A1 & 1 &  04:25:06.8 & +54:56:28 & 150.47 &  3.93 &  12.5 (1.6) &  - &  - &  4.9 (2.2) \\
     G151.4+03.9A1 & 1 &  04:29:51.8 & +54:15:16 & 151.45 &  3.95 &  11.7 (0.3) &  10.1 (0.9) &  2.6 (0.4) &  9.2 (3.5) \\
     G154.0+05.0A1 & 1 &  04:47:14.5 & +53:02:11 & 154.07 &  5.07 &  11.4 (0.3) &  10.2 (0.9) &  2.4 (0.3) &  4.1 (1.7) \\
     G155.45-14.60 & 1 &  03:35:48.5 & +37:40:45 & 155.45 & -14.60 &  11.5 (0.4) &  9.8 (1.3) &  2.7 (0.6) &  7.3 (4.0) \\
     G156.9-08.4A1 & 1 &  04:01:36.3 & +41:28:31 & 156.91 & -8.52 &  12.6 (1.0) &  - &  - &  2.7 (0.8) \\
      G157.12-8.72 & 1 &  04:01:42.6 & +41:11:11 & 157.12 & -8.72 &  13.0 (2.3) &  - &  - &  2.9 (1.7) \\
      G157.93-2.51 & 1 &  04:28:30.7 & +45:06:11 & 157.93 & -2.51 &  12.6 (3.3) &  - &  - &   - \\
     G158.8-21.6A1 & 1 &  03:28:00.0 & +30:07:55 & 158.89 & -21.60 &  10.4 (1.8) &  - &  - &  5.1 (3.5) \\
                   & 2 &  03:27:19.4 & +30:00:40 & 158.84 & -21.78 &  -  &  - &  - &   - \\
     G158.8-34.1A1 & 1 &  02:55:43.0 & +19:52:32 & 158.86 & -34.19 &  12.5 (0.7) &  10.8 (1.8) &  2.6 (0.7) &  3.1 (1.9) \\
     G159.0-08.4A1 & 1 &  04:09:47.6 & +40:07:33 & 158.99 & -8.47 &  13.4 (1.4) &  - &  - &  2.0 (0.7) \\
                   & 2 &  04:10:19.5 & +40:11:51 & 159.02 & -8.35 &  -  &  - &  - &   - \\
     G159.0-08.4A3 & 1 &  04:10:19.5 & +40:11:51 & 159.02 & -8.35 &  -  &  - &  - &   - \\
     G159.1-08.7A1 & 1 &  04:09:21.3 & +39:48:14 & 159.15 & -8.76 &  12.4 (0.7) &  12.0 (1.7) &  2.1 (0.5) &  1.7 (0.8) \\
     G159.21-34.28 & 1 &  02:56:32.6 & +19:38:13 & 159.21 & -34.28 &  12.9 (0.5) &  12.0 (1.6) &  2.3 (0.6) &  3.0 (1.5) \\
     G159.23-34.51 & 1 &  02:56:02.3 & +19:26:13 & 159.23 & -34.51 &  12.1 (0.4) &  10.6 (1.1) &  2.5 (0.4) &  8.7 (3.7) \\
     G159.4-34.3A1 & 1 &  02:56:53.7 & +19:27:53 & 159.41 & -34.37 &  12.3 (0.5) &  10.9 (1.3) &  2.5 (0.5) &  4.5 (2.1) \\
     G159.6-19.6A1 & 1 &  03:36:37.8 & +31:10:58 & 159.76 & -19.63 &  -  &  - &  - &   - \\
     G159.6-19.6A2 & 1 &  03:35:53.7 & +31:12:06 & 159.62 & -19.71 &  12.7 (1.9) &  - &  - &  3.2 (1.6) \\
     G159.6-19.6A3 & 1 &  03:35:53.7 & +31:12:06 & 159.62 & -19.71 &  12.7 (1.9) &  - &  - &  3.2 (1.6) \\
                   & 2 &  03:36:37.8 & +31:10:58 & 159.76 & -19.63 &  -  &  - &  - &   - \\
     G159.7-19.6A1 & 1 &  03:36:37.8 & +31:10:58 & 159.76 & -19.63 &  -  &  - &  - &   - \\
     G160.6-16.7A1 & 1 &  03:48:23.5 & +32:53:39 & 160.64 & -16.72 &  -  &  - &  - &   - \\
     G160.8-09.4A1 & 1 &  04:13:05.7 & +38:08:28 & 160.85 & -9.46 &  12.6 (4.0) &  - &  - &  1.0 (0.9) \\
     G160.8-09.4A2 & 1 &  04:13:05.7 & +38:08:28 & 160.85 & -9.46 &  12.6 (4.0) &  - &  - &  1.0 (0.9) \\
     G161.3-09.3A1 & 1 &  04:15:23.9 & +37:53:03 & 161.36 & -9.33 &  12.5 (1.2) &  12.8 (2.4) &  1.9 (0.7) &  1.0 (0.6) \\
      G161.56-9.29 & 1 &  04:16:15.7 & +37:46:18 & 161.56 & -9.29 &  13.0 (0.7) &  13.3 (1.8) &  1.9 (0.5) &  1.2 (0.5) \\
                   & 2 &  04:16:51.5 & +37:45:42 & 161.65 & -9.22 &  -  &  - &  - &   - \\
     G162.4-08.7A1 & 1 &  04:21:29.0 & +37:34:31 & 162.44 & -8.70 &  12.1 (0.3) &  12.1 (0.7) &  2.0 (0.2) &  5.3 (1.1) \\
      G163.32-8.41 & 1 &  04:25:32.9 & +37:09:00 & 163.32 & -8.41 &  11.7 (0.4) &  11.4 (1.7) &  2.1 (0.5) &  5.7 (3.0) \\
      G163.68-8.33 & 1 &  04:27:06.5 & +36:56:55 & 163.68 & -8.33 &  11.9 (0.4) &  11.2 (1.2) &  2.3 (0.4) &  3.7 (1.5) \\
      G163.82-8.33 & 1 &  04:27:33.2 & +36:50:32 & 163.82 & -8.33 &  11.9 (0.6) &  11.0 (1.9) &  2.3 (0.7) &  3.5 (2.2) \\
     G164.1-08.8A1 & 1 &  04:26:54.5 & +36:14:11 & 164.18 & -8.84 &  11.9 (0.6) &  10.7 (1.6) &  2.5 (0.6) &  2.8 (1.6) \\
      G164.11-8.17 & 1 &  04:29:07.7 & +36:44:35 & 164.11 & -8.17 &  12.3 (0.5) &  11.6 (1.7) &  2.3 (0.6) &  1.8 (1.0) \\
      G164.26-8.39 & 1 &  04:28:50.0 & +36:29:15 & 164.26 & -8.39 &  12.8 (0.5) &  12.1 (1.5) &  2.3 (0.5) &  2.2 (1.0) \\
     G165.1-07.5A1 & 1 &  04:34:50.0 & +36:23:20 & 165.16 & -7.57 &  12.4 (0.4) &  11.9 (1.1) &  2.2 (0.4) &  2.7 (0.9) \\
     G165.3-07.5A1 & 1 &  04:35:42.0 & +36:16:35 & 165.36 & -7.51 &  12.2 (0.3) &  11.6 (0.8) &  2.2 (0.2) &  3.3 (0.8) \\
     G165.6-09.1A1 & 1 &  04:30:53.1 & +34:55:24 & 165.71 & -9.15 &  12.8 (2.3) &  - &  - &  3.0 (1.7) \\
     G167.2-15.3A1 & 1 &  04:14:28.6 & +29:35:25 & 167.23 & -15.33 &  12.2 (0.2) &  12.1 (0.6) &  2.1 (0.2) &  2.7 (0.5) \\
     G168.1-16.3A1 & 1 &  04:13:43.3 & +28:15:15 & 168.10 & -16.38 &  10.6 (0.4) &  9.5 (2.3) &  2.5 (0.7) &  18.4 (15.9) \\
     G169.1-01.1A1 & 1 &  05:12:18.6 & +37:20:28 & 169.15 & -1.15 &  13.6 (1.1) &  17.1 (1.4) &  1.5 (0.2) &  2.1 (0.5) \\
     G170.0-16.1A1 & 1 &  04:20:13.3 & +27:06:19 & 170.00 & -16.13 &  12.3 (2.1) &  - &  - &  1.6 (1.0) \\
     G170.1-16.0A1 & 1 &  04:20:50.2 & +27:03:10 & 170.14 & -16.07 &  11.4 (0.7) &  11.3 (2.9) &  2.0 (0.8) &  5.2 (4.3) \\
                   & 2 &  04:21:17.5 & +26:59:27 & 170.26 & -16.04 &  11.0 (1.0) &  10.8 (3.6) &  2.1 (1.2) &  6.1 (6.1) \\
     G170.8-18.3A1 & 1 &  04:15:23.5 & +25:02:02 & 170.81 & -18.34 &  13.3 (1.0) &  - &  - &  1.9 (0.5) \\
     G170.9-15.8A1 & 1 &  04:24:14.0 & +26:37:19 & 171.00 & -15.80 &  10.6 (0.3) &  10.0 (2.4) &  2.3 (0.6) &  7.5 (6.2) \\
     G171.14-17.57 & 1 &  04:18:50.1 & +25:19:29 & 171.14 & -17.57 &  11.3 (0.2) &  9.8 (0.7) &  2.6 (0.3) &  8.6 (2.7) \\
     G171.8-15.3A1 & 1 &  04:28:06.8 & +26:22:00 & 171.80 & -15.32 &  11.8 (1.0) &  - &  - &  3.5 (1.3) \\
     G172.8-14.7A1 & 1 &  04:33:01.9 & +25:59:55 & 172.84 & -14.74 &  12.2 (0.3) &  12.0 (1.4) &  2.1 (0.4) &  2.8 (1.2) \\
     G173.1-13.3A1 & 1 &  04:38:42.4 & +26:41:50 & 173.14 & -13.32 &  12.6 (1.7) &  - &  - &  1.8 (0.9) \\
     G173.3-16.2A1 & 1 &  04:29:20.5 & +24:37:35 & 173.35 & -16.26 &  12.0 (0.6) &  - &  - &  6.9 (1.3) \\
     G173.9-13.7A1 & 1 &  04:39:27.7 & +25:48:25 & 173.95 & -13.76 &  12.1 (1.3) &  - &  - &  6.4 (2.5) \\
     G174.0-15.8A1 & 1 &  04:32:44.7 & +24:24:09 & 174.05 & -15.82 &  11.3 (0.7) &  - &  - &  8.4 (2.1) \\
     G174.4-15.7A1 & 1 &  04:33:58.6 & +24:10:21 & 174.42 & -15.76 &  12.5 (3.8) &  - &  - &  1.7 (1.5) \\
     G174.7-15.4A1 & 1 &  04:35:41.7 & +24:09:28 & 174.70 & -15.47 &  11.3 (0.2) &  10.6 (0.7) &  2.3 (0.2) &  12.0 (3.3) \\
     G174.8-17.1A1 & 1 &  04:30:43.8 & +22:55:59 & 174.90 & -17.13 &  11.8 (0.4) &  11.1 (0.9) &  2.3 (0.4) &  3.0 (1.0) \\
     G175.2+01.2A2 & 1 &  05:38:56.8 & +33:41:15 & 175.21 &  1.29 &  11.7 (0.7) &  12.5 (1.4) &  1.8 (0.4) &  1.5 (0.6) \\
     G175.4-16.8A2 & 1 &  04:33:25.1 & +22:42:56 & 175.50 & -16.80 &  11.8 (1.0) &  - &  - &  5.3 (1.8) \\
     G177.6-20.3A1 & 1 &  04:27:17.3 & +18:51:43 & 177.64 & -20.37 &  12.9 (0.4) &  12.8 (1.6) &  2.0 (0.5) &  1.7 (0.7) \\
     G181.8+00.3A1 & 1 &  05:51:12.6 & +27:30:15 & 181.86 &  0.32 &  11.6 (0.5) &  11.4 (1.2) &  2.1 (0.4) &  3.6 (1.4) \\
     G201.2+00.4A1 & 1 &  06:31:33.1 & +10:34:06 & 201.27 &  0.45 &  12.5 (7.2) &  - &  - &  1.4 (1.6) \\
      G201.83+2.83 & 1 &  06:41:13.0 & +11:09:56 & 201.83 &  2.83 &  13.4 (2.3) &  - &  - &   - \\
      G202.00+2.65 & 1 &  06:40:53.2 & +10:55:56 & 202.00 &  2.65 &  12.9 (2.4) &  - &  - &  1.6 (1.0) \\
      G202.54+2.46 & 1 &  06:41:10.8 & +10:22:19 & 202.54 &  2.46 &  14.1 (0.8) &  15.0 (1.8) &  1.8 (0.4) &  2.8 (1.0) \\
     G204.4-11.3A2 & 1 &  05:55:43.9 & +02:12:19 & 204.49 & -11.33 &  11.1 (1.9) &  - &  - &  2.6 (1.5) \\
     G204.8-13.8A1 & 1 &  05:47:26.0 & +00:42:47 & 204.83 & -13.86 &  13.2 (1.8) &  - &  - &  5.7 (2.6) \\
     G210.90-36.55 & 1 &  04:35:08.6 & -14:15:27 & 210.90 & -36.55 &  12.0 (0.3) &  11.0 (0.6) &  2.4 (0.2) &  5.3 (1.2) \\
     G215.44-16.38 & 1 &  05:57:01.7 & -09:33:14 & 215.44 & -16.38 &  12.5 (0.4) &  12.1 (0.8) &  2.1 (0.3) &  1.7 (0.4) \\
      G216.76-2.58 & 1 &  06:49:13.6 & -04:34:59 & 216.76 & -2.58 &  10.4 (1.1) &  - &  - &  4.0 (1.7) \\
      G219.13-9.72 & 1 &  06:27:28.1 & -09:54:36 & 219.12 & -9.79 &  12.9 (1.1) &  13.6 (2.3) &  1.8 (0.7) &  1.4 (0.8) \\
                   & 2 &  06:27:42.7 & -09:53:06 & 219.13 & -9.72 &  13.0 (1.6) &  14.1 (4.3) &  1.7 (1.1) &   - \\
     G219.22-10.02 & 1 &  06:26:46.9 & -10:05:45 & 219.22 & -10.02 &  12.0 (2.7) &  - &  - &  1.6 (1.2) \\
      G219.36-9.71 & 1 &  06:28:09.8 & -10:05:11 & 219.36 & -9.71 &  12.9 (0.5) &  13.0 (1.1) &  2.0 (0.3) &  3.4 (1.0) \\
\label{table:PGCC}  
\end{longtable} 
\tablefoot{
$^a$Index within the field, used in the plots of Appendix.~\ref{sect:allmaps}.
}
} 

\subsection{Dust spectrum at small scales} \label{result:small}


We investigate the small-scale dust emission using the combination of
{\it Herschel} 250-500\,$\mu$m and SCUBA-2 850\,$\mu$m data and clumps
extracted with the Fellwalker method (see Sect.~\ref{sect:clumps}). 
We use four clump samples that correspond to: (1) $\theta_{\rm F}=500\arcsec$
and LM source masks, (2) $\theta_{\rm F}=200\arcsec$, SM masks, and $\tau_{\rm
high}$ part of the clumps, (3) $\theta_{\rm F}=200\arcsec$ and M850 masks, and
(4) data reduction with $\theta_{\rm F}=200\arcsec$ and without external
source masks.
The four alternatives are used to examine the robustness of the
results with respect to the details of the data reduction and clump
selection.

After the rejection of clumps further than 5$\arcmin$ from the field
centres or with $S(850\mu{\rm m})$ values below 0.1\,Jy, 
there remain of the order of 200 clumps, the number depending on the
data version. Many SED fits are not reliable and we carry out further
selection based on the $\chi^2$ values of the fits.

\begin{figure*}
\includegraphics[width=18cm]{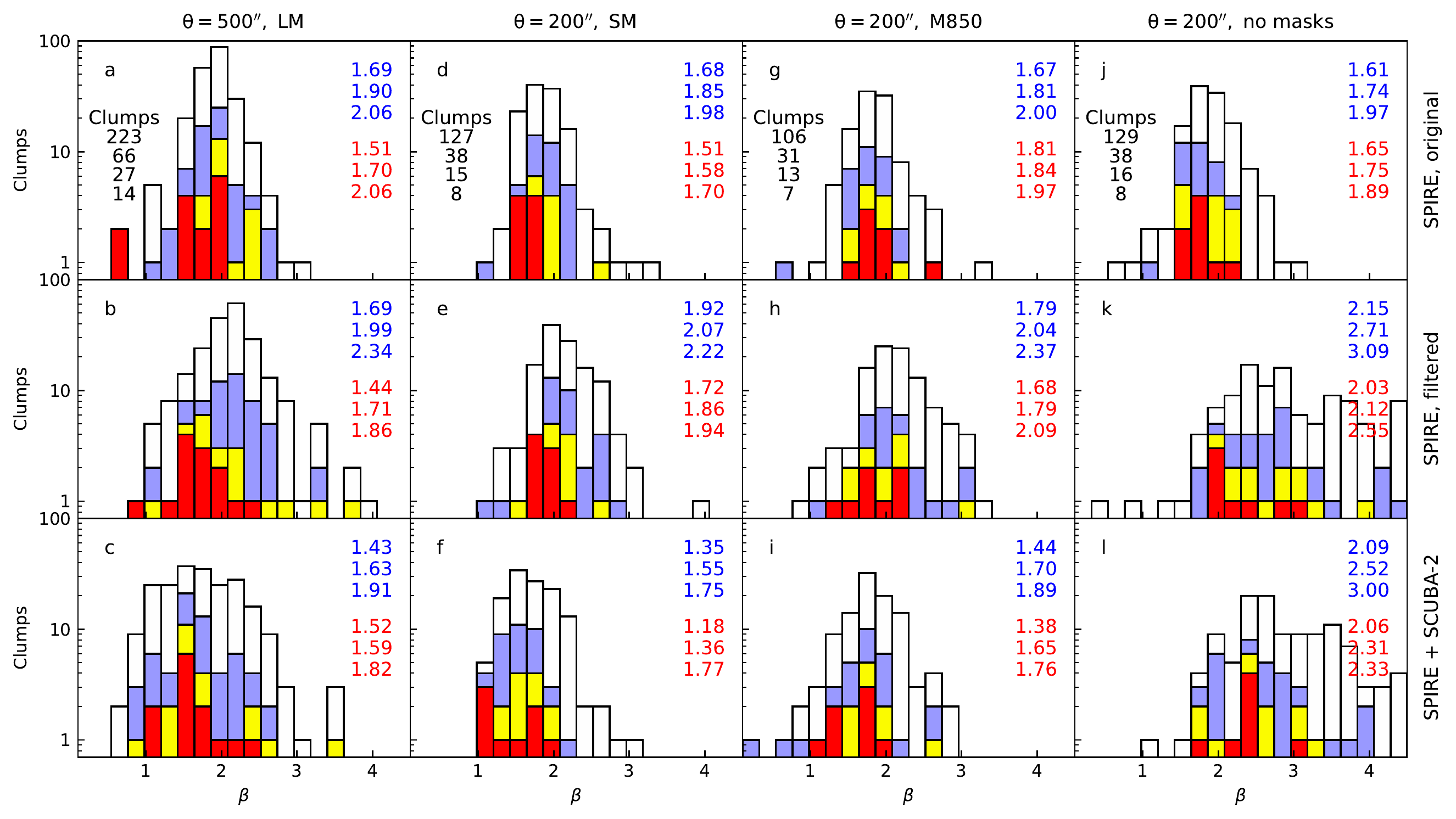}
\caption{
Distributions of $\beta$ values in fits to {\it Herschel} and SCUBA-2
850\,$\mu$m observations of clumps.  Results are shown for original
SPIRE data (first row), filtered SPIRE data (second row) and the
combination of filtered SPIRE and 850\,$\mu$m data (third row). The
columns correspond to different versions of data reduction, as
indicated above the frames. The white, blue, yellow, and red
histograms correspond, respectively, to $P(\chi^2$) values of 85\%,
25\%, 10\%, and 5\% (see text). The number of clumps in each sample
and the quartile values (25\%, 50\% and 75\% percentiles) of the blue
and red histograms are given in the frames.
}
\label{fig:betahis}
\end{figure*}

Figure~\ref{fig:betahis}a-c shows the $\beta$ distributions for the
reduction with $\theta_{\rm F}=500\arcsec$ and LM masks, using the full clump
apertures ($\tau_{\rm all}$). The clumps are divided into subsamples
where the $\chi^2$ values of the fits are within the best
$P(\chi^2)$=85\%, 25\%, 10\%, or 5\%. Thus, $P(\chi^2)$=5\%
corresponds to the strictest selection criteria (least clumps) with
$\chi^2$ values below the 5\% percentile point. The first two frames
show the results for fits to {\it Herschel} 250-500\,$\mu$m bands,
before and after filtering by the SCUBA-2 pipeline. The filtering
could lead to different results, if the filtering procedure introduces
additional uncertainty or bias, or if the true SEDs are affected by
the removal of large-scale emission.

In the following, we refer mainly to the numbers for the
$P(\chi^2)=25$\% samples. For the original SPIRE data (the
$\theta_{\rm F}=500\arcsec$ case), the median value is 
$\beta$=1.90. Based on the distribution of $\beta$ values, the
error of the mean is less than 0.05.
The filtering increases the width of the distributions, partly because
of the larger relative uncertainty of the lower flux values and
possibly via larger background fluctuations. If the densest parts of
the clumps have lower colour temperatures and higher $\beta$ values
than their environment, one could expect the filtering to increase in
the $\beta$ estimates. In Fig.~\ref{fig:betahis}, there are only
weak indications of such a trend.

Figure~\ref{fig:betahis}c shows the results for fits to combined {\it
Herschel} and SCUBA-2 data, with self-consistent filtering across all
bands. The median value has decreased to 
$\beta=1.63$. 
The same trend exists for the $P(\chi^2)=5$\% sample, although it not
very reliable because of the low number of clumps.

\begin{figure*}
\includegraphics[width=18.0cm]{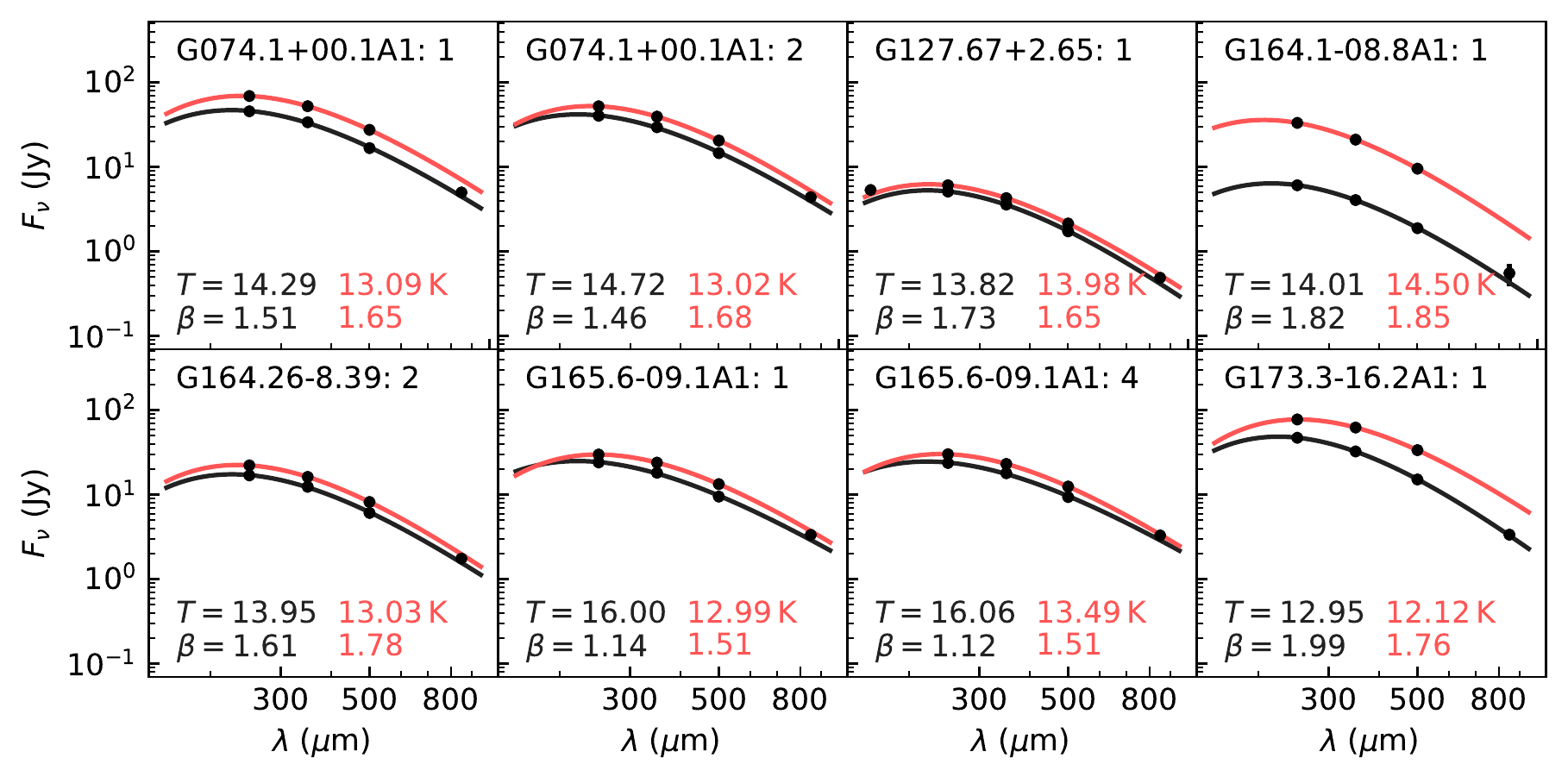}
\caption{
The SEDs of eight sources from the $P(\chi^2)=5$\% subsample of SM
clumps in Fig.~\ref{fig:betahis}f. Plots include fits to the original
SPIRE data (upper red curve and red numbers) and to the filtered SPIRE
data plus the SCUBA-2 850\,$\mu{\rm m}$ point. The clump numbers given
after the field name are the same as in the figures of
Appendix~\ref{sect:allmaps}. The 160\,$\mu$m data points is available
for one of the clumps: it is plotted in the figure but is not part of
the fit. 
}
\label{fig:plot_example_SEDs}
\end{figure*}

Figure~\ref{fig:betahis}d-f shows a second analysis that is a priori
very far from the previous case, using $\theta_{\rm F}=200\arcsec$
filter scale, SM masks, and fluxes integrated over the $\tau_{\rm
high}$ apertures. More of the large-scale emission is being filtered
out and the smaller source mask and clump size further suppress the
contribution of extended emission. The number of clumps is lower than
in the previous case and also the $\beta$ distributions are slightly
more narrow. Figure~\ref{fig:plot_example_SEDs} shows the SEDs for the
clumps in the $P(\chi^2)=5$\% sample. In six out of eight cases, the
addition of the 850\,$\mu$m has decreased the $\beta$ estimate.
Table~\ref{table:SED} lists the fit parameters for the
$P(\chi^2)=25$\% sample.

\longtab{
\begin{landscape}
\begin{longtable}{llcccccccccc}
\caption[]{SED parameters for the $\theta_{\rm F}=500\arcsec$ maps, the LM
masks, the $\tau_{\rm high}$ apertures, and the $P(\chi^2$)=25\%
sample of clumps.}
\label{table:SED}
\\
\hline\hline
Field &  \#$^a$ & RA (J2000) & DEC (J2000) & 
$F_{\nu}(250\,\mu{\rm m})$ & $F_{\nu}(350\,\mu{\rm m})$ &
$F_{\nu}(500\,\mu{\rm m})$ & $F_{\nu}(850\,\mu{\rm m})$ &
$T_{\rm 3}^b$       &   $\beta_{\rm 3}^b$    &
$T_{\rm 4}^c$       &   $\beta_{\rm 4}^c$             \\
 & & & & (Jy) & (Jy) & (Jy) & (Jy) &  (K) &  & (K) &  \\
\hline
\endfirsthead
\caption{continued.}\\
\hline\hline
Field &  \#$^a$  & RA (J2000) & DEC (J2000) & 
$F_{\nu}(250\,\mu{\rm m})$ & $F_{\nu}(350\,\mu{\rm m})$ &
$F_{\nu}(500\,\mu{\rm m})$ & $F_{\nu}(850\,\mu{\rm m})$ &
$T_{\rm 3}^b$  & $\beta_{\rm 3}^b$  &
$T_{\rm 4}^c$  & $\beta_{\rm 4}^c$  \\
 & & & & (Jy) & (Jy) & (Jy) & (Jy) &  (K) &  & (K) &  \\
\hline
\endhead
\hline
\endfoot
G070.7-00.6A1 &  1 & 20 12 02.9 & +32 41 47.0 &  24.8(0.6) &  18.7(0.4) &   8.6(0.2) &   2.9(0.3) &  10.2(1.4) &  2.68(0.50) &  13.5(2.1) &  1.72(0.39) \\
G070.7-00.6A1 &  2 & 20 12 02.9 & +32 41 51.7 &  19.1(0.4) &  14.6(0.3) &   6.9(0.2) &   2.2(0.2) &  10.6(1.6) &  2.49(0.50) &  13.3(1.8) &  1.72(0.35) \\
G070.7-00.6A1 &  3 & 20 12 02.9 & +32 41 46.9 &  24.9(0.6) &  18.6(0.4) &   8.5(0.2) &   2.7(0.3) &  10.0(1.4) &  2.80(0.50) &  12.8(1.9) &  1.91(0.41) \\
G070.7-00.6A1 &  4 & 20 12 02.9 & +32 41 47.4 &  23.0(0.6) &  17.7(0.5) &   8.4(0.3) &   2.6(0.3) &  10.3(1.5) &  2.54(0.51) &  12.9(1.7) &  1.78(0.35) \\
G070.7-00.6A1 &  5 & 20 11 57.6 & +32 41 36.9 &  32.5(0.9) &  23.5(0.6) &  10.5(0.3) &   3.4(0.4) &  10.4(1.5) &  2.71(0.49) &  13.3(2.2) &  1.88(0.41) \\
G074.1+00.1A1 &  1 & 20 17 57.0 & +35 55 41.9 &  45.8(0.9) &  34.0(0.7) &  16.8(0.3) &   5.0(0.5) &  12.0(2.2) &  2.02(0.50) &  14.3(2.0) &  1.51(0.33) \\
G074.1+00.1A1 &  2 & 20 17 57.1 & +35 55 40.6 &  40.6(0.8) &  29.7(0.6) &  14.7(0.3) &   4.4(0.4) &  12.3(2.3) &  1.97(0.49) &  14.7(2.1) &  1.46(0.32) \\
G074.1+00.1A1 &  3 & 20 17 57.2 & +35 55 47.3 &  53.1(1.1) &  40.0(0.8) &  19.9(0.4) &   6.0(0.6) &  11.8(2.0) &  2.07(0.49) &  14.1(1.9) &  1.51(0.32) \\
G074.1+00.1A1 &  4 & 20 17 57.5 & +35 55 50.6 &  51.9(1.1) &  39.1(0.8) &  19.5(0.4) &   6.0(0.6) &  11.8(2.1) &  2.04(0.51) &  14.3(2.1) &  1.48(0.33) \\
G074.1+00.1A1 &  5 & 20 17 54.6 & +35 56 49.1 &  63.2(1.5) &  47.9(1.1) &  23.8(0.5) &   7.4(0.8) &  11.5(1.8) &  2.12(0.48) &  14.1(2.0) &  1.51(0.33) \\
G074.1+00.1A1 &  6 & 20 17 56.6 & +35 56 03.5 &  55.8(1.2) &  41.8(0.9) &  20.6(0.4) &   6.5(0.7) &  11.6(1.8) &  2.14(0.48) &  14.4(2.2) &  1.49(0.34) \\
G074.1+00.1A1 &  7 & 20 17 54.9 & +35 56 41.5 &  68.1(1.4) &  51.4(1.1) &  25.1(0.5) &   8.1(0.8) &  11.3(1.7) &  2.23(0.48) &  14.3(2.2) &  1.49(0.34) \\
G074.1+00.1A1 &  8 & 20 17 53.8 & +35 57 24.0 &  64.4(1.7) &  48.4(1.2) &  23.6(0.6) &   7.4(0.8) &  11.4(1.8) &  2.20(0.49) &  14.1(2.2) &  1.54(0.35) \\
G074.1+00.1A1 &  9 & 20 17 53.1 & +35 57 45.0 &  65.9(1.5) &  50.1(1.2) &  24.1(0.6) &   7.1(0.9) &  10.8(1.6) &  2.39(0.48) &  12.9(1.8) &  1.79(0.36) \\
G074.1+00.1A1 & 10 & 20 17 54.7 & +35 56 46.0 &  63.1(1.4) &  47.6(1.0) &  23.5(0.5) &   7.5(0.8) &  11.6(1.9) &  2.13(0.49) &  14.4(2.2) &  1.47(0.34) \\
G074.1+00.1A1 & 11 & 20 17 55.0 & +35 56 37.6 &  60.8(1.3) &  46.3(1.0) &  22.7(0.5) &   7.3(0.8) &  11.2(1.7) &  2.24(0.48) &  14.0(2.1) &  1.52(0.35) \\
G111.6+20.2A1 &  1 & 20 57 15.1 & +77 36 04.6 &  18.5(0.4) &  14.2(0.3) &   8.0(0.2) &   2.4(0.2) &  15.2(4.9) &  1.16(0.54) &  14.9(2.3) &  1.21(0.32) \\
G127.67+2.65 &  1 & 01 36 52.3 & +65 04 06.5 &   5.1(0.1) &   3.6(0.1) &   1.7(0.0) &   0.5(0.1) &  13.0(2.8) &  1.91(0.51) &  13.8(2.5) &  1.73(0.42) \\
G149.2+03.0A2 &  1 & 04 15 06.3 & +55 11 18.7 &   2.2(0.1) &   1.5(0.0) &   0.9(0.1) &   0.2(0.0) &  84.2(6.4) &  -0.51(0.39) &  17.2(7.7) &  1.16(0.67) \\
G151.4+03.9A1 &  1 & 04 29 42.9 & +54 14 54.9 &   4.8(0.1) &   3.4(0.1) &   1.6(0.1) &   0.3(0.1) &  11.8(2.3) &  2.26(0.55) &  11.8(2.1) &  2.27(0.52) \\
G160.8-09.4A2 &  1 & 04 13 05.8 & +38 10 42.0 &   2.7(0.1) &   2.1(0.1) &   1.2(0.0) &   0.3(0.1) &  14.0(5.1) &  1.34(0.66) &  13.8(3.8) &  1.39(0.56) \\
G162.4-08.7A1 &  4 & 04 21 32.9 & +37 33 56.6 &  40.3(0.9) &  28.7(0.6) &  14.0(0.3) &   4.6(0.5) &  13.0(2.9) &  1.87(0.52) &  16.2(3.5) &  1.29(0.38) \\
G162.4-08.7A1 &  5 & 04 21 31.2 & +37 34 05.6 &  49.5(1.0) &  36.6(0.8) &  17.9(0.4) &   6.8(0.8) &  11.9(2.1) &  2.08(0.51) &  15.7(3.9) &  1.29(0.44) \\
G164.1-08.8A1 &  1 & 04 26 58.4 & +36 12 46.4 &   6.1(0.2) &   4.1(0.1) &   1.9(0.1) &   0.6(0.2) &  13.5(3.4) &  1.93(0.56) &  14.0(3.9) &  1.82(0.55) \\
G164.26-8.39 &  1 & 04 28 43.9 & +36 27 34.1 &  16.9(0.3) &  12.4(0.3) &   6.0(0.1) &   1.8(0.2) &  11.9(2.0) &  2.11(0.48) &  14.1(2.2) &  1.60(0.37) \\
G164.26-8.39 &  2 & 04 28 43.8 & +36 27 33.5 &  17.0(0.3) &  12.5(0.3) &   6.1(0.1) &   1.8(0.2) &  12.1(2.1) &  2.02(0.48) &  13.9(2.0) &  1.61(0.34) \\
G164.26-8.39 &  3 & 04 28 43.6 & +36 27 29.8 &  15.7(0.3) &  11.5(0.2) &   5.6(0.1) &   1.7(0.2) &  12.0(2.1) &  2.06(0.49) &  13.5(2.2) &  1.70(0.41) \\
G165.6-09.1A1 &  1 & 04 30 51.7 & +34 56 38.5 &  24.2(0.5) &  18.3(0.4) &   9.5(0.2) &   3.4(0.3) &  13.0(3.1) &  1.68(0.55) &  16.0(3.5) &  1.14(0.38) \\
G165.6-09.1A1 &  2 & 04 30 52.0 & +34 56 31.7 &  26.1(0.5) &  20.0(0.4) &  10.5(0.2) &   3.7(0.4) &  12.7(3.0) &  1.73(0.55) &  15.7(3.2) &  1.16(0.38) \\
G165.6-09.1A1 &  3 & 04 30 52.1 & +34 56 28.9 &  27.6(0.6) &  21.1(0.4) &  11.0(0.2) &   4.0(0.4) &  12.5(2.4) &  1.80(0.48) &  15.9(3.4) &  1.13(0.38) \\
G165.6-09.1A1 &  4 & 04 30 51.6 & +34 56 40.3 &  23.8(0.5) &  18.0(0.4) &   9.4(0.2) &   3.3(0.3) &  13.0(3.1) &  1.68(0.55) &  16.1(3.4) &  1.12(0.38) \\
G173.3-16.2A1 &  1 & 04 29 24.0 & +24 34 27.1 &  47.4(1.0) &  32.6(0.7) &  15.2(0.3) &   3.4(0.4) &  12.5(2.3) &  2.09(0.48) &  13.0(1.6) &  1.99(0.32) \\
G173.9-13.7A1 &  1 & 04 39 32.4 & +25 47 21.8 &  25.3(1.1) &  16.7(0.8) &   8.2(0.4) &   2.3(0.5) &  16.0(9.1) &  1.43(0.69) &  16.3(7.5) &  1.39(0.61) \\
G201.2+00.4A1 &  5 & 06 31 31.7 & +10 32 55.2 &   7.1(0.2) &   5.6(0.1) &   2.7(0.1) &   1.1(0.1) &  10.0(1.6) &  2.62(0.59) &  14.4(3.5) &  1.38(0.49) \\
G204.4-11.3A2 &  1 & 05 55 38.1 & +02 11 20.8 &  22.7(0.5) &  17.6(0.4) &   9.0(0.2) &   3.3(0.3) &  11.7(2.0) &  2.00(0.50) &  15.4(3.3) &  1.22(0.41) \\
G204.4-11.3A2 &  2 & 05 55 38.8 & +02 11 28.9 &  24.4(0.5) &  19.0(0.4) &   9.8(0.2) &   3.6(0.4) &  11.9(2.0) &  1.94(0.49) &  15.3(3.1) &  1.21(0.39) \\
G204.4-11.3A2 &  3 & 05 55 39.8 & +02 11 43.8 &  26.9(0.6) &  20.9(0.4) &  10.4(0.2) &   4.0(0.4) &  11.1(1.8) &  2.20(0.49) &  14.6(3.0) &  1.37(0.42) \\
G219.13-9.72 &  1 & 06 27 29.7 & -09 54 20.9 &   8.6(0.2) &   6.5(0.2) &   3.3(0.1) &   0.9(0.1) &  12.1(2.5) &  1.93(0.55) &  13.1(1.9) &  1.69(0.37) \\
\end{longtable}
\tablefoot{
$^a$Clump number (see Appendix~\ref{sect:allmaps}).
$^b$Fit to filtered SPIRE bands, 250-500\,$\mu{\rm m}$.
$^c$Fit to filtered SPIRE bands and SCUBA-2 data, 250-850\,$\mu{\rm m}$.
}
\end{landscape}
}

Figure~\ref{fig:betahis}g-i shows a case where maps were made using
M850 source masks and $\theta_{\rm F}=200\arcsec$ filtering scale.
Compared to the previous cases, the clump selection is thus more
directly based on the SCUBA-2 850\,$\mu$m data. However, the result
for the unfiltered SPIRE data show that underlying clump samples are
still relatively similar. The inclusion of the 850\,$\mu$m point again
decreases the median $\beta$ for both the $P(\chi^2)=5$\% and
$P(\chi^2)=25$\% samples.

The final plots in Fig.~\ref{fig:betahis}j-l correspond to maps made
with $\theta_{\rm F}=200\arcsec$ but without any external masks. This should
favour the extraction of the most compact sources although the total
number of sources is not lower than in the previous two cases. When
filtering is applied to SPIRE data, the $\beta$ distribution moves
towards higher values and becomes wider. We interpret this mostly as a
sign of general increase of uncertainty. In the combined fits of SPIRE
and SCUBA-2 data the spectral index values are also much higher than
in the SPIRE-only fits. 

A table of the fit parameters can be found in
Appendix~\ref{sect:allhisto}, where also the uncertainty of the
850\,$\mu$m calibration is briefly discussed.

\begin{figure}
\includegraphics[width=8.8cm]{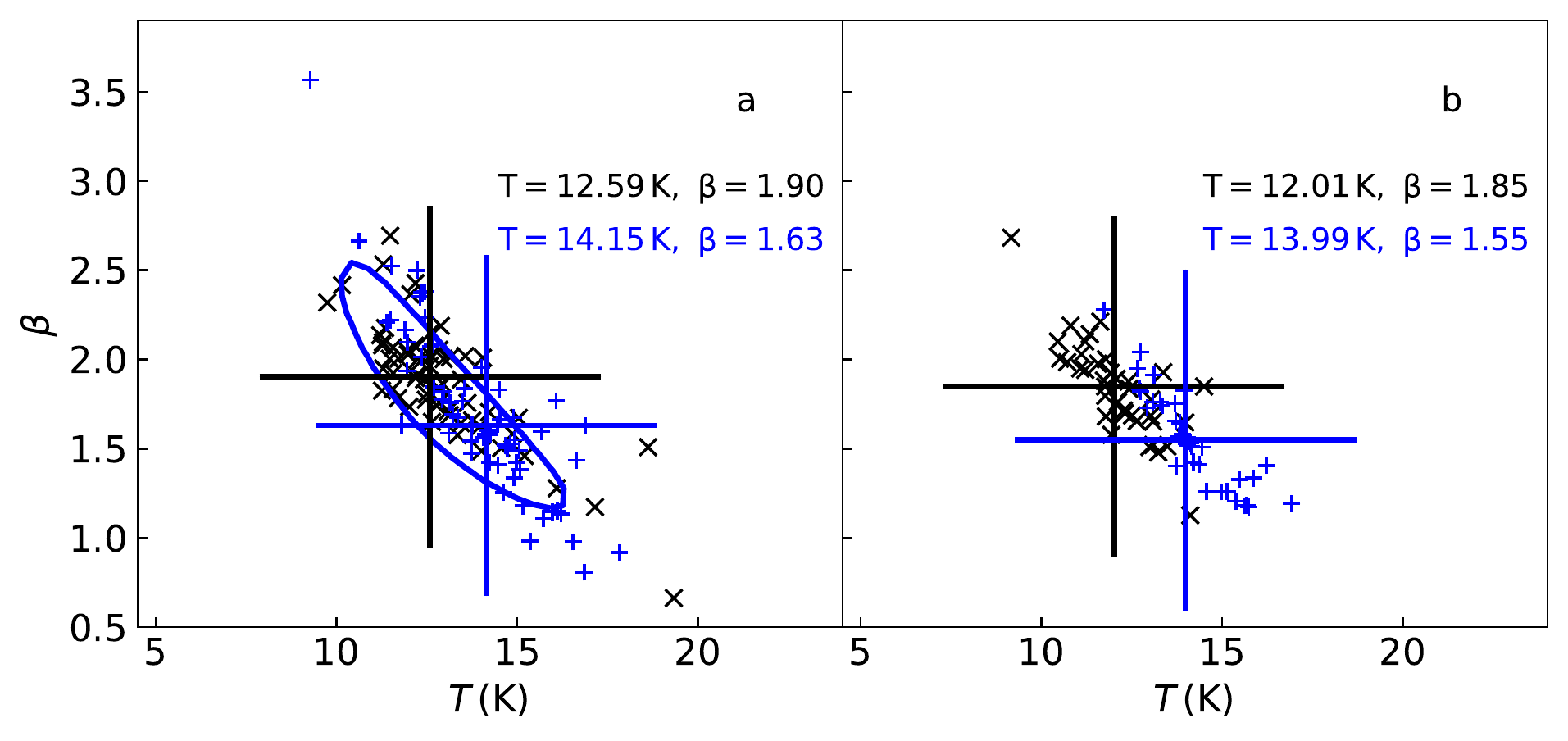}
\caption{
Spectral index $\beta$ vs. colour temperature $T$ for the clump
samples of Fig.~\ref{fig:betahis}a-c (left frame) and
Fig.~\ref{fig:betahis}d-f (right frame). The black symbols correspond
to fits to the original SPIRE data and the blue symbols to the
combination of SCUBA-2 and filtered SPIRE data. Only clumps from the
$P(\chi^2)$=25\% samples are included. The median values are indicated
with large crosses. Frame a includes a contour for a typical 67\%
confidence region in the 4-band fits.
}
\label{fig:TB_1}
\end{figure}

Figure~\ref{fig:TB_1} uses clumps from Fig.~\ref{fig:betahis}a-f and
compares the fits to the original SPIRE data and to the combination of
SCUBA-2 and filtered SPIRE data. Figure~\ref{fig:TB_1}a shows an
example of the confidence regions estimated with the MCMC method. It
indicates the direction of the error ellipses but the size of the
confidence region varies from clump to clump. The figure suggests that
a large fraction of the anticorrelation between $T$ and $\beta$ can be
caused by the observational uncertainties.

The dependence of $\beta$ values on other parameters and directly on
error estimates is examined further in Appendix~\ref{sect:betacorr}
and in Appendix~\ref{sect:allhisto}. The latter section also
quantifies further the potential effects of the 850\,$\mu$m
calibration uncertainty.

\subsection{SEDs with 160\,$\mu$m point}  \label{sect:PACS}

\begin{figure}
\includegraphics[width=8.8cm]{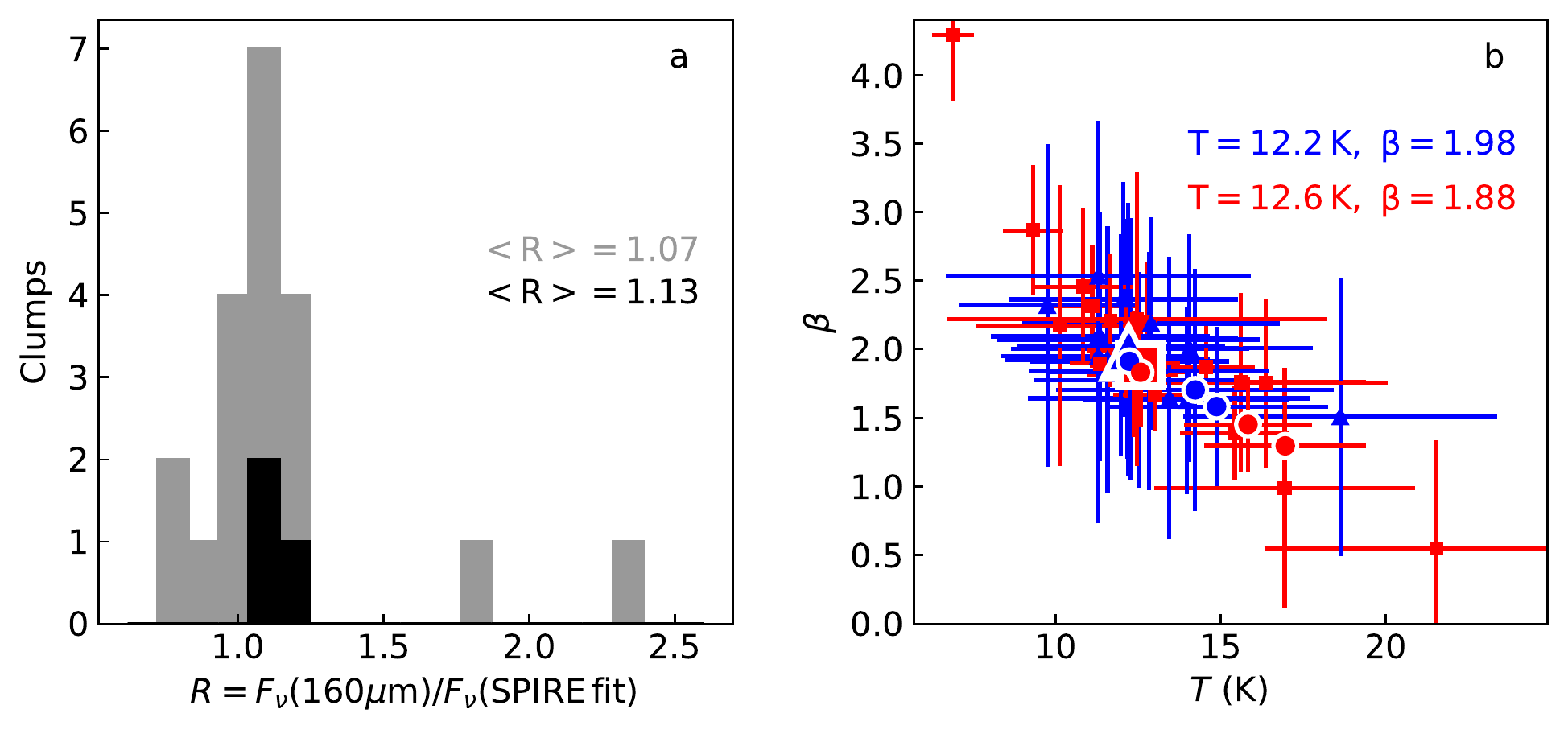}
\caption{
Comparison of {\it Herschel} MBB fits with and without 160\,$\mu$m
data point in the case of the P($\chi^2$)=25\% clump sample with
$\theta_{\rm F}=500\arcsec$ and LM masks. Left frame shows the distribution of
ratios between the 160\,$\mu$m flux density and the value predicted by
the 250-500\,$\mu$m fit. The grey histogram contains all clumps and
the black histogram three clumps with YSO candidates.
Second frame shows ($T$, $\beta$) values for individual clumps for the
250-500\,$\mu$m (blue triangles) and 160-500\,$\mu$m (red squares)
fits. The circles indicate clumps with YSO candidates.
The median values are quoted in the frame and plotted with larger
symbols. 
}
\label{fig:PACS}
\end{figure}

Although 37 fields are partially covered by {\it Herschel} PACS
observations, the number of clumps with 160\,$\mu$m data is limited.
Figure~\ref{fig:PACS} examines the 23 clumps in the $P(\chi^2)=25$\%
sample of clumps from the maps made with $\theta_{\rm F}=500\arcsec$ and LM
masks. 

The 160\,$\mu$m values are typically above the SPIRE fit
(Fig.~\ref{fig:PACS}a), which could be expected if each source
contains a wide range of temperatures. Figure~\ref{fig:PACS}b compares
($T$, $\beta$) values for {\it Herschel} 250-500\,$\mu$m and
160-500\,$\mu$m MBB fits. Both fits cover a similar region in the
($T$, $\beta$) plane except for a few clumps where the inclusion of
the 160\,$\mu$m point leads to much higher temperatures and lower
$\beta$ values. The number of these sources is small enough so that
the shift in the median parameter values remains small with 
$\Delta T$=+0.4\,K and $\Delta \beta=-0.1$. 
For simplicity both fits were carried out without considering error
correlations between the different bands. The SPIRE fits are therefore
not identical to those in Fig.~\ref{fig:betahis}a. The difference in
the median value of $\beta$ is 0.08 units.

\subsection{Young stellar objects} \label{sect:YSO}

We examined the spatial distribution of young stellar objects (YSOs)
using the catalogue of YSO candidates of \citet{Marton2016} that is
based on the analysis of 2MASS \citep{Skrutskie2006} and WISE
\citep{Wright2010} data .

\begin{figure}
\includegraphics[width=8.8cm]{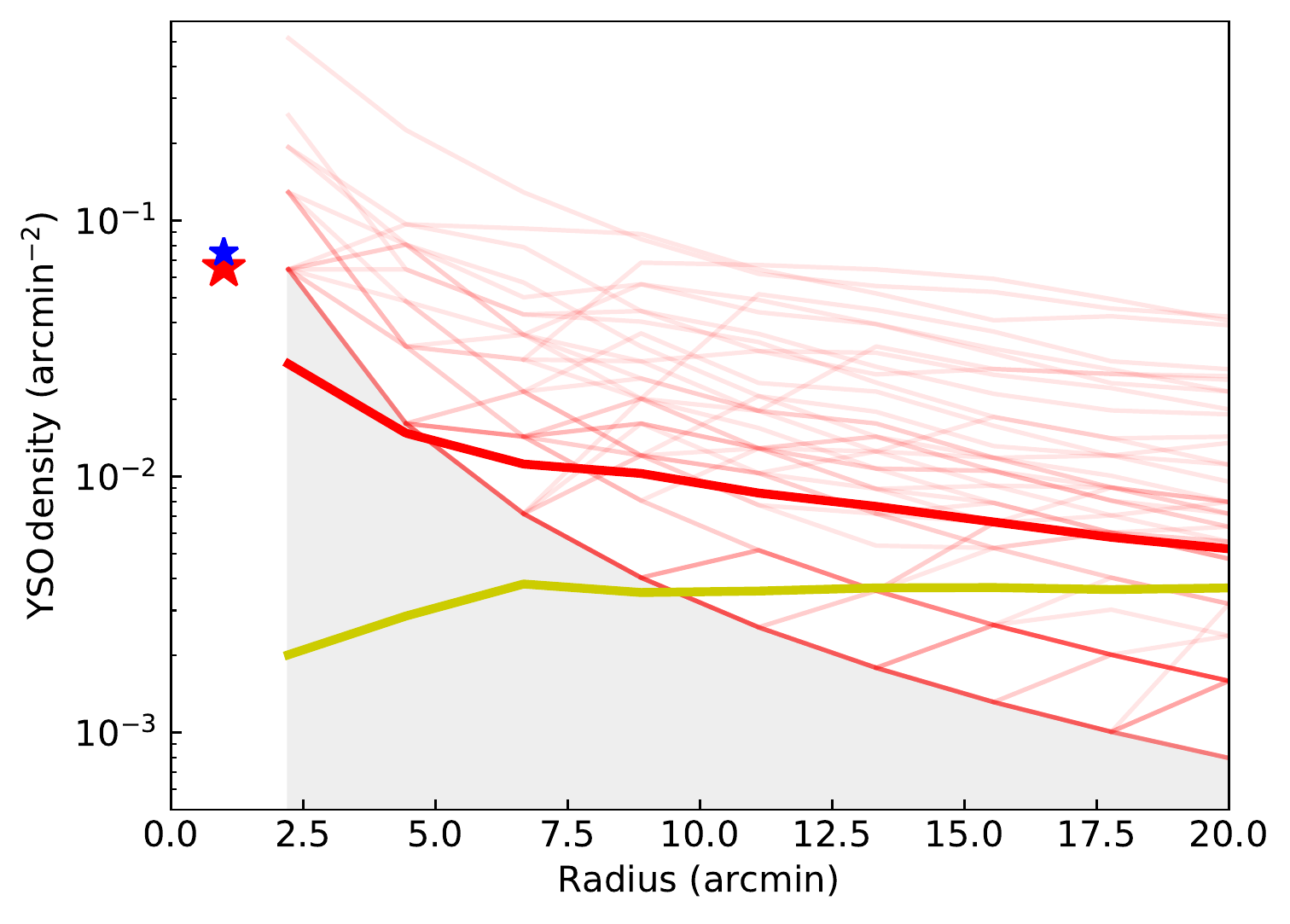}
\caption{
YSO statistics. The surface density of sources is shown as a function
of the distance ($>2\arcmin$) from the field centre. The mean values
over all fields are shown for Class I-II (thick red line) and Class
III (thick yellow line) candidates. The thin red lines correspond to
Class I-II sources in individual fields. The shading correspond to a
region with less than one source for a given distance. The two
overlapping stars indicate the density of Class I-II candidates in all
clumps (all fields) selected by the Fellwalker algorithm (red star)
and in clumps belonging to the $P(\chi^2)=85$\% sample (blue star).
}
\label{fig:YSO_stat}
\end{figure}

Figure~\ref{fig:YSO_stat} shows the surface density as a function of
the distance from the centre of the SCUBA-2 fields. The density of
Class III candidates is practically constant, about 3-4$\times
10^{-3}$\,arcmin$^{-2}$. The density drops only within the innermost
7$\arcmin$, possibly because of the observational bias resulting from
the increasing extinction. Apart from that, Class III sources are not
correlated with the clumps. The total number of Class I-II sources is
higher. Their density also increases significantly towards the centre
of the fields, reaching $2.8\times 10^{-2}$arcmin$^{-2}$ within the
innermost $2\arcmin$ radius.
The highest concentration of Class I-II candidates is found in the
field G130.1+11.0A1 where, unfortunately, the {\it Herschel} coverage
is incomplete.

We calculated separately the YSO surface densities for all clumps and
for the subsample selected for the SED analysis (see
Sect.~\ref{result:small}). For Class I-II candidates, the trend seen
in radial profiles continues and they are found preferentially inside
the clumps (see Fig.~\ref{fig:YSO_stat}).  The surface density is
$7.7\times 10^{-2}$\,arcmin$^{-2}$
within the $P(\chi^2)$=85\% clump sample. It is also significant that
none of the Class III candidates falls inside a clump.

In the LM case, the $P(\chi^2)$=25\% sample includes 56 clumps of
which 5 had Class I-II YSO candidates. In SPIRE fits, the YSO clumps
have median values of
$T=14.5 \pm 0.5$\,K and $\beta=1.62 \pm 0.14$ while starless clumps have 
$T=11.9 \pm 0.9$\,K and $\beta=2.1 \pm 0.3$, where the quoted
uncertainties are median absolute deviations of the distributions.
For the 250-850\,$\mu$m fits the corresponding values are
$T=14.8 \pm 1.0$ and $\beta=1.58 \pm 0.08$  vs. 
$T=14.0 \pm 1.2$ and $\beta=1.59 \pm 0.23$.
There is thus a tendency for the protostellar clumps to be warmer
while the evidence of lower $\beta$ values is not altogether
consistent. The sample of protostellar source candidates also
comprises only six sources.
In Fig.~\ref{fig:PACS} there were only three clumps with YSO
candidates that also had PACS observations at 160\,$\mu$m. Also for
this small sample the clump temperatures were higher than the average.
The ratio between the 160\,$\mu$m measurement and the fit to the
longer wavelength bands was also higher for clumps with YSO
candidates.  However, the effects of YSO heating appear to be small
(Fig.~\ref{fig:PACS}a).

The observed differences in the apparent ($T$, $\beta$) values do not
necessarily mean changes in the intrinsic dust properties. Large
temperature variations, such as caused by internal heating, could 
lower the $\beta$ values estimated from the SEDs.

\subsection{Clump masses} \label{sect:masses}

\begin{figure}
\includegraphics[width=8.8cm]{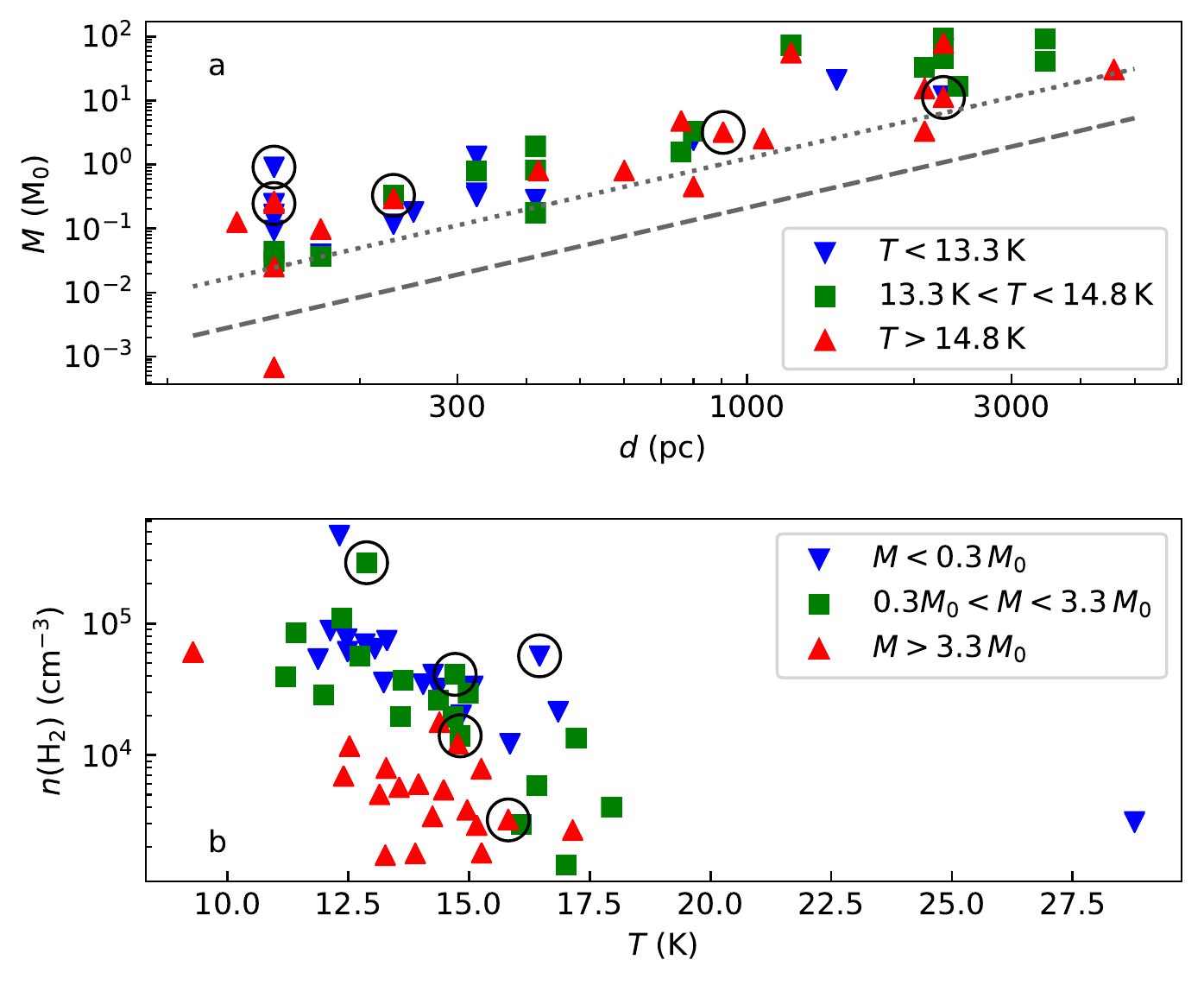}
\caption{
Clump mass vs. field distance (frame a) and average volume density vs.
temperature (frame b). The plots correspond to the LM data version and
the $P(\chi^2)$=25\% clump sample. The fluxes and temperatures are
from the 250-850\,$\mu$m fits. The data are further divided into three
samples based on temperature (frame a) or mass (frame b).  The clumps
with YSO Class I or II candidates are marked with circles. In frame a
the thresholds corresponding to a 17\,mJy per 850\,$\mu$m beam (dashed
line) and to a 850\,$\mu$m flux density of 0.1\,Jy f(dotted line) are
shown (see text). 
}
\label{fig:clump_masses}
\end{figure}

Most fields have distance estimates in Table~\ref{table:distances},
which allows the calculation of the physical size, mass, and volume
density of the clumps. Figure~\ref{fig:clump_masses} shows results for
the LM data, the $P(\chi^2)$=85\% clump sample. The masses correspond
to fluxes inside measurement apertures of area $A$ and the volume
densities are calculated for a spherical volume with an effective
radius $r$ for which $A=\pi r^2$.

The mass and distance are correlated, because the clump detection are
affected by our sensitivity to low flux densities. In
Fig.~\ref{fig:clump_masses}a the dashed line corresponds to a mass
threshold that assumes a 3.4\,mJy/beam noise level and a 5-$\sigma$
detection over a single 850\,$\mu$m beam. The dotted line corresponds
to a 850\,$\mu$m flux density of 0.1\,Jy, both for $T=13$\,K and
$\beta=1.8$ spectrum. The correlation in Fig.~\ref{fig:clump_masses}a
thus mainly reflects the flux density threshold combined with the 
geometrical effect of the distance. All sources are relatively close
to the threshold and we do not have nearby sources that would also be
very massive. This is natural because low-mass objects are generally
much more numerous and, on the other hand, the spatial filtering of
the SCUBA-2 data and the clump detection algorithm both limit the
possibility to detect very extended sources. The mass-distance
relation is also affected by the large uncertainty of the distances
and the functional dependence between the distance and the mass
estimates.

The temperature is not dependent on the field distance but there is
some negative correlation between the volume density and the clump
temperature. However, negative errors in temperature will naturally
lead to positive errors in column density, mass, and volume density.
Higher temperature also directly enables the detection of lower
density objects. Furthermore, the fluxes were measured from apertures
that were extended 20$\arcsec$ beyond the original SCUBA-2 clump
detections in order to accommodate the lower resolution of the SPIRE
data. The modified clump size can introduce some bias that is also
dependent of the temperature. For the same sample of clumps the
correlation between the temperature and the column density remains
very weak.

We have marked in the plot the clumps with Class I or II YSO
candidates. The median distance of YSO sources is below that of the
full clump sample but this is not significant considering the small
number of YSO candidates. As noted in Sect.~\ref{sect:YSO}, there is 
some tendency for protostellar clumps to be warmer than clumps on
average. In Fig.~\ref{fig:clump_masses}, the protostellar clumps do
not stand out either based on their mass or their density.

We note that while clumps with embedded protostars may be typically
among the warmer cores in the sample, they obey a similar
temperature/density relation as the starless clumps.  This is in
contrast to the behaviour seen on the scale of individual star-forming
cores (with sizes $\sim 10^{-2}$\,pc): combined SCUBA-2 and Herschel
observations of starless cores in the Taurus molecular cloud
\citep{Ward-Thompson2016} and in the Cepheus Flare \citep{Pattle2017}
have shown that cores heated by nearby or embedded protostars are
significantly warmer than cores of the same density which are
uninfluenced by protostars. The lack of a systematic elevation in
temperature in {\it Planck} clumps with embedded protostars relative
to starless {\it Planck} clumps of similar densities suggests that the
heating effect of a protostar on its environment is limited to a
relatively small volume surrounding the protostar itself, and so that
embedded protostars do not significantly raise the average dust
temperature of their parent clumps.


\section{Discussion}  \label{sect:discussion}

We have investigated submillimetre dust emission in 96 fields observed
with SCUBA-2 instrument at 850\,$\mu$m and with at least partial {\it
Herschel} coverage. In the following, we discuss the results obtained
at large-scales, based on IRAS, {\it Planck} and {\it Herschel} data,
and at small scales, based on the combination of {\it Herschel} and
SCUBA-2 observations.

\subsection{Dust SEDs}

\subsubsection{SEDs of PGCC host clouds} \label{dis:large}

PGCC sources represent the coldest part of the ISM and, therefore, the
environments examined in Sect.~\ref{result:large} should consist
mostly of parts of molecular clouds. 
This is confirmed by the column densities. Based on IRAS and {\it
Planck} data 100-550\,$\mu$m and the dust opacity adopted in
Sect.~\ref{sect:SEDfit}, the average values are ${\rm N}({\rm
H}_2)=3.5\times 10^{21}$\,cm$^{-2}$ within the $r=8\arcmin$ annuli but
only $1.1 \times 10^{21}$\,cm$^{-2}$ immediately outside this region.
Typical SEDs measured with IRAS and {\it Planck} data correspond to a
colour temperature of $T=15.7$\,K and a dust opacity spectral index of
$\beta=1.7$, with some variations depending on the actual bands used
(see Fig.~\ref{fig:IP_8am}). Most colour temperature estimates are
within 1\,K of the quoted median values.

When IRAS 100\,$\mu$m point is replaced with 250-550\,$\mu$m {\it
Herschel} measurements, the median colour temperature is about 14.7\,K
and the spectral index close to $\beta=1.8$ (see
Fig.~\ref{fig:HP_8}a). 
Bootstrapping gives purely statistical uncertainties of 0.13\,K in
temperature and 0.015 in $\beta$ (sample median values).
The difference between the fits could be affected by very small grain
emission in the 100\,$\mu$m band or by the fact that wavelengths below
the emission peak are more sensitive to warm dust and, in the presence
of temperature variations, lead to lower $\beta$ estimates. When the
analysis was repeated with background-subtracted flux measurements
(Fig.~\ref{fig:HP_8}b), the spectral index estimate remained similar,
which could support the latter explanation. However, we also repeated
the fits using only {\it Planck} data. In spite of the absence of
short wavelengths $\lambda<350$\,$\mu$m, these resulted in lower
spectral index values $\beta \sim 1.6$ (see
Appendix~\ref{sect:NO_IRAS}). That result is, of course, affected by
larger uncertainties. One can note that also in
\citet{Malinen2014}, the use of different wavelength ranges (from
100-500\,$\mu$m to 160-1380\,$\mu$m) did not result in significant
changes in the ($T$, $\beta$) estimates of the protostellar and
starless cores of the cloud LDN~1642.

The colour temperatures of the selected fields are lower and the
spectral indices higher than in the general ISM. For example,
\citet{planck2013-XIV} reported in the 100-850\,$\mu$m range values
$\beta=1.54$ for regions dominated by atomic gas and $\beta=1.66$ for
mostly molecular clouds \citep[see also][]{Planck-2015-X}. The fact
that our SEDs are somewhat steeper (with higher $\beta$) reflects the
fact that the selected fields represent the densest parts of molecular
clouds. \citet{GCC-VI} found $T=16.1$\,K and $\beta=1.84$ for a sample
of PGCC field, also using the combination of IRAS and {\it Planck}
bands 100-850\,$\mu$m. The values are comparable to those derived for
the present sample, although the median spectral index in
Fig.~\ref{fig:IP_8am} is slightly lower, $\beta$=1.73. The values also
depend on the relative weights given for the fitted bands, which are
not identical between these two studies. In the present study the
weight of the 857\,GHz and 545\,GHz relative to both IRAS and lower
frequency {\it Planck} bands was higher than in \citet{GCC-VI}.

The results of Fig.~\ref{fig:HP_8}a and Fig.~\ref{fig:HP_8}b could be
affected by temperature variations within the apertures. By removing
the warmer diffuse component, background subtraction should make the
remaining emission more isothermal, which could in turn increase the
observed $\beta$ values. However, no such effect was observed.

\subsubsection{SEDs of SCUBA-2 clumps} \label{dis:small}

Dust emission of dense clumps and cores was examined with {\it
Herschel} and SCUBA-2 data. Depending on the $P(\chi^2)$ value, the
average clump area of the samples is 2.1-2.5\,arcmin$^2$ or
0.19-0.3\,pc$^2$. 
The spectral index distributions were examined in
Fig.~\ref{fig:betahis}, for alternative reductions, clump selections,  
and wavelength ranges. Such comparisons help to establish the full
uncertainty of the results that may not be captured by the a formal
photometric and calibration errors alone.
In spite of possible differences in the reliability (partially
reflected in the width of the distributions), the main results are
relatively robust. Most estimates of the submillimetre spectral index
are in the range $\beta=1.6-1.9$. The 850\,$\mu$m data point has a
clear effect on the spectral index estimates. While the fits to the
SPIRE data gave median $\beta$ values slightly above $1.8$, the joint
fits of SPIRE and SCUBA-2 850\,$\mu$m decreased these to $\beta \sim
1.6$. However, as demonstrated by Fig.~\ref{fig:betahis}l, there can
be error sources that are not yet all fully understood.

The SPIRE fits can be compared with the results of \citet{GCC-VI}.
There the spectral index values were higher at the scale of
10$\arcmin$. The median estimates were $T=14.9$\,K and $\beta=2.04$
compared to the values $T=14.7$\,K and $\beta=1.8$ in
Fig.~\ref{fig:HP_8}a. \citet{GCC-VI} did not report statistics on
dense clumps but presented $T$ and $\beta$ maps, which sometimes
showed clear spatial $\beta$ variations with maximum values rising
above $\beta=2.0$. This was true mostly for well-resolved nearby
clumps. The median distance of the fields in \citet{GCC-VI} was
620\,pc. In this paper the median distance of the fields is only
230\,pc but, as shown by Fig.~\ref{fig:clump_masses} most of the
detected clumps are at larger distances. The nearby fields tend to be
at higher Galactic latitudes, have lower column densities, and have
thus fewer SCUBA-2 detections. Based on \citet{GCC-VI} we can expect
significant dispersion in the $\beta$ histograms. The $\beta$
estimates are in the range $1.0 < \beta < 2.5$ and we found no real
evidence for extreme values of either $\beta \le 1$ or $\beta \ge 3$.

Data at wavelengths below 250\,$\mu$m were mostly excluded from the
analysis, because of the incomplete coverage of the SCUBA-2 fields and
because of the potential effects of temperature mixing. In
Sect.~\ref{sect:PACS} the addition of the 160\,$\mu$m point changed
the median values by $\Delta T=+0.4$\,K and $\Delta \beta$=-0.1. The
exact values again depend significantly on the relative weighting of
the different bands \citep[see e.g.][]{Shetty2009a}.

The $\beta$ values derived in this work can be compared with recent
results on starless and protostellar clumps.
\citet{Sadavoy2013} analysed {\it Herschel} and SCUBA-2 observations
of Perseus (clump B1) and found typical spectral index values of 
$\beta\approx 2$. Towards protostellar sources the values were lower
but still mostly above $\beta=1.6$. The values are thus similar to
this study, although we did not observe significant difference between
clumps with and without YSO candidates that could be interpreted as
signs of systematic dust evolution (Fig.~\ref{fig:PACS}).
\citet{Chen2016_Perseus} investigated the Perseus cloud further and
found notable differences between the various cloud regions. The
pixel-by-pixel histograms covered a range $\beta \sim 1.0-2.7$. The
lower $\beta$ values were found to be correlated with local peaks of
colour temperature but also with CO outflows and thus directly with
YSOs. This still raises the question to what extent the changes in the
observed $\beta$ values are the result of dust evolution or of the
more complex temperature structure of the sources. 
\citet{Bracco2017} mapped spectral indices of the Taurus B213 filament,
including both prestellar and protostellar clumps. The observations
consisted of 1.15\,mm and 2.0\,mm ground-based observations that were
analysed together with temperature constraints obtained from {\it
Herschel}. A prestellar core was found to have a high spectral index of
$\beta \sim 2.4$ that also remained radially constant. Because the
values resulted from a 3D inversion, they are already corrected for the
line-of-sight temperature mixing and should thus be somewhat higher than
values derived from the SEDs directly. In the case of protostellar
sources, the inversion of the radial profiles suggested a decrease from
$\beta \sim 1.5-2.0$ in the outer parts to values 0.5 lower in the
warmer centre. The lowest values at a core centre were $\beta \sim 1.0$
but the values derived from the basic SED analysis would be expected to
remain significantly higher.
\citet{Li2017_Class0} examined specifically a sample of Class 0 YSOs
with interferometric observations from submillimetre to centimetre
wavelengths. Most sources exhibited values $\beta<1.7$ (down to $\beta
\sim$ 0.6). The authors noted that, in addition to grain growth, the
result may be affected by a strong temperature mixing when the
high-resolution observations probe the emission from the circumstellar
disk. The comparison with the {\it Herschel} and SCUBA-2 results is also
affected by the difference of the wavelength ranges (if submillimetre
SEDs are generally steeper) and the interferometric observations
filtering out much of the extended emission. 

In summary, our results of $T\sim 14$\,K and $\beta\sim$1.6 for the
joint SPIRE and SCUBA-2 fits are very similar to the previous studies.
However, in our sample the difference between starless sources and
protostellar sources (clumps with YSO candidates) is small. Our YSO
statistics may also be incomplete and affected by the sensitivity and
resolution of the WISE survey relative to the data used in the other
studies.

In our study, data reductions with different {\it Herschel}-based
source masks and different filtering scales in
Fig.~\ref{fig:betahis}a-i gave relatively consistent results. Larger
filter scales and larger source masks resulted in the recovery of more
850\,$\mu$m extended emission. The maps generally agreed well with the
SPIRE observations that are sensitive to much lower column densities.
The use of $\theta_{\rm F}=500\arcsec$ and large LM masks lead to more than
two times as many clump detections as the combination of
$\theta_{\rm F}=200\arcsec$ and SM. On the other hand, large masks can
sometimes increase random fluctuations in the maps. For example, in
the figure in Appendix.~\ref{sect:allmaps} the main clump of
G167.215.3A1 is associated with a negative feature that is located
inside the LM mask. The feature disappears if a smaller filter scale
(200$\arcsec$ instead of 500$\arcsec$) or a more tight source mask (SM
instead of LM) is used. While large filter scale and liberal source
masks work well in statistical studies, one may want to be more
conservative when studying individual objects.

\subsubsection{Wavelength dependence of spectral index} \label{dis:wvl}

One of the main goals of this paper was to investigate the potential
flattening of dust SEDs towards millimetre wavelengths. In this
respect, the most interesting comparisons are the fits with and
without the {\it Planck} 217\,GHz band (1380\,$\mu$m) and, at smaller
scales, the fits with and without the SCUBA-2 850\,$\mu$m band.

According to the IRAS and {\it Planck} analysis in
Fig.~\ref{fig:IP_8am}, the 100-1380\,$\mu$m fit lead to a marginally
lower value of the spectral index ($\beta=1.69$) when compared to the
fits at $\lambda \le 550\,\mu$m 
($\beta=1.74$) or $\lambda \le 850\,\mu$m ($\beta=1.73$). 
The corrections for line emission in the 217\,GHz and 353\,GHz bands
was identified as the main source of uncertainty. Based on
Fig.~\ref{fig:IP_8am}, such small differences in $\beta$ could be
explained by a systematic 10\% error in the CO corrections. The
comparison with the sum of the ground-based $^{12}$CO(2-1) and
$^{13}$CO(2-1) measurements showed that the {\it Planck} CO estimates,
which were used to correct the continuum data, are correct to within
$\sim$30\% and do not show any significant bias
(Fig.~\ref{fig:compare_CO}a). However, the possibility of systematic
errors at 10\% level can not be excluded. The calibration accuracy of
ground-based observations is also not much better than this.
Considering the uncertainty of the line contamination, which may
include contributions of also other lines \citep{Planck-2015-X}, and
because of other potential systematic errors, the change of $\beta$ as
a function of wavelength was not significant.

When the same analysis was based on the combination of {\it Herschel}
and {\it Planck} data, the 217\,GHz band caused no significant change
in $\beta$ (Fig.~\ref{fig:HP_8}a). Furthermore, in the
background-subtracted measurements the $\beta$ value was actually {\em
higher} when the 217\,GHz band was included. This shows that if
$\beta$ does change with wavelength, the effect is at most at the
level of the observational uncertainties.

The analysis of {\it Herschel} and SCUBA-2 data gave clear indications
that the 850\,$\mu$m point would favour lower $\beta$ values
(Figs.~\ref{fig:betahis}, \ref{fig:plot_example_SEDs}).  The
difference with and without the 850\,$\mu$m point resulted in a change
of the order of $\Delta \beta \sim $0.2 but the result may still
depend on the details of the data reduction, as indicated by
Fig.~\ref{fig:betahis}j-l. On the other hand, a relative calibration
error of 10\% between SPIRE and SCUBA-2 (or similar CO correction of
the 850\,$\mu$m band) would correspond in our fits only to a change
of $\Delta \beta=\pm 0.03$.

\subsubsection{Relation between $T$ and $\beta$} \label{dis:TB}

The dependence between dust temperature and spectral index is 
important because it is related to dust evolution during the
star-formation process and is a potential tracer of the past history
and current physical state of clumps. The simultaneous determination
of $T$ and $\beta$ is difficult because observational noise produces
anticorrelation between the parameters. Moreover, temperature
variations within the beam make the observed colour temperature a
biased estimate of the physical grain temperature, and spectral index
estimates will be similarly biased.

It is therefore not surprising that the observed relations in
Figs.~\ref{fig:IP_8am}, \ref{fig:HP_8}, and \ref{fig:TB_1} all exhibit
some negative correlation between $T$ and $\beta$. In the following we
examine if these can be caused by noise alone. It is more difficult to
make the final connection between the apparent and the real dust
properties. This would require detailed modelling and is outside the
scope of this paper.

We first investigate the SEDs of IRAS and {\it Planck} data shown in
Fig.~\ref{fig:IP_8am}a. Fits of a function $\beta(T)=A \times T^B$
\citep[see][]{Desert2008, GCC-VI} give for $B$ values of -0.53 and
-0.69 for the 100-550\,$\mu$m, and 100-1380\,$\mu$m data,
respectively. We can examine whether the data are consistent with a
constant $\beta$ value, the negative correlation being caused by noise
alone. We fit the data of each field with a MBB function $B_{\nu}(T)
\times {\nu}^{1.7}$ and rescale the flux values onto this SED curve. In
other words, at this point the simulated photometry matches exactly
the assumed $\beta(T)$ law. We then add photometric noise to the flux
density values (see below), and determine the apparent $T$ and $\beta$
values with new fits of the simulated SEDs. In the above we assumed
$B_{\nu}(T) \times {\nu}^{1.7}$ spectra but because the goal is to
simulate the relative difference between the true and the observed
SEDs, the procedure is not sensitive to the selected value of $\beta$.
The results remain essentially identical if a value of $\beta=1.4$ or
$\beta=2.0$ is used instead $\beta=1.7$

Because direct comparison between observations and simulations would
require precise knowledge of the uncertainties, we include the
noise level as an additional parameter $k{\rm (noise)}$. 
Thus we use the same error covariance matrices as in
Sect.~\ref{sect:SED_IP} but scale all the matrix elements with the
square of $k{\rm (noise)}$.
We examine how $k{\rm (noise)}$ affects the $\beta(T)$ relation and,
on the other hand, the scatter around the fitted $\beta = A\times
T^{B}$ curve. Comparison with the observed scatter should provide the
most direct empirical {\it upper} limit on the noise effects. It is an
upper limit because the real emission cannot be expected to follow a
single $\beta(T)$ relation and thus only part of the observed scatter
is caused by noise.

\begin{figure}
\includegraphics[width=9cm]{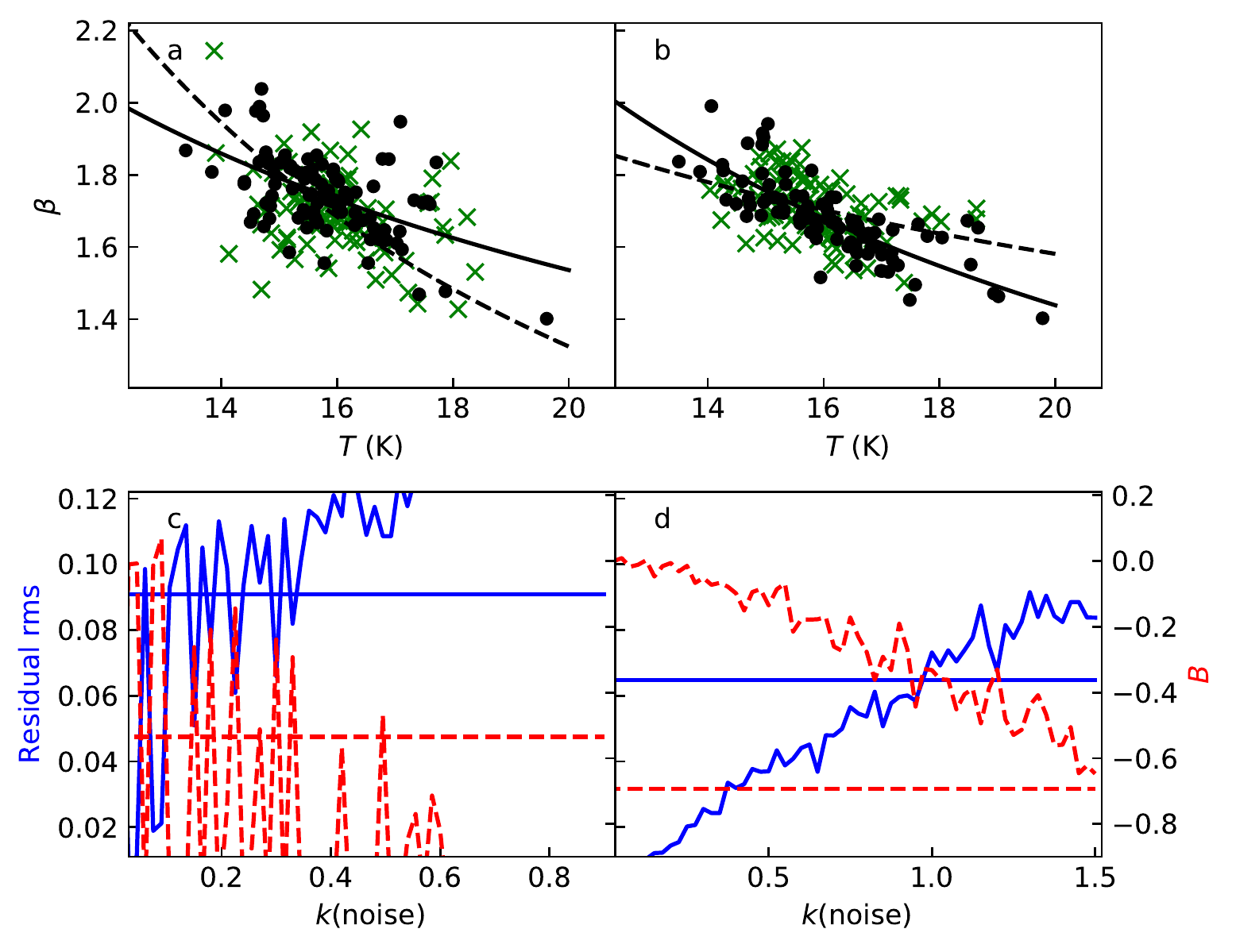}
\caption{
Comparison of ($T$, $\beta$) relations in observations and in
simulations with a constant value of $\beta$=1.7. The observations
consist of the same IRAS and {\it Planck} data as in
Fig.~\ref{fig:IP_8am} and represent the average emission of the
fields. Left and right frames show results for 100-550\,$\mu$m and
100-1380\,$\mu$m fits, respectively. In the upper frames, black
symbols show the real observations and green crosses show one
realisation of simulations with $k{\rm (noise)}$ equal to 0.4 (left
frame) and 1.0 (right frame). The black lines are fits $\beta=A\times
T^B$ to the observations (solid line) and to the green crosses (dashed
line).
The lower frames show how the rms value of the residuals of the
$\beta(T)$ fit (solid blue lines, left axis) and the value of $B$
(dashed red lines and right axis) change as a function of $k{\rm
(noise)}$. The horizontal lines correspond to the values in the actual
observations.
}
\label{fig:test_TB_IP}
\end{figure}

For the 100-550\,$\mu$m fits (Fig.~\ref{fig:test_TB_IP}a,c) the
simulations match the observed scatter over a range of
$k{\rm (noise)}\sim$ values where the value of the $B$ parameter of the
$\beta(T)$ relation is also similar to the observed value. The data
are thus compatible with a constant spectral index. The situation is
different for the fits to 100-1380\,$\mu$m data, which have smaller
uncertainty on the $\beta$ parameter (Fig.~\ref{fig:test_TB_IP}b,d).
The observed scatter is matched with $k{\rm (noise)}\sim$1.0, at which
point $B$ is -0.35 instead of the value $B=-0.69$.
The result would thus suggest that the intrinsic $\beta(T)$ relation
is indeed a decreasing function of temperature. Half of the apparent
decrease would be caused by noise but only assuming that the intrinsic
$\beta(T)$ relation itself has no scatter. The result remains
qualitatively the same, albeit with different $k{\rm (noise)}$ values,
if the relative weighting of the bands is modified (such as doubling
the error estimates of the IRAS band or of the two longest wavelength
{\it Planck} bands).
Figure~\ref{fig:test_TB_IP} contains some inconsistency because the
100-550\,$\mu$m fits suggest a lower value of $k{\rm (noise)}$.
However, if the relative error estimates of the 353\,GHz and 217\,GHz
bands are increased, it is possible to fit both 100-550\,$\mu$m and
100-1380\,$\mu$m ranges even with the same value of $k{\rm (noise)}$. 

\begin{figure}
\includegraphics[width=9cm]{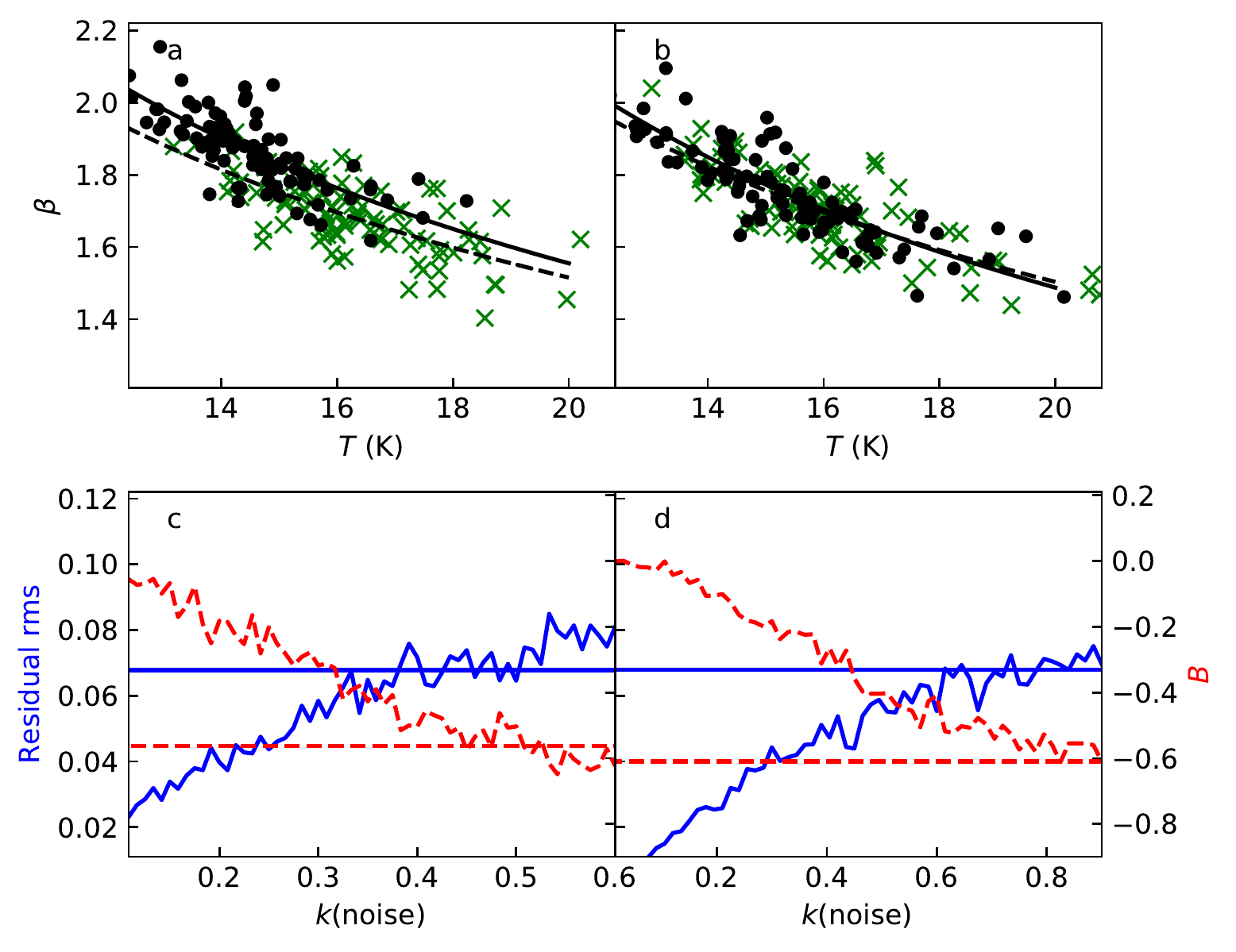}
\caption{
As Fig.~\ref{fig:test_TB_IP} but for the combined {\it Herschel} and
{\it Planck} data from Fig.~\ref{fig:HP_8}. The fitted data are
250-850\,$\mu$m (frames a and c) and 250-1380\,$\mu$m (frames b and
d). The green crosses correspond to one realisation with $k{\rm
(noise)}$=0.4 (frame a) and $k{\rm (noise)}$=0.7 (frame b).
}
\label{fig:test_TB_HP}
\end{figure}

In Fig.~\ref{fig:test_TB_HP} we show the same test applied to the
combination of {\it Herschel} and {\it Planck} data (cf.
Fig.~\ref{fig:IP_8am}a). 
In the simulations the noise is generated using the same error
covariance matrices as in Sec.~\ref{result:large} but scaling the
elements with $k{\rm (noise)}$ squared.
For this shorter wavelength baseline
250-850\,$\mu$m the observed scatter is reproduced around $k{\rm
(noise)}\approx 0.45$, at which point the predicted $B$ is only
slightly above its observed value.
For the longer wavelength baseline 250-1380\,$\mu$m, the scatter and
observed $B$ can both be reproduced with $k{\rm (noise)}=0.8$ and the
data are consistent with a flat intrinsic $\beta(T)$ relation. This
does not completely exclude a decreasing $\beta(T)$ function, if the
clumps do have some real scatter relative to the average $\beta(T)$
relation.

For the clumps, we can only use the shorter wavelength baseline
provided by {\it Herschel} and SCUBA-2 data. The error estimates that
were used to derive the confidence regions in Fig.~\ref{fig:TB_1} may
have been too large because the 67\% contour already covers almost all
of the observations. Nevertheless, the figure shows that the observed
distribution of points is consistent with the orientation of the error
ellipse and the observed distribution of ($T$, $\beta$) points could
in principle be caused by noise.

Figures~\ref{fig:test_TB_IP}-\ref{fig:test_TB_HP} give only weak
support for the $T-\beta$ anticorrelation, although, by attributing
all the observed ($T$, $\beta$) scatter to observational errors, we 
underestimate the probability of a decreasing $T(\beta)$ relation. The
results at least exclude the possibility of an increasing (apparent)
$\beta(T)$ relation.

If the intrinsic $\beta$ of dust grains is constant, the apparent
values (derived from the SEDs) could still appear to be positively
correlated. This is the case for samples of externally heated clumps
of different masses \citep[see e.g.][Fig. 6]{Juvela2012_Tmix}. If the
clumps have temperature variations caused by protostellar sources,
those tend to cause negative correlation between the observed
(apparent) $T$ and $\beta$ values \citep{Shetty2009a, Malinen2011,
Juvela2012_Tmix}. Once the noise effects have been excluded, if we
were to conclude that the observation exhibit negative correlation
between the apparent $T$ and $\beta$ values, this would still not
constitute a proof of similar behaviour for the actual dust opacity
spectral index of the dust grains. Figure~\ref{fig:test_TB_IP}b gave
the strongest indications of apparent $T-\beta$ anticorrelation but,
because of the inclusion of the 100\,$\mu$m data, the result may be
affected by temperature variations within the aperture or more
directly by the emission of very small grains.

The effect of noise on the $\beta(T)$ relation and the joint
estimation of the noise level and the $\beta(T)$ relation are examined
further in Appendix~\ref{sect:simu}.

\section{Conclusions}  \label{sect:conclusions}

We have used IRAS, {\it Planck}, {\em Herschel}, and SCUBA-2
observations to examine the dust emission spectra in 96 fields that
were originally selected based on the presence of prominent PGCC cold
clumps.  SCUBA-2 data, in conjunction with {\it Herschel} SPIRE
observations, was used to examine the submillimetre spectra of the
densest clumps and cores. The study led to the following conclusions:
\begin{itemize}
\item The fields have typical colour temperatures of 14-18\,K and
spectral index values of $\beta$=1.5-1.9. 
\item The clumps extracted from SCUBA-2 maps are characterised by median
values of $T \sim $13\,K and $\beta \sim$ 1.7. Different options in
the data reduction lead to uncertainty at a level of $\delta\beta \sim
\pm$0.2. 
\item The use of large source masks and large filter scales 
was successful in recovering more extended emission in the 850\,$\mu$m
SCUBA-2 maps. This also resulted in a larger number of clump
detections. Because very extended masks can lead to undesirable map
fluctuations in individual maps, they may be most useful in
statistical studies.
\item At large scales, there is no consistent proof for the 
dust spectrum becoming flatter at millimetre wavelengths. We are
limited by the uncertainty of the corrections for CO line emission in
the 353\,GHz and 217\,GHz {\it Planck} bands. 
\item 
At clump scales, the SCUBA-2 850\,$\mu$m flux measurements tend to be
above the SED fitted to the shorter wavelength SPIRE bands. The effect
on the $\beta$ values in the joint fits is $\Delta \beta \sim $0.2.
The result is not entirely robust with respect to the decisions made
in the data reduction.
\item 
The positions of the SCUBA-2 submillimetre clumps are positively
correlated with Class I and II YSO candidates. For Class III sources
the correlation is negative.
\item 
Compared to starless clumps, protostellar clumps have 1--2\,K higher
median colour temperature and marginally lower $\beta$ values. The two
populations are, however, mostly overlapping. This applies also to the
mass and volume density distributions.
\item 
Most of the observed $T$-$\beta$ anticorrelation may be caused by
noise. Evidence for the anticorrelation is found mainly at large
scales.
\end{itemize}

\begin{acknowledgements}
The James Clerk Maxwell Telescope is operated by the East Asian
Observatory on behalf of The National Astronomical Observatory of
Japan, Academia Sinica Institute of Astronomy and Astrophysics, the
Korea Astronomy and Space Science Institute, the National Astronomical
Observatories of China and the Chinese Academy of Sciences (Grant No.
XDB09000000), with additional funding support from the Science and
Technology Facilities Council of the United Kingdom and participating
universities in the United Kingdom and Canada.  Additional funds for
the construction of SCUBA-2 were provided by the Canada Foundation for
Innovation.
This research made use of Montage, funded by the National Aeronautics
and Space Administration's Earth Science Technology Office,
Computational Technologies Project, under Cooperative Agreement Number
NCC5-626 between NASA and the California Institute of Technology. The
code is maintained by the NASA/IPAC Infrared Science Archive.
The data presented in this paper are partly based on the ESO-ARO 
programme ID 196.C-0999(A).
M. Juvela and V.-M. Pelkonen acknowledge the support of the Academy of
Finland Grant No. 285769.
T. Liu is supported by KASI fellowship and EACOA fellowship.
J. Malinen acknowledges the support of ERC-2015-STG No. 679852
RADFEEDBACK.
V.-M. Pelkonen acknowledges the financial support from the European
Research Council, Advanced Grant No. 320773 entitled Scattering and
Absorption of Electromagnetic Waves in Particulate Media (SAEMPL). 
\end{acknowledgements}

\bibliography{biblio_with_Planck}

\begin{appendix}

\section{Colour corrections} \label{sect:CC}

\begin{figure*}
\sidecaption
\includegraphics[width=12cm]{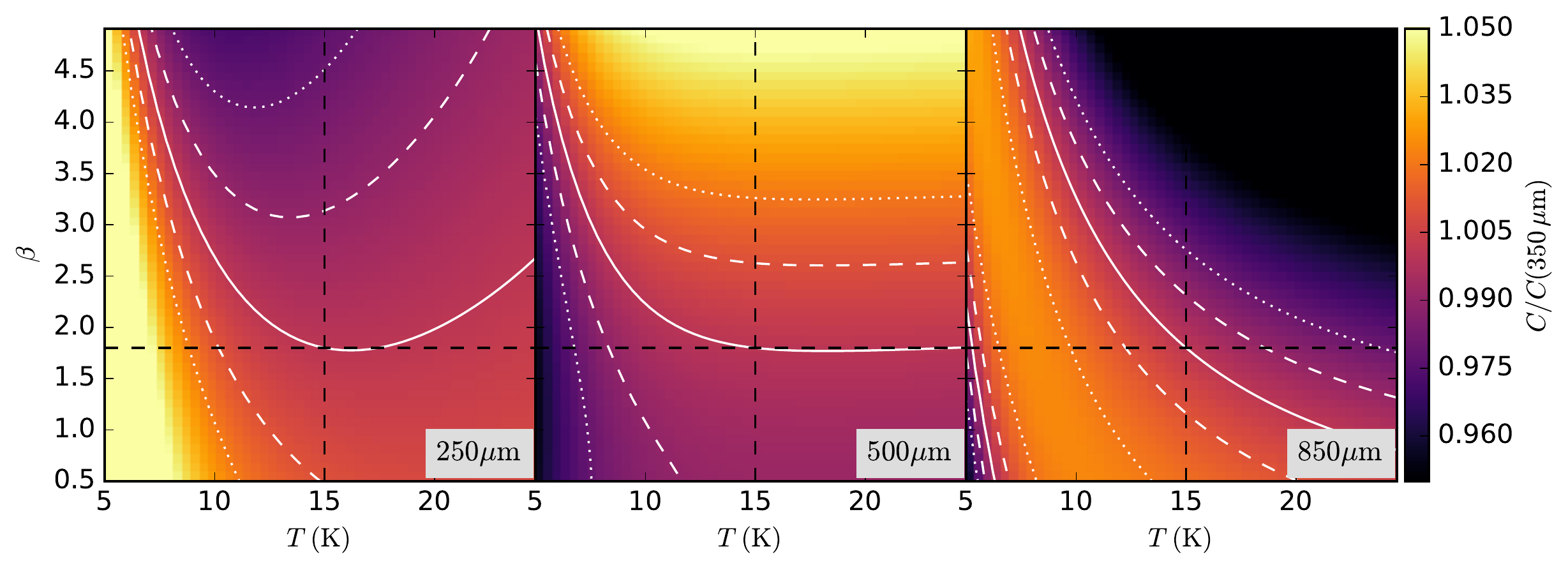}
\caption{
The colour correction factors at 250\,$\mu$m, 500\,$\mu$m, and
850\,$\mu$m bands relative to the same at 350\,$\mu$m.  The ratios are
normalised to one at $T=15$\,K and $\beta$=1.5. The contours denote
the ratio one (solid lines) and deviations from one by $\pm 1$\%
(dashed lines) and $\pm$2\% (dotted lines).
}
\label{fig:CC}
\end{figure*}

The maps were colour-corrected pixel-by-pixel, using the MBB
fits $B(T) \times \nu^{1.8}$ to the original, unfiltered
SPIRE data. Because the fits that include the 850\,$\mu$m point can
lead to different SEDs, we examined how the colour correction factors
vary over the plausible parameter range. Figure~\ref{fig:CC} compares
the ratios of colour correction factors\footnote{Colour correction
factor is here defined such that original data divided by the factor
results in an estimate of the monochromatic value at the reference
wavelength.} relative to these factors at 350\,$\mu$m. The
normalisation with the 350\,$\mu$m values is included because we are
interested only in the SED shape, not its normalisation. In the plot,
the ratio is further normalised so that it is one for $T=15.0$\,K and
$\beta$=1.5. If the ratios significantly change between the position
where colour correction was estimated and the true SED shape, it would
bias the fit results.

According to Fig.~\ref{fig:CC}, the relative variations are of the
order of 2\% or less over most of the plausible parameter space. Since
the differences between the fitted and the unknown true SED shape
should be smaller and, if the differences are caused by noise, they
are mostly along the line between the high $T$ and low $\beta$ and the
low $T$ and high $\beta$ combinations. Thus, the colour correction
uncertainties are likely to be at most $\sim$2\% and thus remain much
smaller than the typical photometric errors that are closer to 10\%. A
2\% relative error between 350\,$\mu$m and 850\,$\mu$m corresponds to
a spectral index error less than $\Delta \beta$=0.05.

\section{CO contamination in 850\,$\mu$m data} \label{sect:CO}

We do not have high-resolution observations of the CO $J$=3-2 lines to
directly correct the SCUBA-2 850\,$\mu$m data for line contamination
but we can use lower transition data to estimate the possible impact.

\begin{figure}
\sidecaption
\includegraphics[width=8cm]{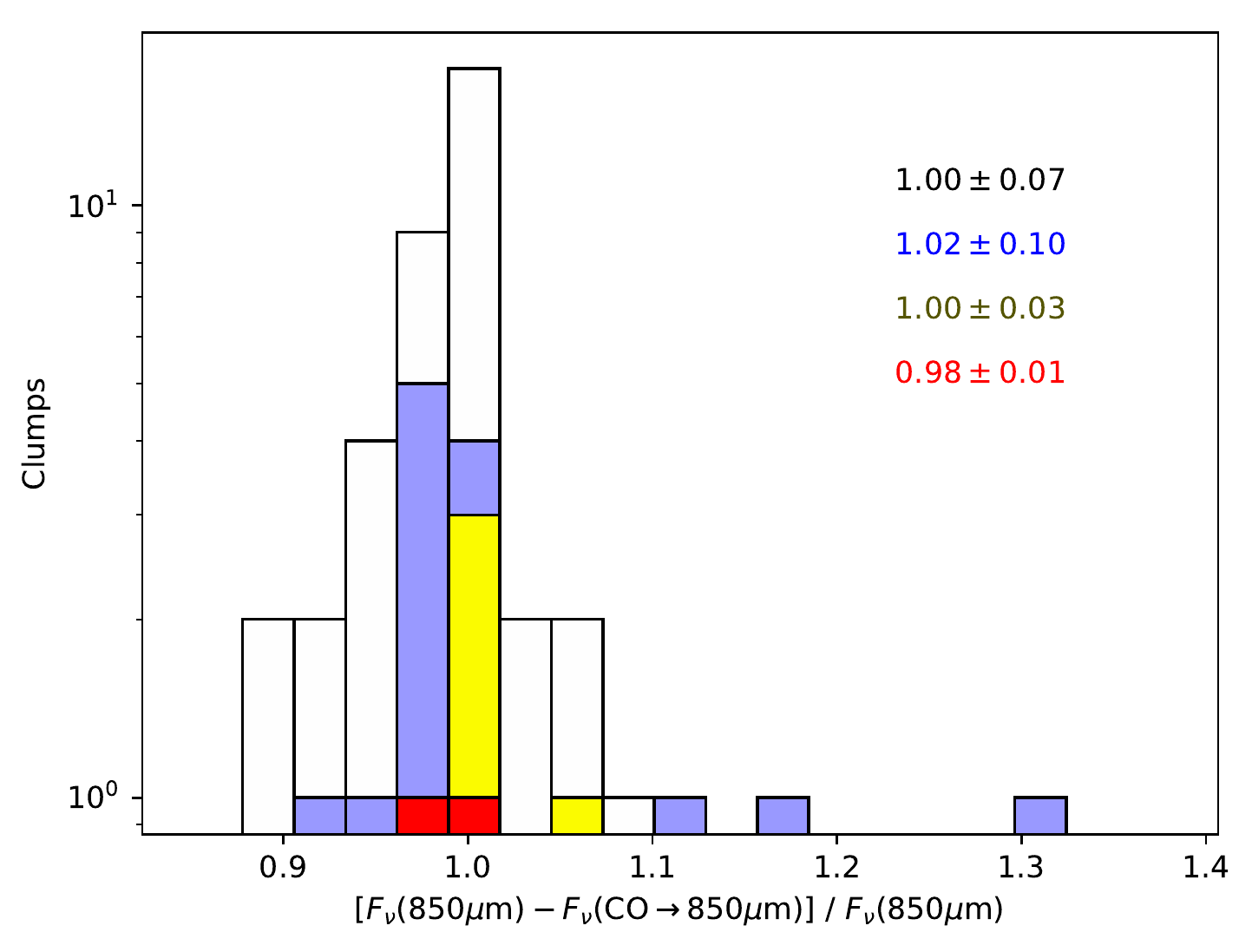}
\caption{
Ratio of CO-corrected and original 850\,$\mu$m flux values. The values
are for a subset of clumps from the $P(\chi^2)$ 85\%, 25\%, 10\%, and
5\% samples (colours white, blue, yellow, and red, respectively) with
the LM data version. The mean and standard deviation values are listed
in the figure.
}
\label{fig:CO}
\end{figure}

The $^{12}$CO(1-0) maps of \citet{Meng2013} and \citet{Zhang2016}
cover a number of SCUBA-2 targets. The CO observations were made with
the Purple Mountain Observatory telescope, which at 115\,GHz has a
beam size of 52$\arcsec$. This is sufficiently close to 40$\arcsec$,
the resolution of the maps where flux densities were measured, so that
a direct comparison is meaningful. The ratio between the $J$=3-2 and
$J$=1-0 lines is not known. The line ratios were discussed in
\citet{GCC-VI}, where they were found to be mostly $\sim$0.3 or below,
both based on {\it Planck} measurements of the velocity-integrated CO
emission and based on examples of ground-based measurements. The low
line ratios are explained by the nature of our sources, PGCC being a
selection of the coldest dust emission sources on the sky. This does
not mean that the line ratios could not be much higher in individual,
actively star-forming clumps. In the following we assume a line ratio
of one, which thus very likely overestimates the typical CO
contamination. For example, Fig.~\ref{fig:CO} suggested that the line
ratios (2-1)/(1-0) would be {\em lower} than 0.5.

We scale the CO emission to 850\,$\mu$m flux density with a constant
0.70\,mJy\,beam$^{-1}$ per K\,km\,s$^{-1}$ \citep{Drabek2012} and,
similar to continuum photometry, calculate the difference between
clump apertures and reference annuli. The exercise is done using the
LM data version. We do not process the CO maps through the SCUBA
pipeline. This should lead the CO effects to be overestimated rather
than underestimated because we ignore the loss of CO intensity caused
by the spatial filtering.

Figure~\ref{fig:CO} shows the ratios between CO-corrected and 
original 850\,$\mu$m flux densities. The mean value is close to one
and for many clumps the CO correction would increase the 850\,$\mu$m
flux density estimates. This means that the CO correction was larger
outside the clump, which is not entirely surprising given that the
cold clumps are not strong CO emitters. Molecular depletion and the
anticorrelation with local heating sources can both lead to a
situation where the CO emission is stronger outside the cold clump.

Figure~\ref{fig:CO} suggests that, even assuming a line ratio
(3-2)/(1-0) of one, the effect of the CO contamination on the
850\,$\mu$m flux density would typically be couple of per cent. Even
if the CO correction was a systematic -5\% for most clumps, the net
effect on the median $\beta$ of the 250-850\,$\mu$m fits would be only
$\Delta \beta=+0.10$.

\section{Plots of {\it Herschel} and SCUBA-2 maps} \label{sect:allmaps}

Figures~\ref{fig:mmap1}-\ref{fig:mmap9} show maps of 250\,$\mu$m
optical depth and 850\,$\mu$m surface brightness for all the 96
fields. 

The optical depths $\tau(250\,\mu{\rm m})$ are based on MBB fits of
the SPIRE 250\,$\mu$m, 350\,$\mu$m, and 500\,$\mu$m data, without
background subtraction. The intensity zero points of the input maps
(obtained from HSA) have been set through a comparison with {\it
Planck} data. For optical depths, our SED fits use a constant value of
$\beta=1.8$ and a map resolution of 40$\arcsec$.
The SCUBA-2 850\,$\mu$m maps correspond the LM version of data. (see
Sect.~\ref{sect:obs_SCUBA}). These are shown at the full resolution of
14$\arcsec$ and are scaled to surface brightness units MJy\,sr$^{-1}$.

The contours on the $\tau(250\,\mu{\rm m})$ maps indicate the extent
of the LM and M850 source masks that were used in the SCUBA-2
pipeline. The clumps extracted with the Fellwalker method are
indicated on the 850\,$\mu$m maps with white contours. Only the
sources with dashed black contours were used in investigations of the
spectral index values.

\begin{figure*}
\begin{minipage}{.9\textwidth}\centering
\includegraphics[width=7.8cm]{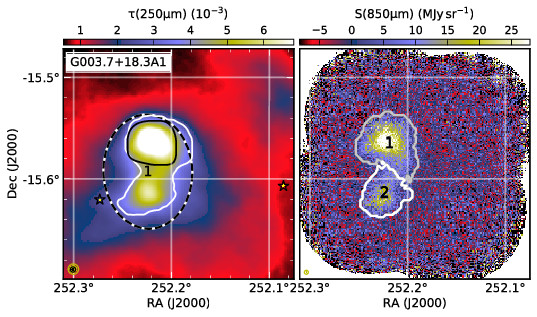}
\includegraphics[width=7.8cm]{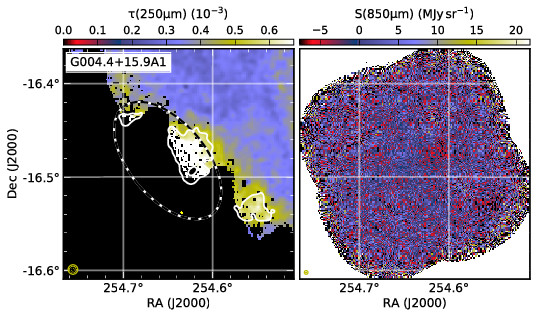}
\includegraphics[width=7.8cm]{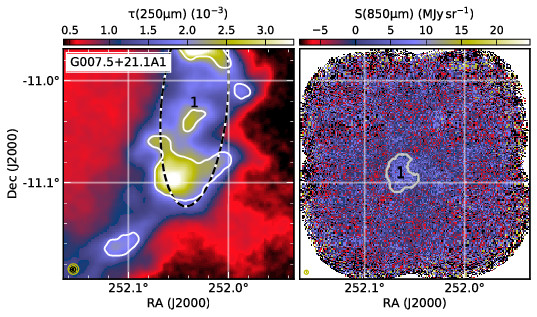}
\includegraphics[width=7.8cm]{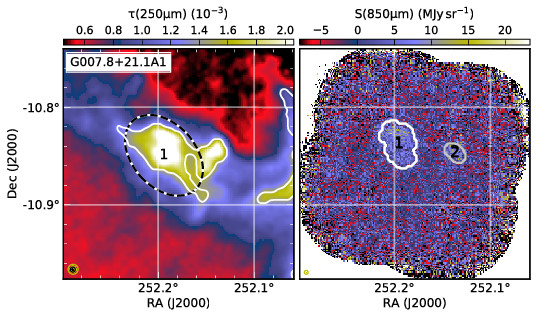}
\includegraphics[width=7.8cm]{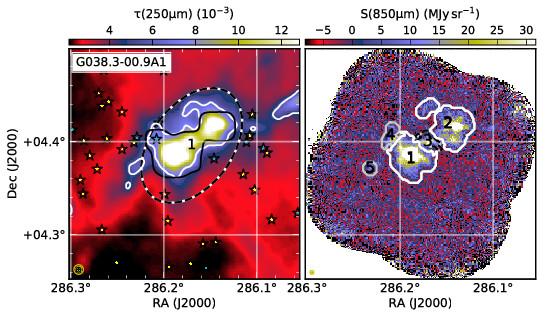}
\includegraphics[width=7.8cm]{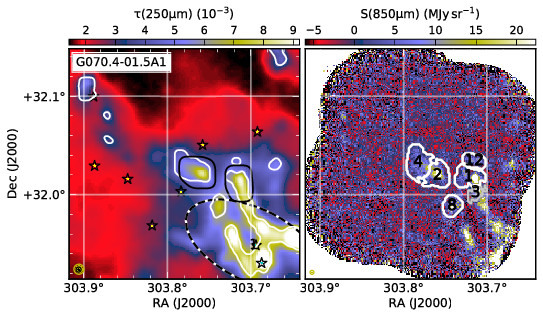}
\includegraphics[width=7.8cm]{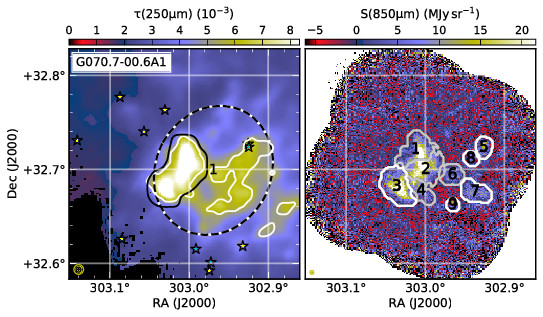}
\includegraphics[width=7.8cm]{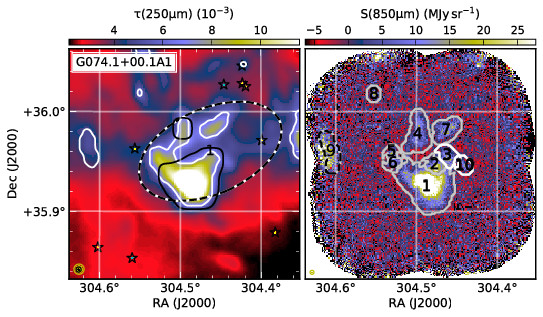}
\includegraphics[width=7.8cm]{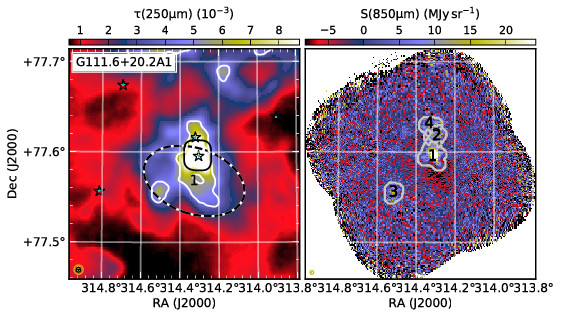}
\includegraphics[width=7.8cm]{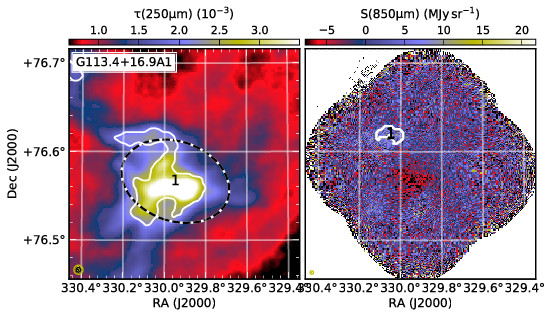}
\end{minipage}
\caption{
Maps of 250\,$\mu$m optical depth and 850\,$\mu$m surface brightness.
On $\tau(250\mu{\rm m})$ maps, white and black contours show,
respectively, the extent of the LM and M850 source masks. Dashed
ellipses show the FWHM size of PGCC clumps (numbers at the PGCC centre
coordinates refer to the numbering in Table~\ref{table:PGCC}). The
stars stand for the Class I-II (cyan stars) and Class III (yellow
stars) YSO candidates \citep{Marton2016}. On the $S(850\mu{\rm m})$
maps, the clumps of the $P(\chi^2)$=85\% sample are indicated with
white contours and the other clumps with grey contours.
}
\label{fig:mmap1}
\end{figure*}

\begin{figure*}
\begin{minipage}{.9\textwidth}\centering
\includegraphics[width=8.0cm]{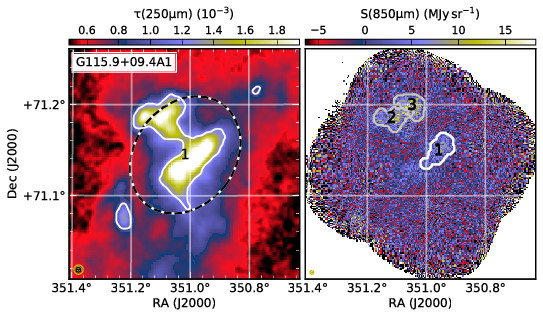}
\includegraphics[width=8.0cm]{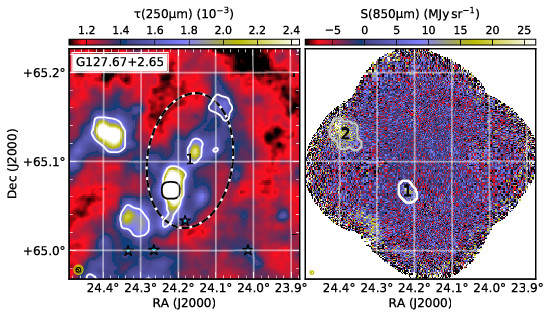}
\includegraphics[width=8.0cm]{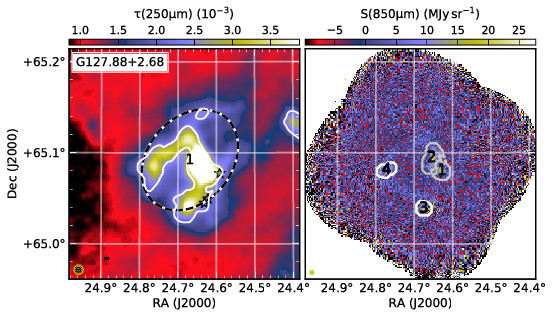}
\includegraphics[width=8.0cm]{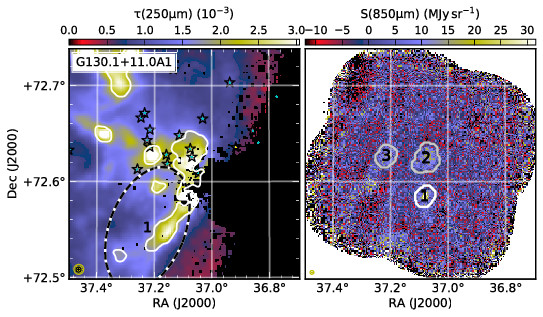}
\includegraphics[width=8.0cm]{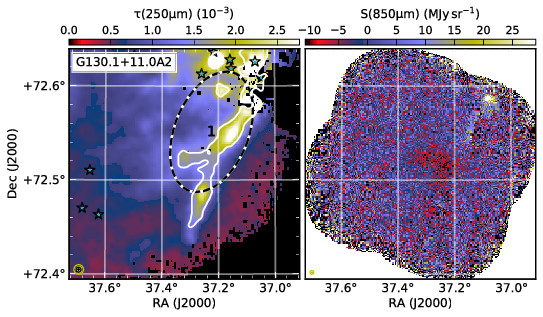}
\includegraphics[width=8.0cm]{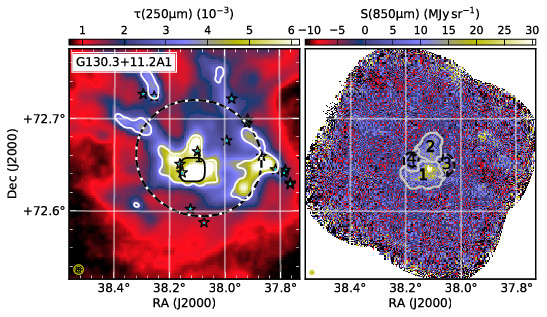}
\includegraphics[width=8.0cm]{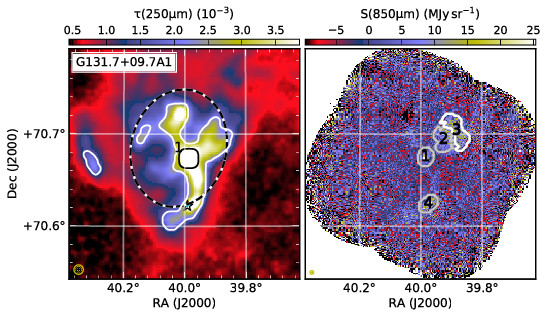}
\includegraphics[width=8.0cm]{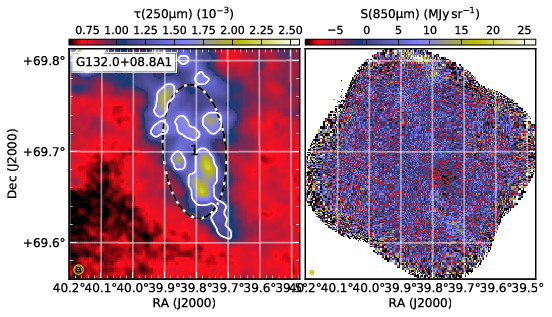}
\includegraphics[width=8.0cm]{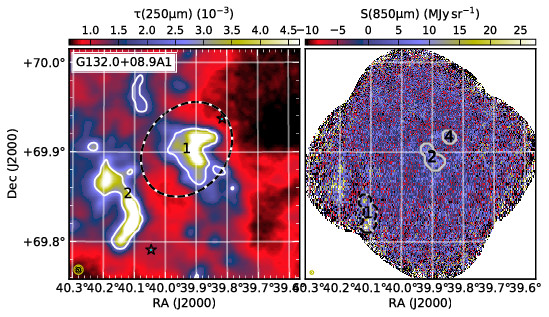}
\includegraphics[width=8.0cm]{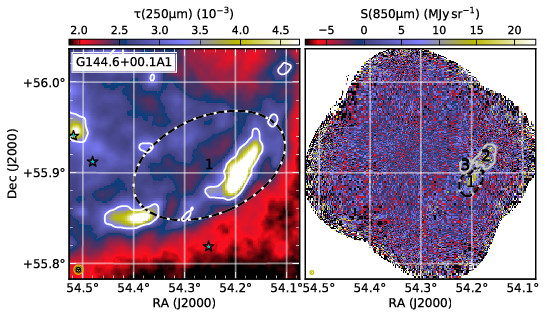}
\end{minipage}
\caption{
continued.
}
\end{figure*}

\begin{figure*}
\begin{minipage}{.9\textwidth}\centering
\includegraphics[width=8.0cm]{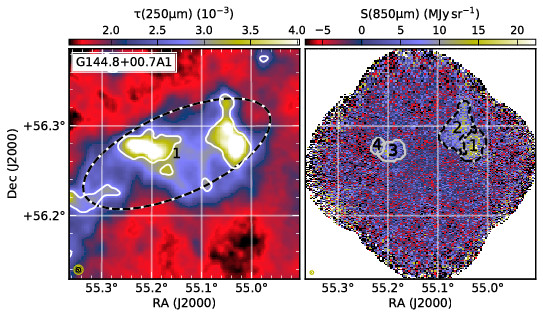}
\includegraphics[width=8.0cm]{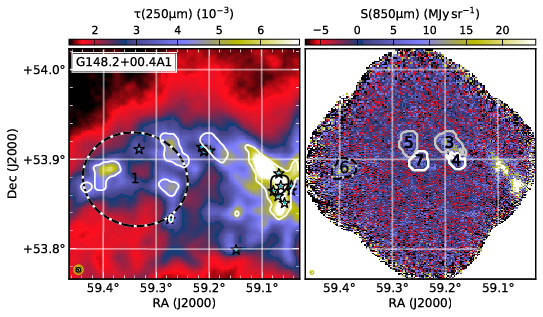}
\includegraphics[width=8.0cm]{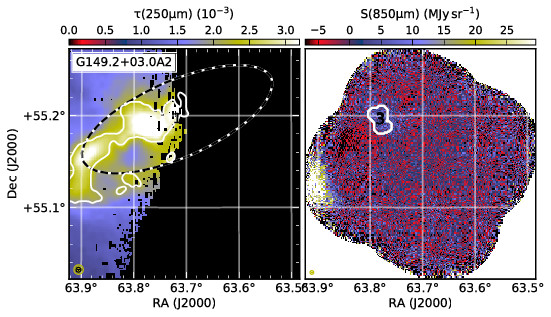}
\includegraphics[width=8.0cm]{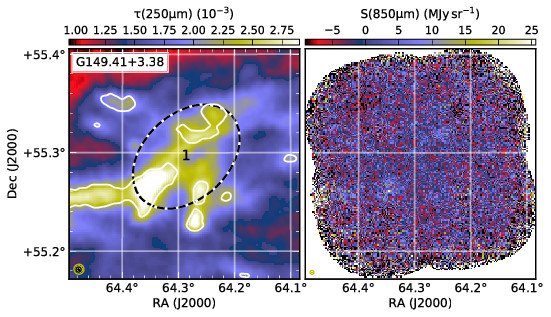}
\includegraphics[width=8.0cm]{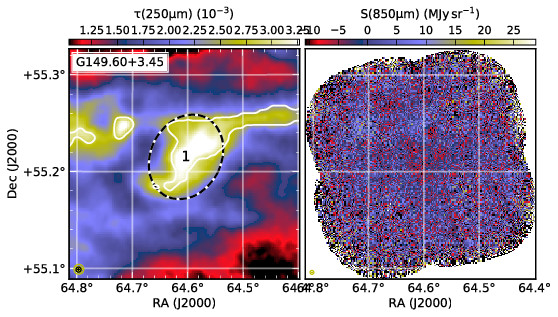}
\includegraphics[width=8.0cm]{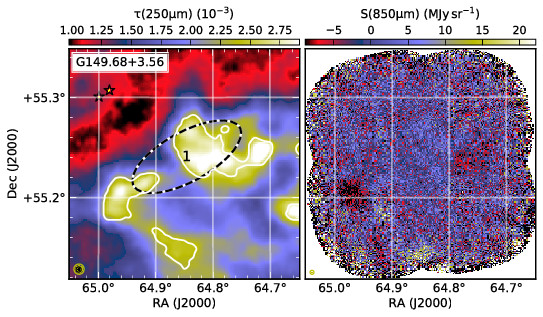}
\includegraphics[width=8.0cm]{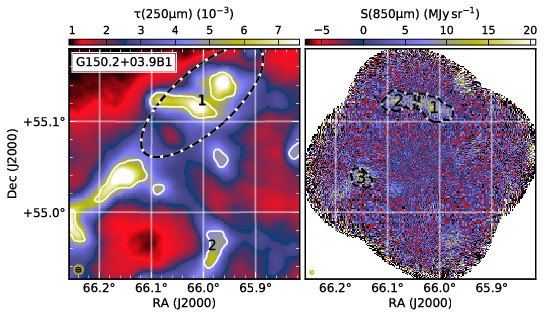}
\includegraphics[width=8.0cm]{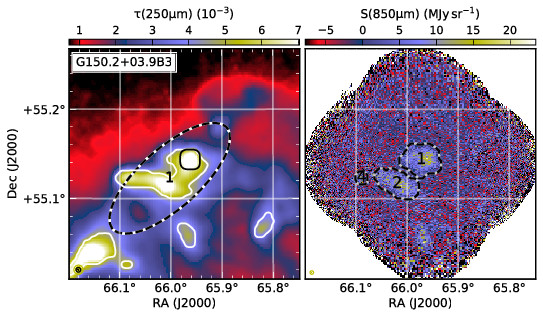}
\includegraphics[width=8.0cm]{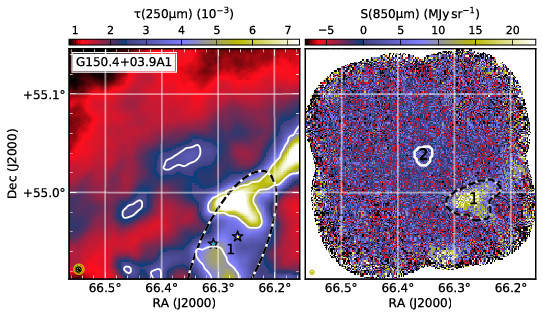}
\includegraphics[width=8.0cm]{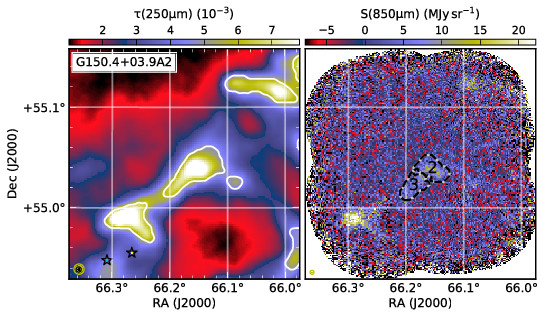}
\end{minipage}
\caption{
continued.
}
\end{figure*}

\begin{figure*}
\begin{minipage}{.9\textwidth}\centering
\includegraphics[width=8.0cm]{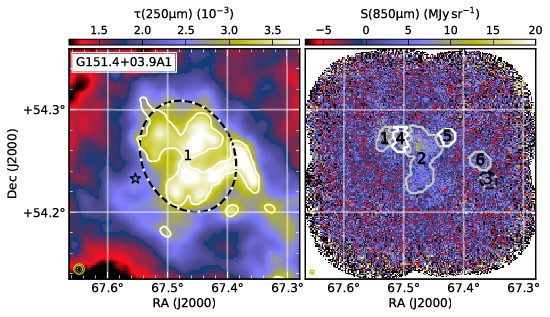}
\includegraphics[width=8.0cm]{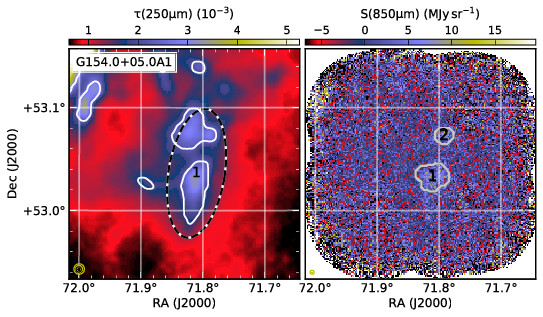}
\includegraphics[width=8.0cm]{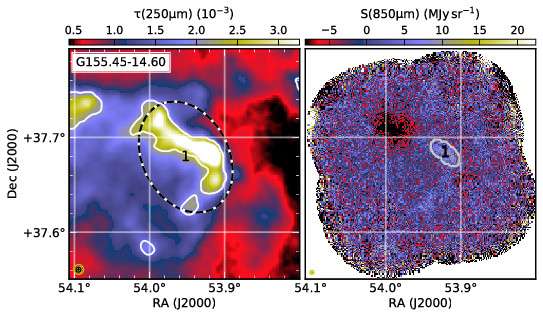}
\includegraphics[width=8.0cm]{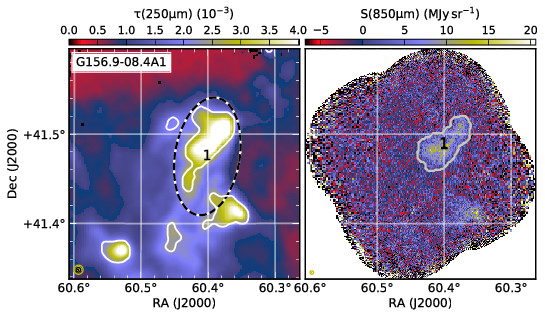}
\includegraphics[width=8.0cm]{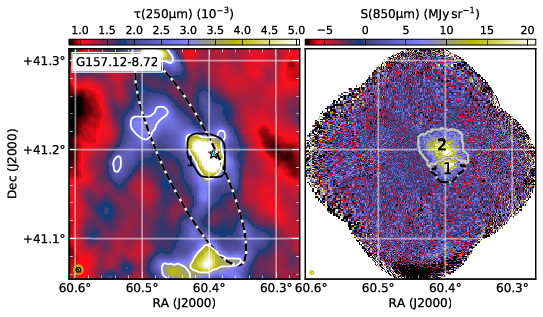}
\includegraphics[width=8.0cm]{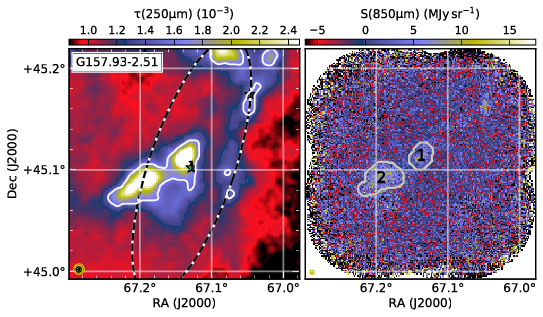}
\includegraphics[width=8.0cm]{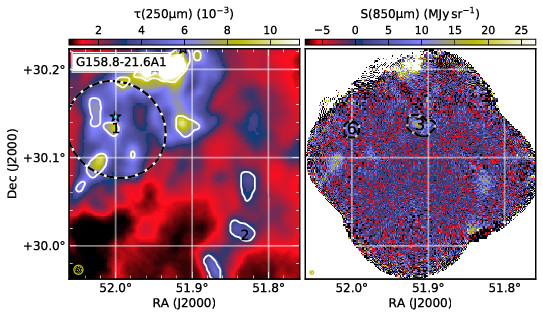}
\includegraphics[width=8.0cm]{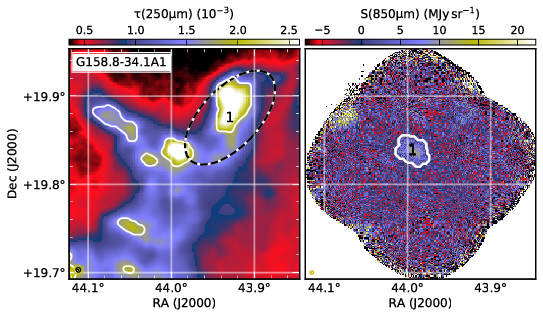}
\includegraphics[width=8.0cm]{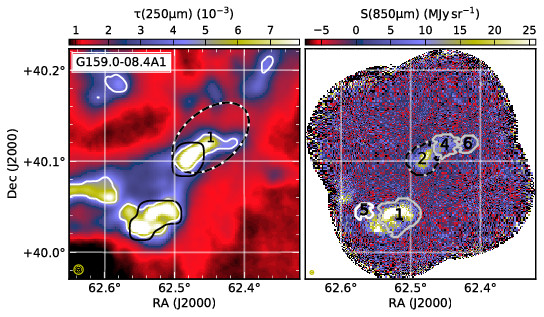}
\includegraphics[width=8.0cm]{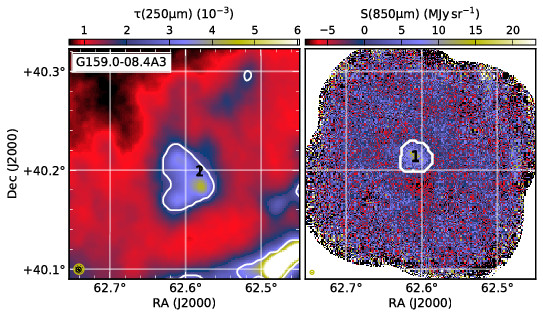}
\end{minipage}
\caption{
continued.
}
\end{figure*}

\begin{figure*}
\begin{minipage}{.9\textwidth}\centering
\includegraphics[width=8.0cm]{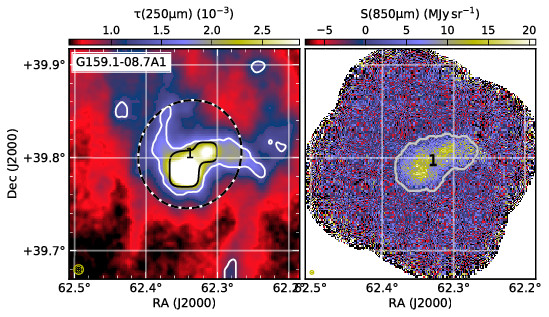}
\includegraphics[width=8.0cm]{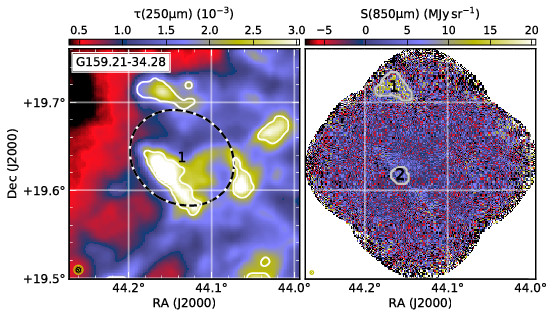}
\includegraphics[width=8.0cm]{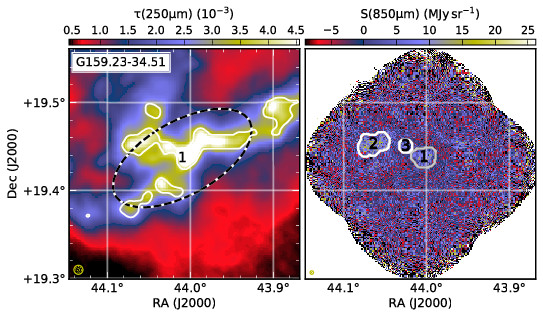}
\includegraphics[width=8.0cm]{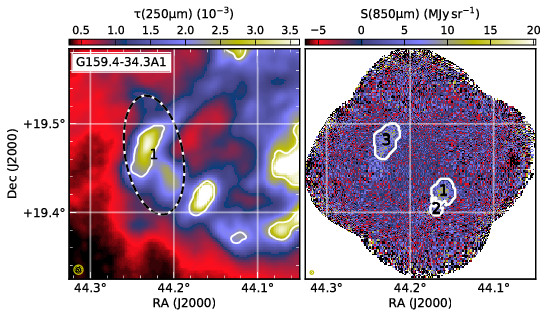}
\includegraphics[width=8.0cm]{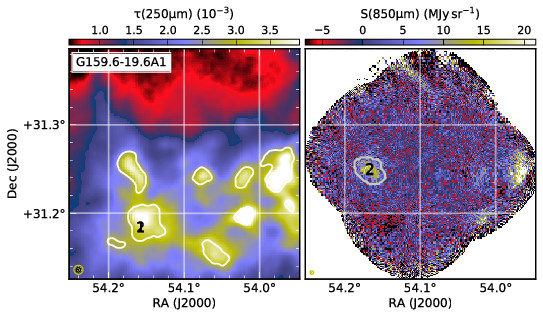}
\includegraphics[width=8.0cm]{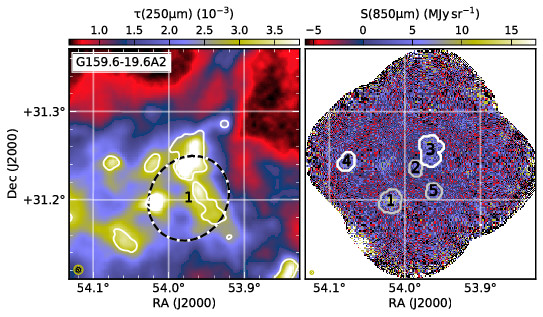}
\includegraphics[width=8.0cm]{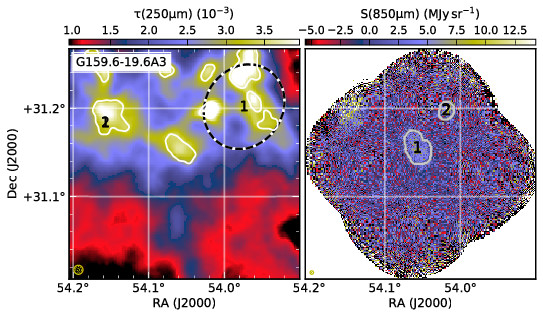}
\includegraphics[width=8.0cm]{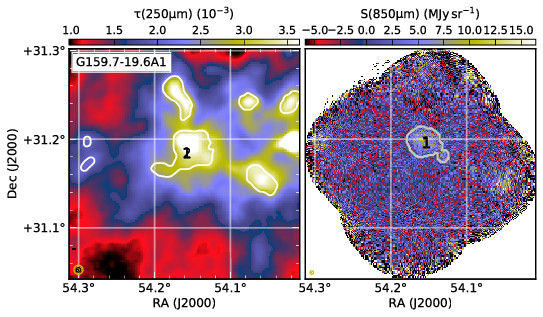}
\includegraphics[width=8.0cm]{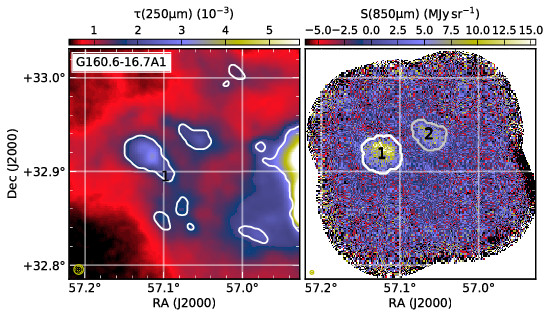}
\includegraphics[width=8.0cm]{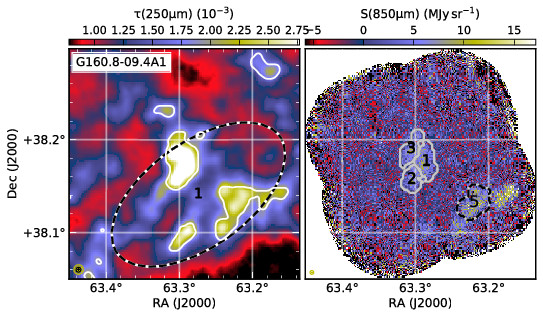}
\end{minipage}
\caption{
continued.
}
\end{figure*}

\begin{figure*}
\begin{minipage}{.9\textwidth}\centering
\includegraphics[width=8.0cm]{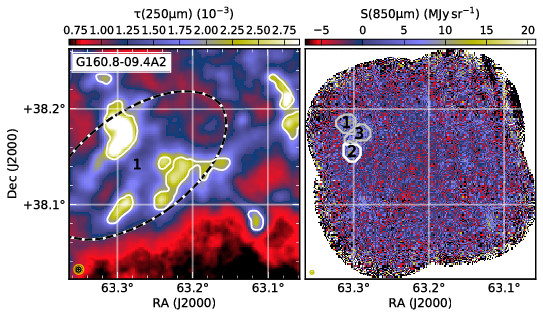}
\includegraphics[width=8.0cm]{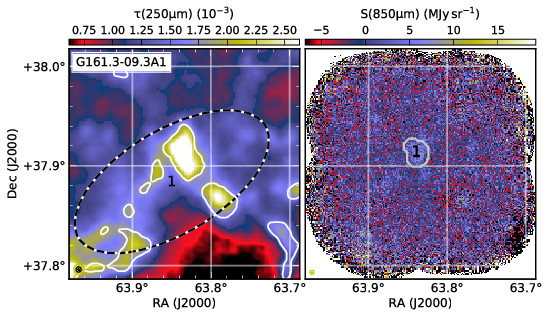}
\includegraphics[width=8.0cm]{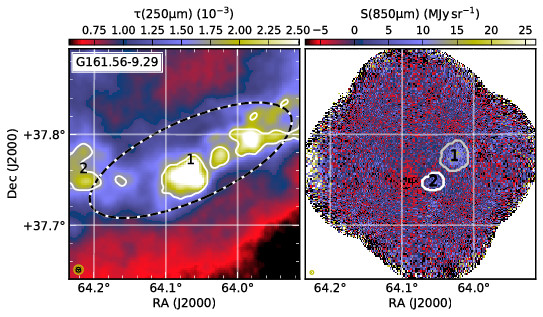}
\includegraphics[width=8.0cm]{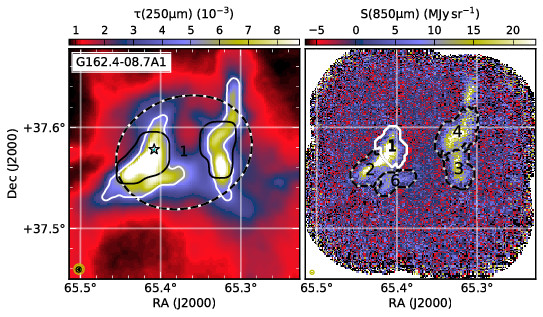}
\includegraphics[width=8.0cm]{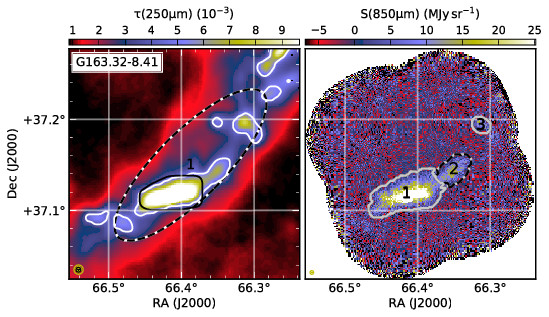}
\includegraphics[width=8.0cm]{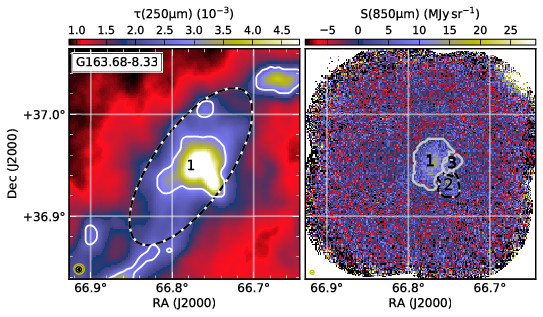}
\includegraphics[width=8.0cm]{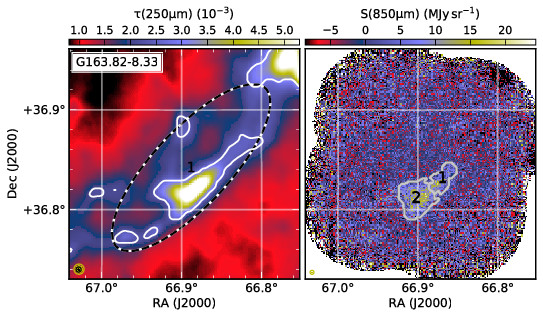}
\includegraphics[width=8.0cm]{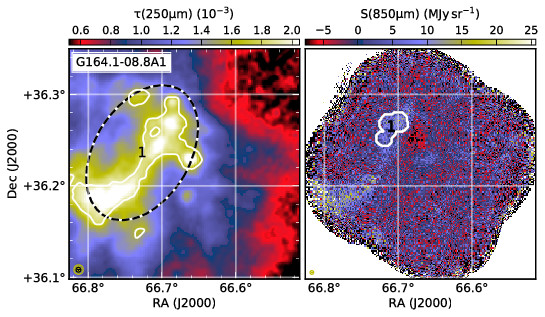}
\includegraphics[width=8.0cm]{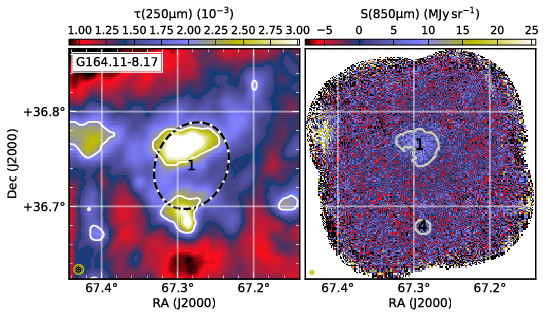}
\includegraphics[width=8.0cm]{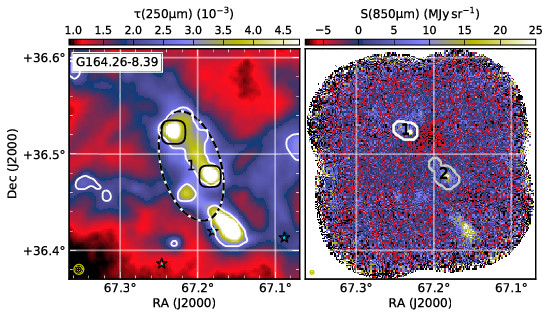}
\end{minipage}
\caption{
continued.
}
\end{figure*}

\begin{figure*}
\begin{minipage}{.9\textwidth}\centering
\includegraphics[width=8.0cm]{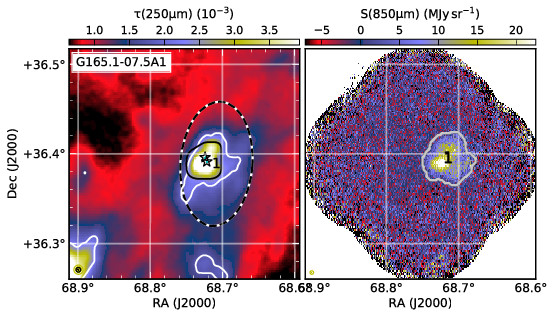}
\includegraphics[width=8.0cm]{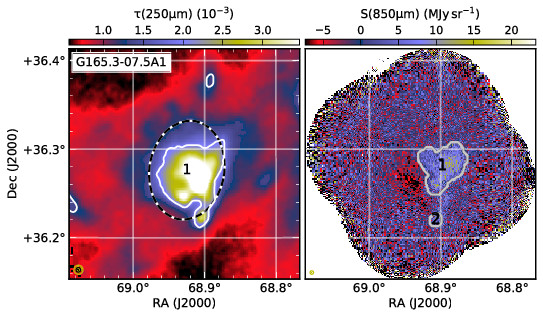}
\includegraphics[width=8.0cm]{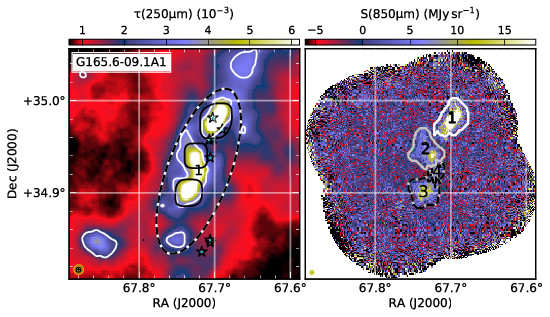}
\includegraphics[width=8.0cm]{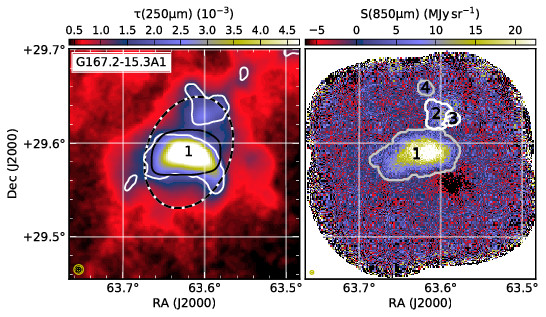}
\includegraphics[width=8.0cm]{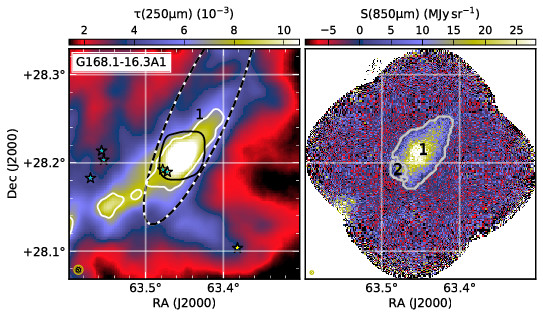}
\includegraphics[width=8.0cm]{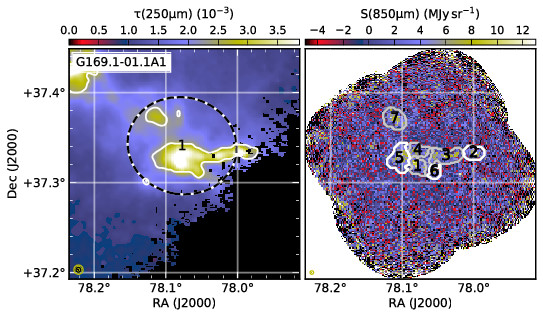}
\includegraphics[width=8.0cm]{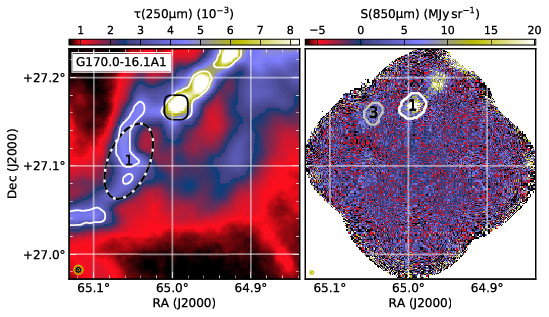}
\includegraphics[width=8.0cm]{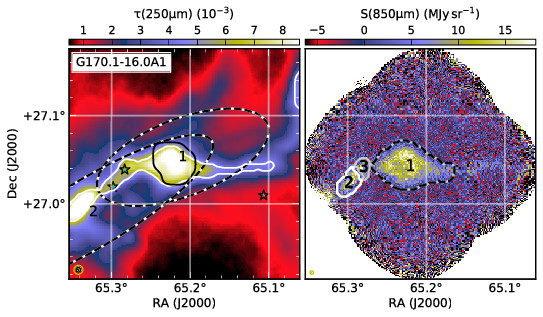}
\includegraphics[width=8.0cm]{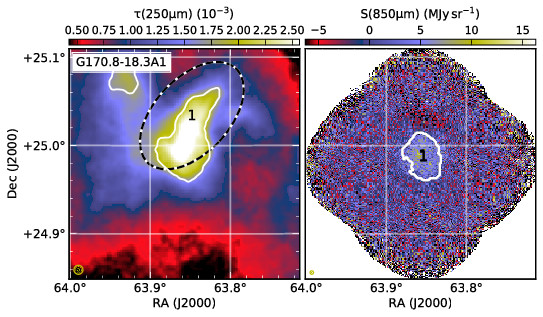}
\includegraphics[width=8.0cm]{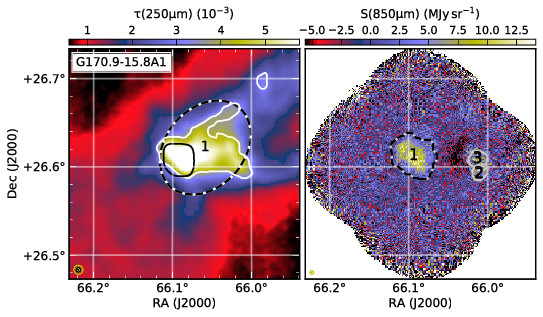}
\end{minipage}
\label{fig:x}
\caption{
continued.
}
\end{figure*}

\begin{figure*}
\begin{minipage}{.9\textwidth}\centering
\includegraphics[width=8.0cm]{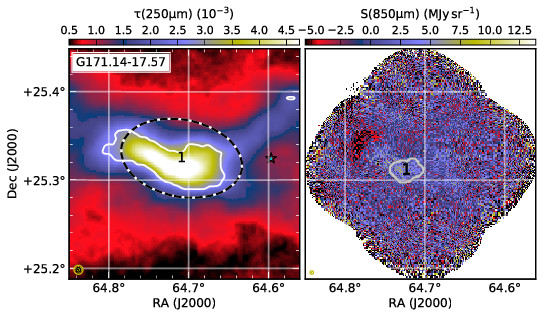}
\includegraphics[width=8.0cm]{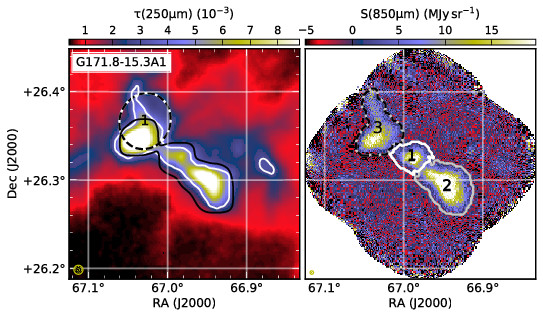}
\includegraphics[width=8.0cm]{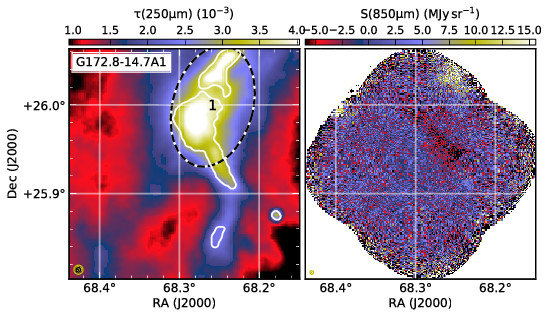}
\includegraphics[width=8.0cm]{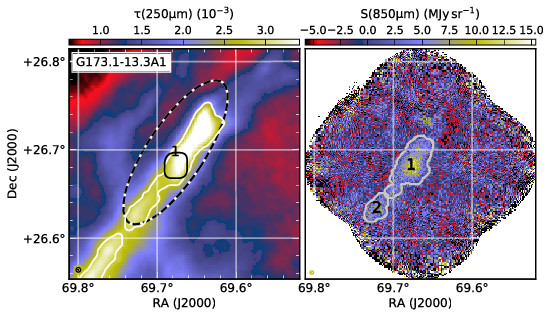}
\includegraphics[width=8.0cm]{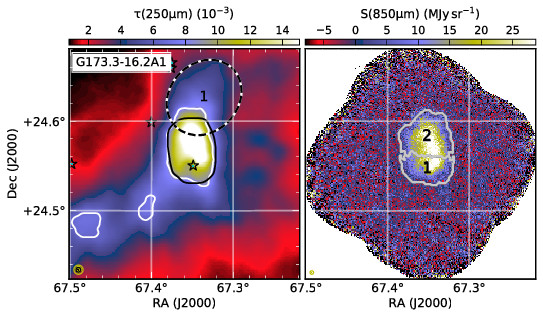}
\includegraphics[width=8.0cm]{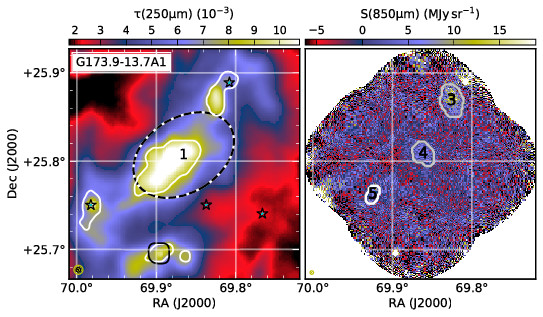}
\includegraphics[width=8.0cm]{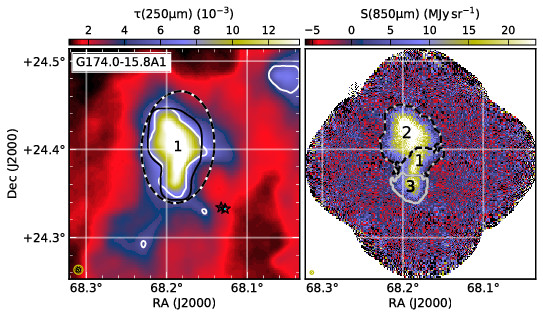}
\includegraphics[width=8.0cm]{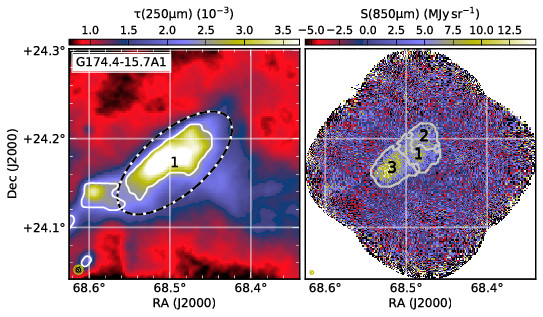}
\includegraphics[width=8.0cm]{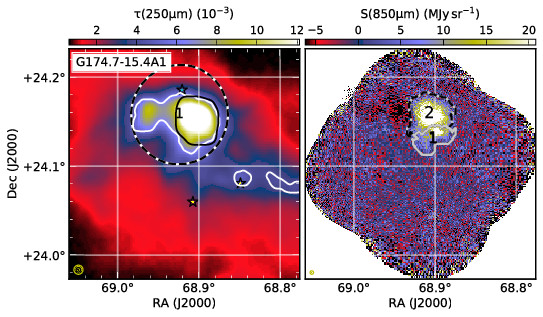}
\includegraphics[width=8.0cm]{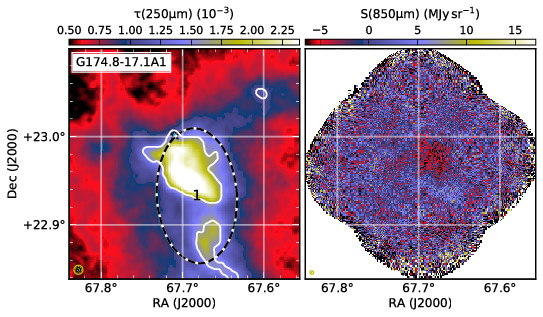}
\end{minipage}
\caption{
continued.
}
\end{figure*}

\begin{figure*}
\begin{minipage}{.9\textwidth}\centering
\includegraphics[width=8.0cm]{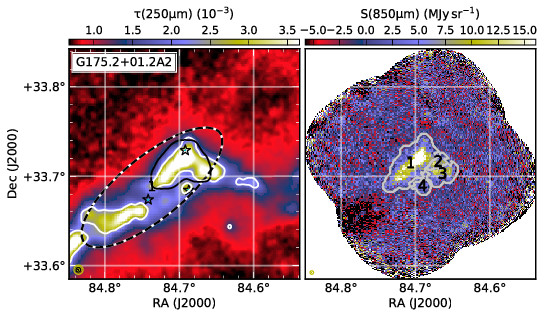}
\includegraphics[width=8.0cm]{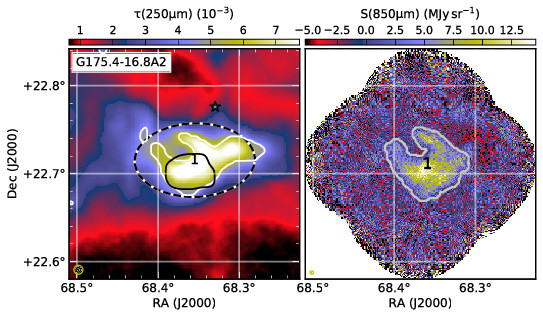}
\includegraphics[width=8.0cm]{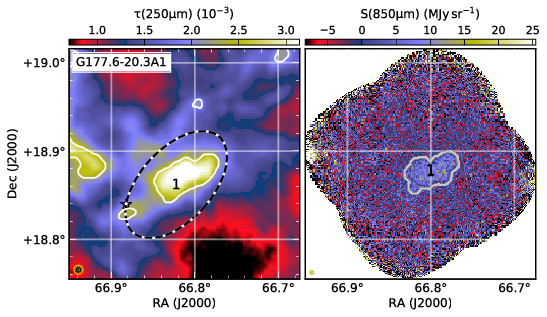}
\includegraphics[width=8.0cm]{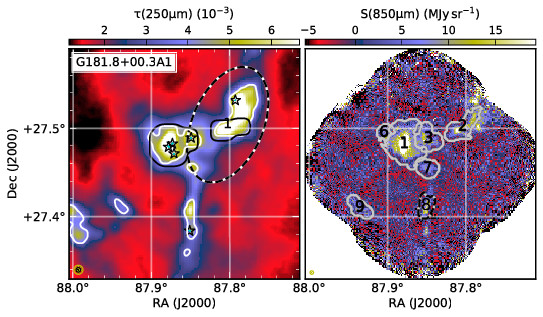}
\includegraphics[width=8.0cm]{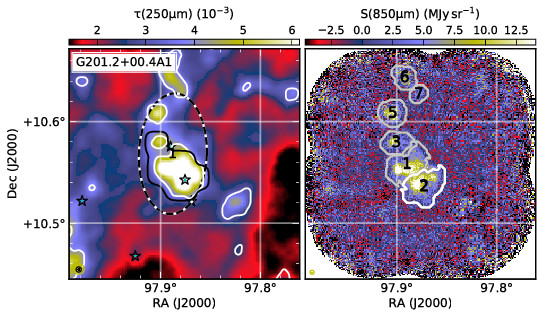}
\includegraphics[width=8.0cm]{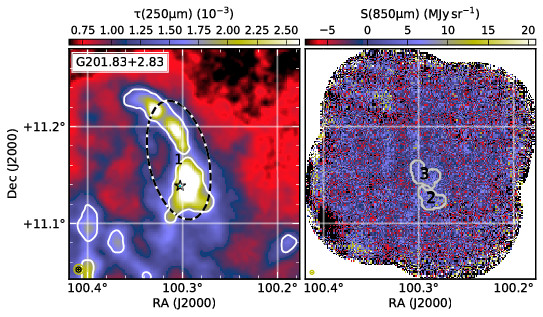}
\includegraphics[width=8.0cm]{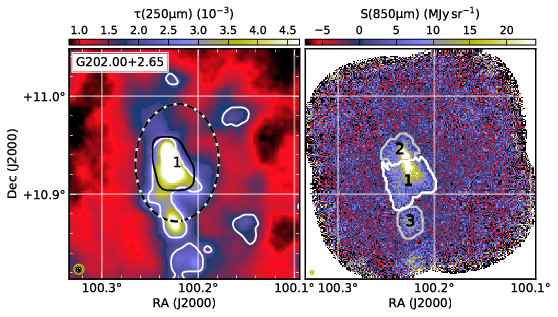}
\includegraphics[width=8.0cm]{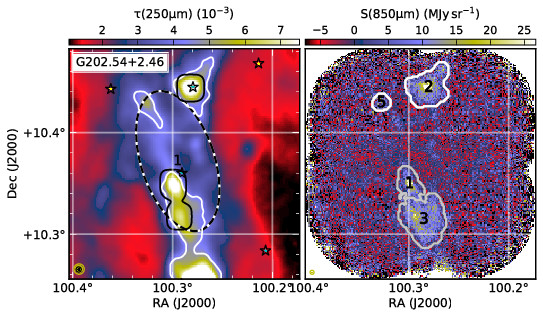}
\includegraphics[width=8.0cm]{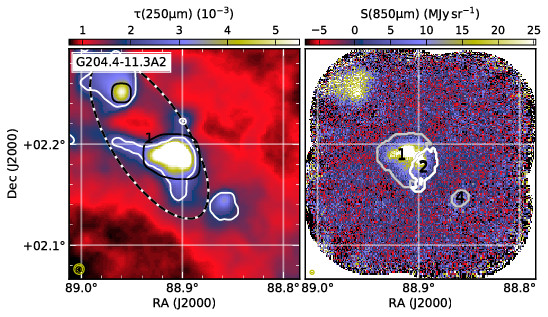}
\includegraphics[width=8.0cm]{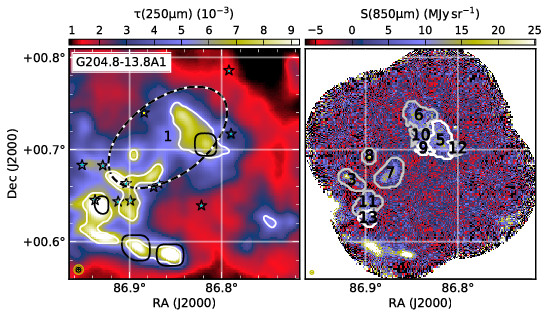}
\end{minipage}
\caption{
continued.
}
\end{figure*}

\begin{figure*}
\begin{minipage}{.9\textwidth}\centering
\includegraphics[width=8.0cm]{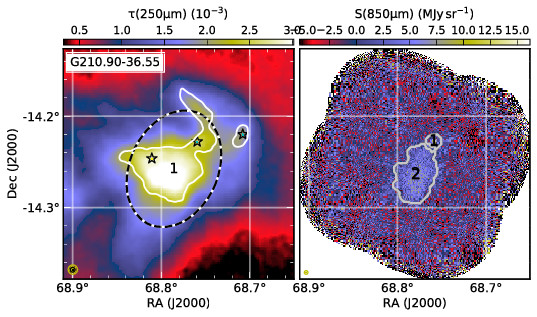}
\includegraphics[width=8.0cm]{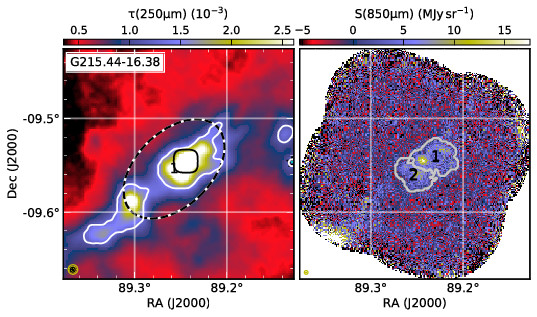}
\includegraphics[width=8.0cm]{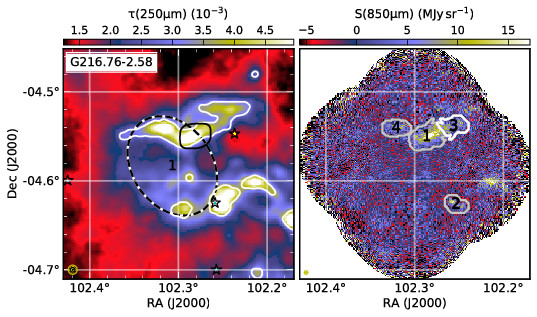}
\includegraphics[width=8.0cm]{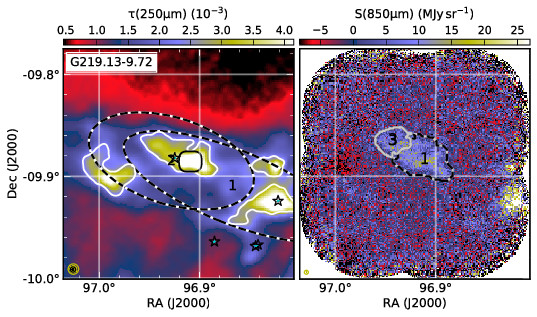}
\includegraphics[width=8.0cm]{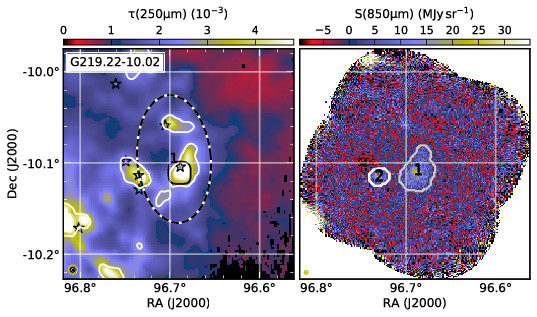}
\includegraphics[width=8.0cm]{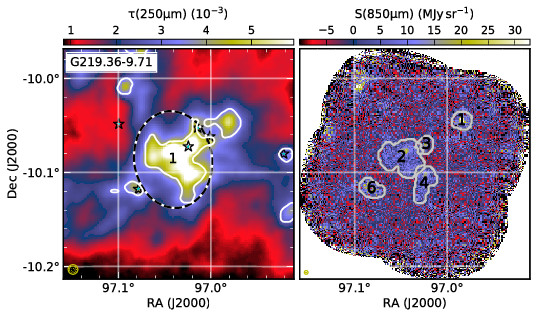}
\end{minipage}
\caption{
continued.
}
\label{fig:mmap9}
\end{figure*}

\section{SED fits of {\it Planck} data alone} \label{sect:NO_IRAS}

\begin{figure}
\includegraphics[width=8.8cm]{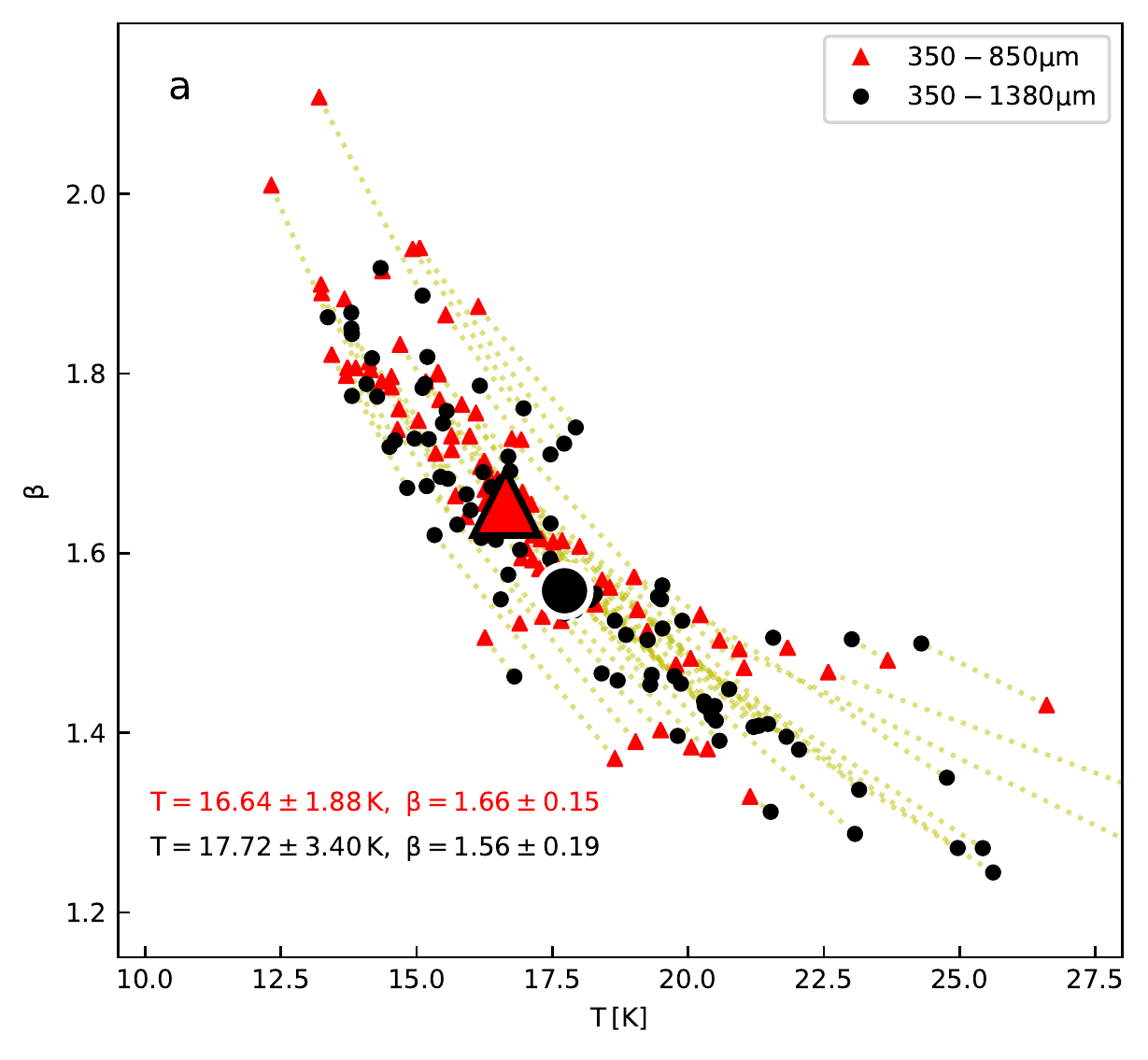}
\caption{
MBB fits to $Planck$ data averaged within 8$\arcmin$ of the centre of
the SCUBA-2 fields. The wavelength range is either 350-850\,$\mu$m or
350-1380\,$\mu$m. Each marker corresponds to one field and the dotted
lines connect estimates of the same field. The median values are
plotted with large symbols and they are quoted in the frame together
with 1-$\sigma$ dispersion estimated from the interquartile ranges.
The plot assumes the default CO corrections.
}
\label{fig:NO_IRAS}
\end{figure}

In Sect.~\ref{sect:SED_IP} the mean emission of the fields was fitted
with IRAS and {\it Planck} data. The shortest {\it Planck} wavelength
of 350\,$\mu$m is near the maximum of cold dust emission spectrum.
Nevertheless, one can attempt SED fits also without the IRAS data
point. Figure~\ref{fig:NO_IRAS}

Although the scatter is significantly larger than in
Fig.~\ref{fig:IP_8am}, the median values are similar to the fits where
also the IRAS 100\,$\mu$m point was used. There is no evidence that
the latter would be significantly affected by very small grain
emission or by temperature variations inside the measurement aperture.
As in Fig.~\ref{fig:IP_8am}a, the extension of the fitted wavelength
range to 1380\,$\mu$m slightly decreases the median $\beta$ value.

\section{Dependence between dust parameters and error estimates} \label{sect:betacorr}

\begin{figure*}
\includegraphics[width=17.8cm]{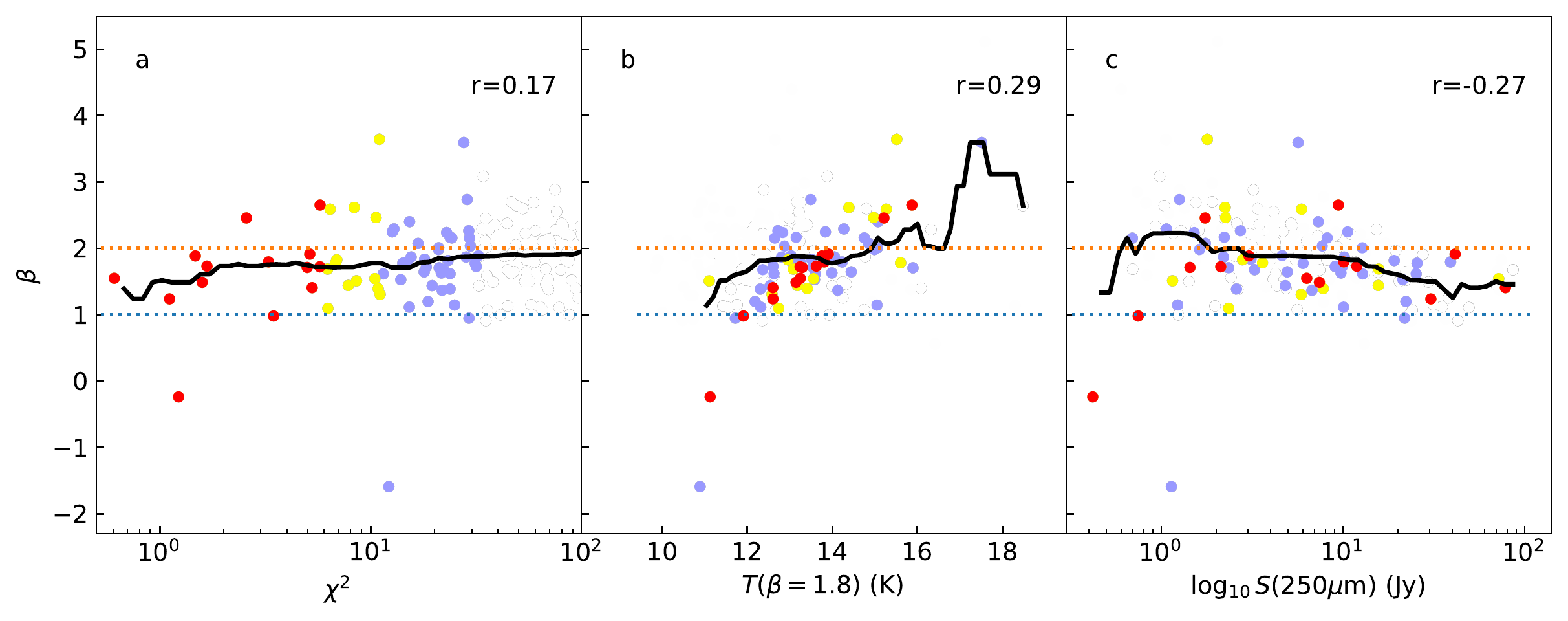}
\caption{
Dependence of $\beta$ estimates on $\chi^2$ values (frame a), 
SPIRE-derived colour temperatures for $\beta = 1.8$ (frame b), and
250\,$\mu$m flux densities (frame c). The colours are the same as in
Fig.~\ref{fig:betahis}. The linear correlation coefficients $r$ and
median-$\beta$ curves are shown for the full set of plotted points.
}
\label{fig:betacorr}
\end{figure*}

The dependence of $\beta$ estimated from SPIRE and SCUBA-2 data is
compared to some other parameters in Fig.~\ref{fig:betacorr}. The
estimates are almost independent of the (non-reduced) $\chi^2$ values
of the fits and weakly correlated with the colour temperature
that is estimated from 250-500\,$\mu$m data using a fixed value of
$\beta = 1.8$. Finally, Fig.~\ref{fig:betacorr}c shows some negative
correlation with the source brightness. However, none of the
correlations are statistically significant.

\begin{figure}
\includegraphics[width=8.8cm]{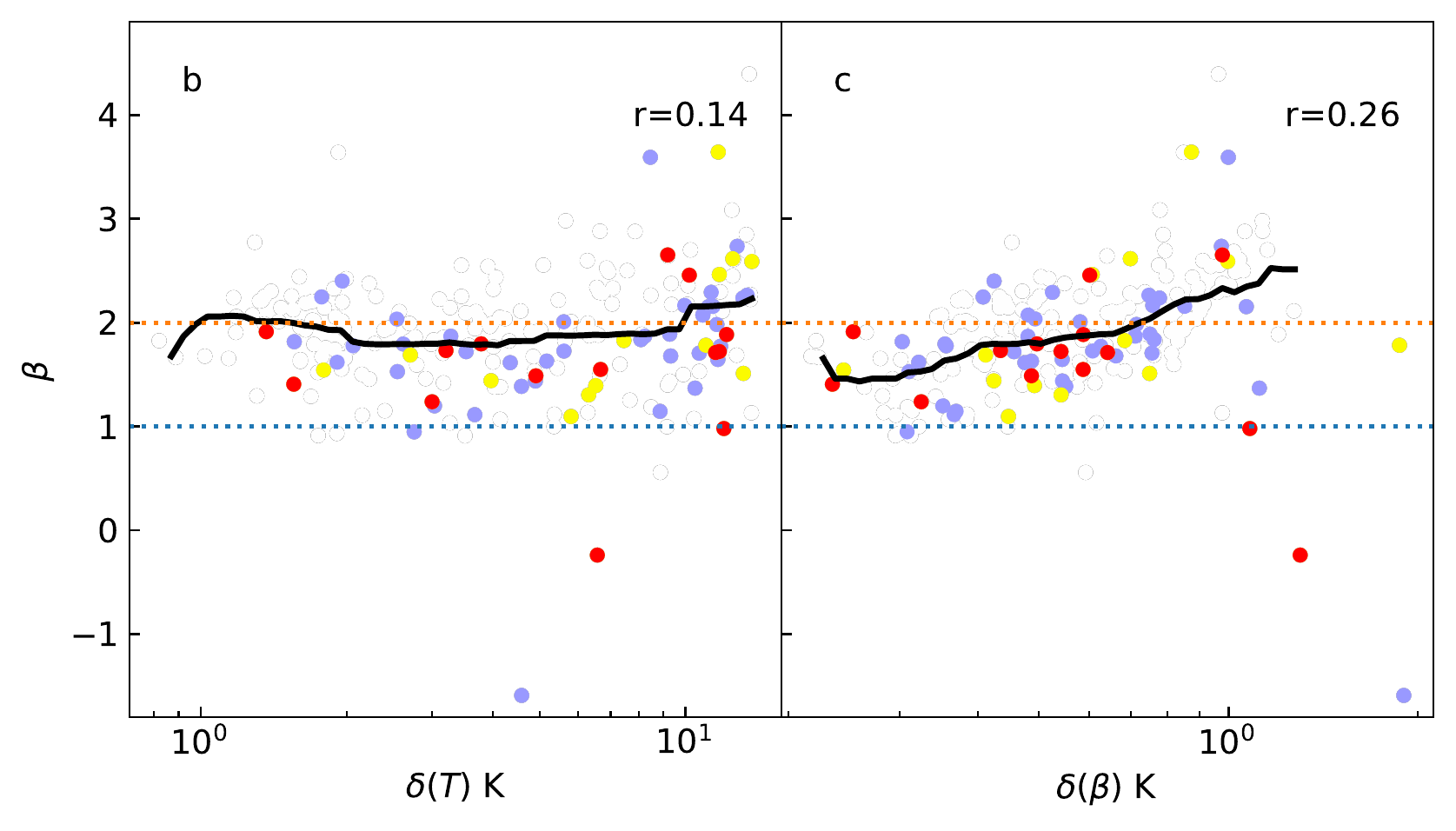}
\caption{
Estimates of $\beta$ as the function of the $T$ and $\beta$ MCMC error
estimates. The symbol colours are as in Fig.~\ref{fig:betacorr} and
the black lines show the median values as the function of the x-axis
variable. The linear correlation coefficients $r$ are given in the
figure.
}
\label{fig:betacorr2}
\end{figure}

It would seem natural to analyse $\beta$ distributions using the
sources with the lowest $\beta$ error estimates. However, the ($T$,
$\beta$) anticorrelation leads to a narrow, curved $\chi^2$ valley in
the ($T$, $\beta$) plane. In the case of large uncertainties, it runs
at low $T$ runs parallel to the $\beta$ axis and at high $T$ parallel
to temperature axis. Thus, the selection of sources with low $\beta$
uncertainty would set preference to warm sources with low $\beta$
values. The effect is illustrated in Fig.~\ref{fig:betacorr2} where
the error estimates $\delta(T)$ and $\delta(\beta)$ correspond to half
of the interquartile range estimated from MCMC runs. As expected, the
figure shows a correlations between $\beta$ and the error estimate of
the $\beta$ uncertainty. Subsamples selected based on the $\chi^2$
values of the fits are not significantly correlated with $\delta(T)$
or $\delta(\beta)$.

\section{Effect of map making and clump selection on $\beta$}
\label{sect:allhisto}

The obtained $\beta$ distributions depend on our decisions regarding
the map making and the data calibration. In map making, the options
were to use 200$\arcsec$ or 500$\arcsec$ filtering scale. Furthermore,
maps could be produced using either large or small source masks based
on {\it Herschel} 500\,$\mu$m data (SM and LM, respectively) or masks
based on SCUBA-2 850\,$\mu$m data itself (M850). For illustration of
the calibration uncertainty of the 850\,$\mu$m data, we compared the
result for the flux conversion factor FCF=2.41\, Jy\,(pW)$^{-1}$,
estimated from calibration measurements near the observing times, to
the results obtained with the default calibration value of FCF=2.34\,
Jy\,(pW)$^{-1}$ arcsec$^{-2}$.

The following table describes the $\beta$ distributions. These include
the $\beta$ distributions calculated using SPIRE data only (unfiltered
and filtered versions), because also the clump selection changes from
case to case.

The data show that the two cases of the 850\,$\mu$m calibration differ
by less than $\Delta \beta=0.1$, except for the M850 case, where the
larger FCF factor results in a median spectral index value lower by
0.2 units.

\section{Effect of noise on the $\beta(T)$ relation} \label{sect:simu}

In Sect.~\ref{dis:TB} we used simulations to test the significance of
the observed negative correlation between the colour temperature and
the spectral index. The noise was kept as a free parameter because
otherwise the result would critically depend on the precision of the a
priori estimates of photometric errors. Here we examine the method
further with purely synthetic observations.

We assume an intrinsic dependence of $\beta(T) = (T/15\,{\rm K})^B$
and simulate observations at 250\,$\mu$m, 350\,$\mu$m, 500\,$\mu$m,
and 850\,$\mu$m. The default photometric errors are 4\% for the first
three and 10\% for the 850\,$\mu$m band. We generate a sample of 100
clumps. The temperatures follow a normal distribution with a mean of
15\,K and a standard deviation of 3\,K. Together with the $\beta(T)$
relation this defines the source fluxes to which we add normally
distributed photometric errors. 
The fitting of a set of synthetic observation gives the observed
values of $B$ and the rms dispersion with respect to the fitted
$\beta(T)$ curve. 

The analysis of these synthetic observations proceeds as described in
Sect.~\ref{dis:TB}. We generate a similar sample of observations under
the assumption of a flat $\beta(T)$ relation. This is obtained by
fitting the syntetic observations with a fixed value of $\beta$ and
replacing the flux density values by the values predicted by the fit.
We then examine the $B$ parameter and the rms dispersion of $\beta$
vs. the fitted $\beta(T)$ relation of this simulated sample as the
function of the noise level. The noise level is again scaled by the
parameter $k {\rm (noise)}$. When the rms noise in the simulation
matches the rms noise of the (this time synthetic) observations, we
can check if the corresponding value of $B$ in the simulation is
higher than the value of $B$ in the observations. If the difference is
positive and significant, we could conclude that the observations
exhibit a non-constant $\beta(T)$ relation ($\beta$ decreasing with
temperature) that cannot be explained by the noise.

\begin{figure*}
\sidecaption
\includegraphics[width=12cm]{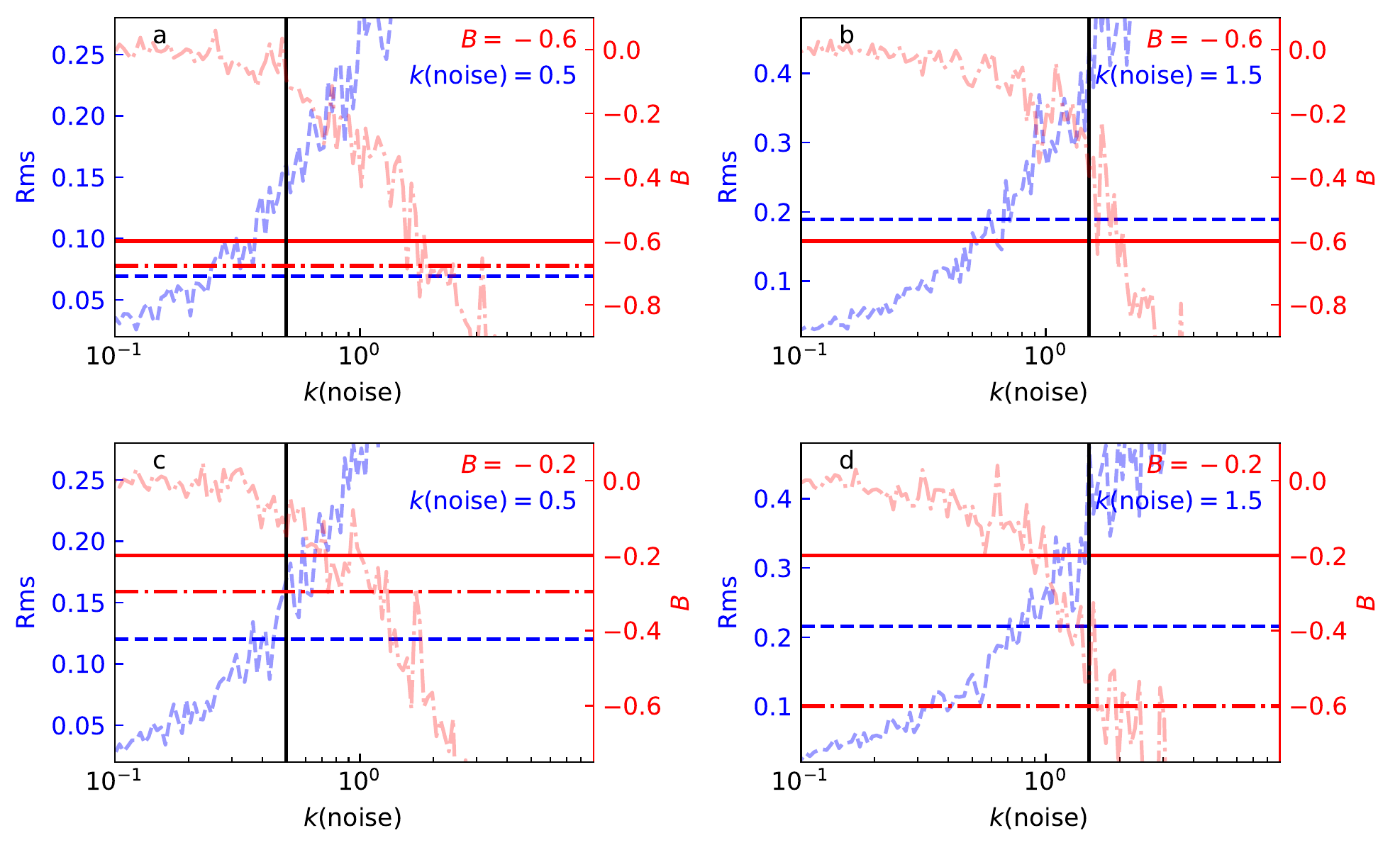}
\caption{
Analysis of a set of synthetic observations that follow an intrinsic
$\beta(T) \propto T^{B}$ relation. The parameters $B$ and $k{\rm
(noise)}$ of the input simulation are given in the frames and
indicated by solid vertical and horizontal lines.
The horizontal dot-dashed and dashed lines correspond, respectively,
to the values of $B$ and the $\beta$ rms dispersion which are
estimated from one realisation of synthetic observations. The other
curves show $B$ (dot-dashed red line) and the rms dispersion (blue
dashed line) from simulations as a function of the noise scaling
$k{\rm (noise)}$.
}
\label{fig:simu}
\end{figure*}

\begin{figure*}
\sidecaption
\includegraphics[width=12cm]{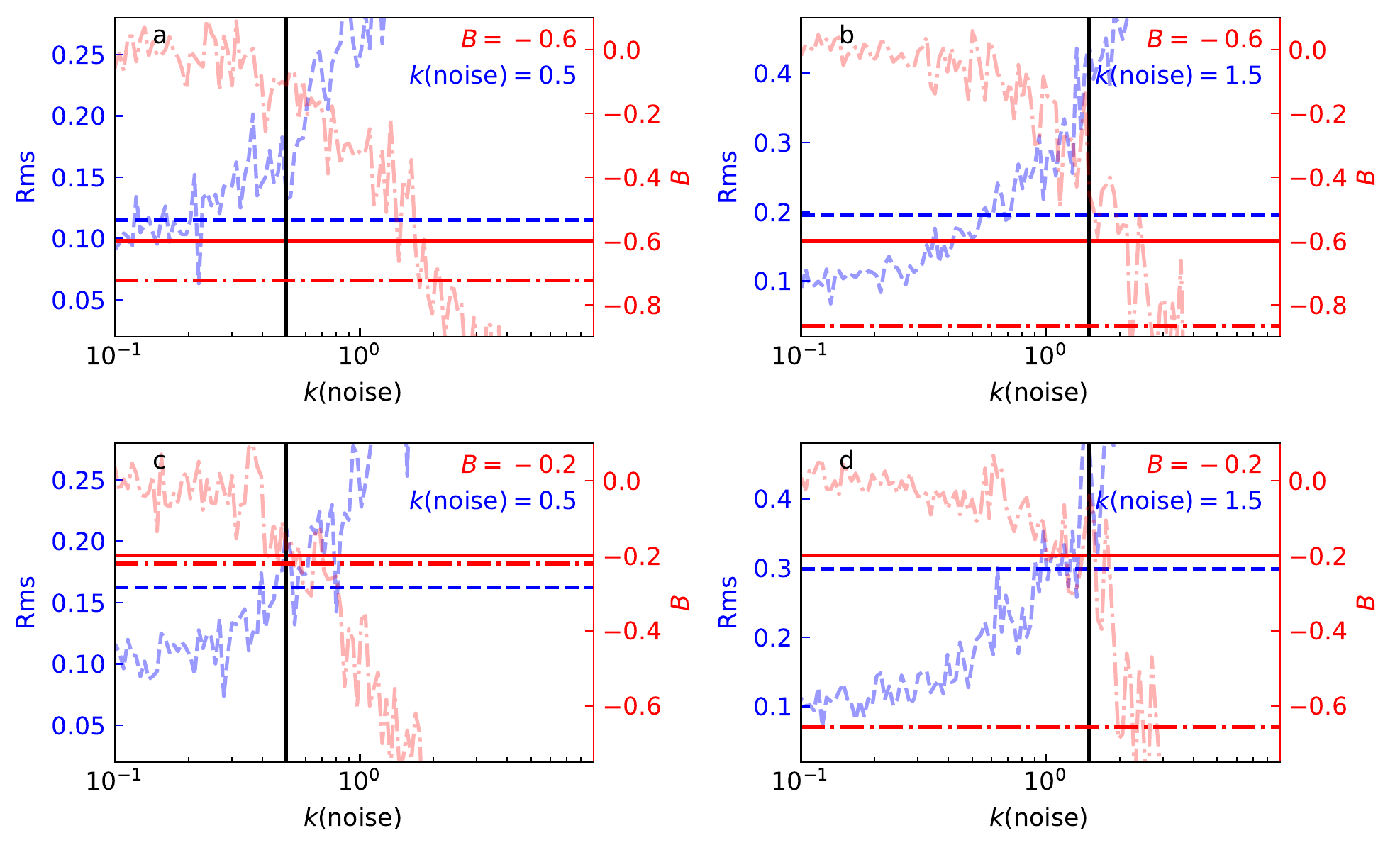}
\caption{
As Fig.~\ref{fig:simu} but allowing for a dispersion
$\sigma(\beta)=0.1$ of the intrinsic $\beta(T)$ relation.
}
\label{fig:simu2}
\end{figure*}

Figure~\ref{fig:simu} shows the results for two values of $B$ and
$k{\rm (noise)}$. The simulations are done for 100 values between
$k{\rm (noise)}$=0.1 and $k{\rm (noise)}$=2.0, which thus also
indicate the dispersion between different realisations. The left hand
frames correspond to cases where the observational noise is half of
the default. The observed rms dispersion of $\beta$ values is matched
at a point where the simulated $B$ is still significantly above the
observed value. The test thus shows that the $T-\beta$ anticorrelation
is real in the (synthetic) observations. Comparison of frames a and c
shows that the significance of the detection (of non-constant
$\beta(T)$) decreases as the intrinsic relation becomes flatter.

The frames b and d correspond to cases where the observational noise
is 50\% higher than the default values. In frame d, with $B$=-0.2, the
simulations match the observed values at the exactly the same $k{\rm
(noise)}$ value, which means that the data are compatible with a flat
$\beta(T)$ model. Only when the intrinsic relation is steeper with
$B=-0.6$ the detection of $T-\beta$ anticorrelation is still found to
be statistically significant (frame b). In such cases, one could
proceed with further simulations to quantify the $B$ value of the
intrinsic $\beta(T)$ relation. However, plots like Fig.~\ref{fig:simu}
already indicate which fraction of $B$ is caused by noise.

The recovered $k{\rm (noise)}$ value (when the rms value of 
$\beta$ values vs. the fitted $\beta(T)$ relation matches the rms
value in the observations) is a biased estimator of the photometric
errors of the observations. For example, in Fig.~\ref{fig:simu}b the
difference is a factor of 1.5 between the input value $k{\rm
(noise)}$=1.5 and the fitted value $k{\rm (noise)}\approx$1.0. The
effect is smaller for small absolute values of $B$ because the
simulations explicitly start with the $B=0$ assumption. 

Figure~\ref{fig:simu} suggests that the precision of SPIRE and SCUBA-2
observations should be sufficient to make a detection of the
anticorrelation, assuming that the default error estimates are correct
and the value of $B$ is -0.2 or lower. However, the simulations are
highly idealised. They do not take into account the possibility of
systematic or non-Gaussian errors or the effects of temperature
mixing. One of the assumptions is that all sources would follow the
same $\beta(T)$ relation. In Fig.~\ref{fig:simu2} we repeat the
previous analysis but allow a dispersion of $\sigma(\beta)=0.1$ with
respect to the $\beta(T)$ relation in both the synthetic observations
and int the subsequent simulations. This increases the noise in the
plotted relations but does not significantly affect the main features
of Fig.~\ref{fig:simu}.

\begin{table*}
\centering \caption[]{Statistics of spectral index $\beta$
distributions for different versions of the data. The listed $\beta$
values are the 25\%, 50\%, and 75\% percentiles for the clumps in the
$P(\chi^2)$=25\% samples.} 
\label{table:allbetas}
\begin{tabular}{ccccccc}
\hline\hline
$\theta_{\rm F}$\tablefootmark{a} ($\arcsec$)& Mask\tablefootmark{b} &
Aperture\tablefootmark{c} & $\beta$(3 bands, U\tablefootmark{d}) &
$\beta$(3 bands, F\tablefootmark{e})
&  $\beta$(4 bands, F) & Clumps \\
\hline
\multicolumn{3}{l}{FCF=2.41\,Jy\,(pW)$^{-1}$ arcsec$^{-2}$} &  &  &  &  \\
500 & LM & $\tau_{\rm all}$ &  1.72,  1.94,  2.07 &  1.69,  1.99,  2.29 &  1.38,  1.59,  1.86  &  27 \\
200 & LM & $\tau_{\rm all}$ &  1.73,  1.90,  2.04 &  1.75,  2.09,  2.50 &  1.53,  1.83,  2.12  &  29 \\
500 & SM & $\tau_{\rm all}$ &  1.69,  1.92,  2.04 &  1.73,  1.91,  2.22 &  1.38,  1.69,  1.84  &  17 \\
200 & SM & $\tau_{\rm all}$ &  1.74,  1.89,  2.06 &  1.83,  1.99,  2.41 &  1.51,  1.77,  1.91  &  18 \\
500 & LM & $\tau_{\rm high}$ &  1.64,  1.77,  1.93 &  1.70,  1.88,  2.17 &  1.40,  1.52,  1.65  &  22 \\
200 & LM & $\tau_{\rm high}$ &  1.62,  1.74,  1.97 &  1.88,  2.03,  2.29 &  1.41,  1.58,  1.76  &  23 \\
500 & SM & $\tau_{\rm high}$ &  1.54,  1.75,  1.93 &  1.55,  1.72,  1.92 &  1.23,  1.46,  1.61  &  14 \\
200 & SM & $\tau_{\rm high}$ &  1.68,  1.85,  1.98 &  1.92,  2.07,  2.22 &  1.31,  1.50,  1.72  &  15 \\
200 & M850 & $\tau_{\rm all}$ &  1.68,  1.82,  2.00 &  1.81,  2.04,  2.32 &  1.42,  1.67,  1.86  &  13 \\
200 & No & $\tau_{\rm all}$ &  1.61,  1.75,  1.98 &  2.12,  2.69,  3.08 &  2.11,  2.49,  2.85  &  16 \\
\multicolumn{3}{l}{FCF=2.34\,Jy\,(pW)$^{-1}$ arcsec$^{-2}$} &  &  &  &  \\
500 & LM & $\tau_{\rm all}$ &  1.69,  1.90,  2.06 &  1.69,  1.99,  2.34 &  1.43,  1.63,  1.91  &  27 \\
200 & LM & $\tau_{\rm all}$ &  1.73,  1.90,  2.04 &  1.75,  2.09,  2.50 &  1.58,  1.84,  2.12  &  29 \\
500 & SM & $\tau_{\rm all}$ &  1.63,  1.92,  2.04 &  1.73,  1.89,  2.17 &  1.48,  1.72,  1.88  &  17 \\
200 & SM & $\tau_{\rm all}$ &  1.69,  1.89,  2.06 &  1.76,  1.98,  2.40 &  1.53,  1.78,  1.96  &  18 \\
500 & LM & $\tau_{\rm high}$ &  1.64,  1.79,  1.93 &  1.70,  1.89,  2.30 &  1.44,  1.56,  1.66  &  22 \\
200 & LM & $\tau_{\rm high}$ &  1.62,  1.75,  1.99 &  1.88,  2.03,  2.38 &  1.48,  1.62,  1.78  &  23 \\
500 & SM & $\tau_{\rm high}$ &  1.53,  1.73,  1.89 &  1.54,  1.68,  1.89 &  1.28,  1.46,  1.64  &  14 \\
200 & SM & $\tau_{\rm high}$ &  1.68,  1.85,  1.98 &  1.92,  2.07,  2.22 &  1.35,  1.55,  1.75  &  15 \\
200 & M850 & $\tau_{\rm all}$ &  1.67,  1.81,  2.00 &  1.79,  2.04,  2.37 &  1.44,  1.70,  1.89  &  13 \\
200 & No & $\tau_{\rm all}$ &  1.61,  1.74,  1.97 &  2.15,  2.71,  3.09 &  2.09,  2.52,  3.00  &  16 \\
\hline
\end{tabular}
\tablefoot{
\tablefoottext{a}{Filtering scale in the SCUBA-2 map making}
\tablefoottext{b}{Small or large masks made based on {\it Herschel} data. See Appendix~\ref{sect:allmaps}}
\tablefoottext{c}{Apertures corresponding to Fellwalker clump
footprints ($\tau_{\rm all}$) or, within them, only pixels above the
median $\tau(250\mu{\rm m})$ value ($\tau_{\rm high}$)
}
\tablefoottext{d}{U refers to original SPIRE data}
\tablefoottext{e}{F refers to SPIRE maps processed through the SCUBA-2 pipeline}
}
\end{table*}

\end{appendix}

\end{document}